\documentclass[journal]{IEEEtran}  %%% remove: onecolumn,10pt!!!!!!!!!!!!!!!!!!!!!!!!!!!!!
\usepackage[utf8]{inputenc} 
\usepackage[T1]{fontenc}
\usepackage{url}
\usepackage{ifthen}
\usepackage{cite}
\usepackage[cmex10]{amsmath} % Use the [cmex10] option to ensure complicance
\usepackage{amssymb,mathtools}
\usepackage{graphicx}
\usepackage{multirow}
\usepackage{amsthm}
\usepackage{cuted}

\usepackage{array}
\newcolumntype{M}[1]{>{\centering\arraybackslash}m{#1}}
\newcolumntype{N}{@{}m{0pt}@{}}
\newcommand\Tstrut{\rule{0pt}{5ex}}       % "top" strut
\newcommand\Bstrut{\rule[-0.9ex]{0pt}{0pt}} % "bottom" strut
\newcommand{\TBstrut}{\Tstrut\Bstrut} % top&bottom struts

\usepackage{caption}
\DeclareCaptionLabelFormat{cont}{#1~#2\alph{ContinuedFloat}}
\usepackage[caption = false]{subfig}

\usepackage{framed}
\usepackage[framemethod=TikZ]{mdframed}

\usepackage{enumerate}

\usepackage{tikz}
\usetikzlibrary{shapes,snakes}
\usetikzlibrary{tikzmark}
\usetikzlibrary{fit,backgrounds}
\usetikzlibrary{matrix,decorations.pathreplacing,angles,quotes}
\usetikzlibrary{patterns}
\usetikzlibrary{calc}
\usetikzlibrary{positioning}
\tikzstyle{block} = [rectangle, draw, 
    minimum width=4em, text centered, rounded corners, minimum height=1em]
\tikzstyle{rec} = [rectangle, draw]
\tikzstyle{line} = [draw, -latex]

\tikzset{snake arrow/.style=
{
decorate,
decoration={snake,amplitude=.4mm,segment length=2mm,post length=1mm}}
}

\usetikzlibrary{intersections}

\usepackage{color}

\newtheorem{lemm}{Lemma}

\newtheorem{theo}{Theorem}
\newtheorem{prop}{Proposition}
\newtheorem{prot}{Protocol}

\newtheorem{coro}{Corollary}

\theoremstyle{definition}
\newtheorem{remark}{Remark}
\newtheorem*{remark*}{Remark}

\newcommand*{\QEDB}{\hfill\ensuremath{\square}}%

\DeclareMathOperator*{\rank}{\mathrm{rank}}
\DeclareMathOperator*{\Ima}{\mathrm{Im}}

\DeclareMathOperator*{\Tr}{\mathrm{Tr}}
\DeclareMathOperator*{\spann}{\mathrm{span}}
\DeclareMathOperator*{\supp}{\mathrm{supp}}
\DeclareMathOperator*{\tr}{\mathrm{tr}}

\DeclareMathOperator{\pr}{Pr}

\DeclareMathOperator{\cA}{{\mathcal{A}}}
\DeclareMathOperator{\cB}{{\mathcal{B}}}
\DeclareMathOperator{\cH}{{\mathcal{H}}}

\DeclarePairedDelimiterX\bkk[2]{\langle}{\rangle}{#1 \delimsize\vert #2}
\DeclarePairedDelimiterX\bk[2]{\langle}{\rangle}{#1 \delimsize\vert #1}
\DeclarePairedDelimiterX\kbb[2]{\vert}{\vert}{#1 \rangle\langle #2}
\DeclarePairedDelimiterX\kb[1]{\vert}{\vert}{#1 \rangle\langle #1}
\DeclarePairedDelimiter\px{\{}{\}}
\DeclarePairedDelimiter\paren{(}{)}

\DeclarePairedDelimiter\floor{\lfloor}{\rfloor}

\newcommand\vD{\mathsf{d}}
\newcommand\vN{\mathsf{n}}
\newcommand\vT{\mathsf{t}}
\newcommand\vF{\mathsf{f}}
\newcommand\vM{\mathsf{m}}

\newcommand\qq{q}
\newcommand\qprot{\Psi_{\mathrm{QPIR}}^{(\vM)}}
\newcommand\VV{\mathrm{V}}
\newcommand\ww{W}
\newcommand\Fq{\mathbb{F}_q}

\begin{document}

\title{Capacity of Quantum Private Information Retrieval with Colluding Servers}

%%% Several authors with up to three affiliations:
\author{%
Seunghoan Song,~\IEEEmembership{Member,~IEEE}, % <-this % stops a space
 and~Masahito~Hayashi,~\IEEEmembership{Fellow,~IEEE}% <-this % stops a space
\thanks{This article was presented in part at Proceedings of 
 2020 IEEE International Symposium on Information Theory \cite{SH20}.}% <-this % stops a space
\thanks{S. Song is with Graduate school of Mathematics, Nagoya University, Nagoya, 464-8602, Japan
(e-mail: m17021a@math.nagoya-u.ac.jp).}
\thanks{M. Hayashi is with 
Shenzhen Institute for Quantum Science and Engineering, Southern University of Science and Technology,
Shenzhen, 518055, China,
Guangdong Provincial Key Laboratory of Quantum Science and Engineering,
Southern University of Science and Technology, Shenzhen 518055, China,
Shenzhen Key Laboratory of Quantum Science and Engineering, Southern
University of Science and Technology, Shenzhen 518055, China,
and Graduate School of Mathematics, Nagoya University, Nagoya, 464-8602, Japan
(e-mail:hayashi@sustech.edu.cn).}
% \thanks{M. Hayashi is also with 
% Shenzhen Institute for Quantum Science and Engineering, Southern University of Science and Technology
% and the Centre for Quantum Technologies, National University of Singapore, Singapore.}
\thanks{
% SS is grateful to Dr. Hsuan-Yin Lin for helpful discussions and comments.
SS is supported by JSPS Grant-in-Aid for JSPS Fellows No. JP20J11484 and Lotte Foundation Scholarship.  
MH is supported in part by Guangdong Provincial Key Laboratory (Grant No. 2019B121203002),
a JSPS Grant-in-Aids for Scientific Research (A) No.17H01280 and for Scientific Research (B) No.16KT0017, and Kayamori Foundation of Information Science Advancement.
}
% \thanks{Copyright (c) 2017 IEEE. Personal use of this material is permitted.  However, permission to use this material for any other purposes must be obtained from the IEEE by sending a request to pubs-permissions@ieee.org.}
}

\maketitle

\begin{abstract}
Quantum private information retrieval (QPIR) is a protocol in which 
a user retrieves one of multiple files from $\vN$ non-communicating servers by downloading quantum systems without revealing which file is retrieved.
As variants of QPIR with stronger security requirements,
	symmetric QPIR is a protocol in which 
	{no other files than}
% 	the files except for 
	the target file are leaked to the user,
	and 
	$\vT$-private QPIR is a protocol in which the identity of the target file is kept secret even if at most $\vT$ servers may collude to reveal the identity.
The QPIR capacity is the maximum ratio of the file size to the size of downloaded quantum systems,
	and we prove that the symmetric $\vT$-private QPIR capacity is $\min\px*{1,2(\vN-\vT)/\vN}$ for any $1\leq \vT< \vN$.
We construct a capacity-achieving QPIR protocol by the stabilizer formalism
	and prove the optimality of our protocol.
The proposed capacity is greater than the classical counterpart.
\end{abstract}

\section{Introduction}

\subsection{Classical and quantum private information retrieval}

When a user retrieves information from databases, 
it is often required to protect the privacy of the user. 
Private information retrieval (PIR) is the {task in which} a user retrieves one of $\vF$ files from $\vN$ non-communicating servers without revealing to each individual server which file is downloaded.
A PIR protocol is constructed by sending queries from the user to the servers and downloading answers in the opposite direction.
Chor, Goldreich, Kushilevitz, and Sudan \cite{CGKS98} originally considered this problem for one server 
	and proved that the optimal communication complexity is linear 
% 		to the size of all files, 
		{with respect to the combined total size of all files,}
	i.e., downloading all files in the server is optimal.
They also considered a PIR protocol with multiple servers, and 
along with the following works \cite{BS03,Yekanin07,DGH12}, the communication complexity has been significantly improved.
Furthermore, 
when the protocol has multiple servers and the file size is allowed to be arbitrarily large,
Sun and Jafar \cite{SJ17} defined the PIR capacity as the maximum rate of the file size over the download size.
Although the communication complexity counts the cost of two-way communication between the user and the servers,
	the PIR capacity counts only the download size
	since the upload (query) size can be neglected by reusing the same query many times for retrieving a large file.
The method of downloading all files asserts that the PIR capacity is greater than $1/\vF$,
and the paper \cite{SJ17} proved that the PIR capacity is $(1-1/\mathsf{n})/(1-(1/\vN)^{\vF})$, which approaches $1$ as the number of servers $\vN$ increases.
PIR capacities have also been derived in many other settings \cite{CHY15,SJ17-2, SJ18, BU18, FHGHK17, KLRG17, LKRG18, TSC18, TSC18-2, WS17, WS17-2,  Tandon17,  BU19}.

Quantum PIR (QPIR) has also been studied \cite{KdW03,KdW04, Ole11,BB15, LeG12,KLLGR16, ABCGLS19, SH19, SH19-2, AHPH20} as a method to overcome the limitations of the classical PIRs, even when the file is composed of classical information.
When two-way quantum communication is allowed,
	Le Gall \cite{LeG12} proposed a one-server QPIR protocol whose communication complexity is the square root of the size of all files.
% as been a growing interest in quantum communication as a mean to overcome the limitations of existing communication technologies. Above all, there has been a great deal of interest in enhancing security and efficiency, and thus, the quantum PIR (QPIR) has also been studied \cite{KdW03,KdW04, Ole11,BB15, LeG12, ABCGLS19, SH19, SH19-2}.
Song and Hayashi \cite{SH19, SH19-2} discussed the QPIR capacity, which is the maximum ratio of the file size to the size of the downloaded quantum systems when the upload from the user to the servers is classical, the download is quantum, and the servers share entanglement before the protocol starts.
They showed that the QPIR capacity is $1$ and proposed a capacity-achieving QPIR protocol with only two servers.
Their approach clearly differs from the other papers \cite{Ole11,BB15, LeG12, KLLGR16, ABCGLS19}, which considered the one-server QPIR,
and the papers \cite{KdW03,KdW04, Ole11,BB15, LeG12, KLLGR16, ABCGLS19}, which considered the communication complexity when two-way quantum communication is allowed.

% % The original problem setting of PIR has been extended in order to guaranteeing higher security.
% There are several variants of the PIR with stronger security constraints.
% One is the PIR with server secrecy.
In PIR, the user may obtain some information on the $\vN-1$ non-targeted files.
{Therefore, it is desirable to consider PIR with server secrecy in which the user also obtains no information other than the target file.}
PIR with the server secrecy is called {\em symmetric PIR}, which is also called {\em oblivious transfer} \cite{Rabin81} in the one-server case.
In the classical case,
	the paper \cite{GIKM00} proved that shared randomness among servers is necessary to construct symmetric PIR protocols even with multiple servers.
Assuming shared randomness among servers, the paper \cite{SJ17-2} derived that symmetric PIR capacity is $1-{\vN}^{-1}$.
In the quantum case, the paper \cite{KdW04} proved that symmetric QPIR can be implemented if two-way quantum communication is available. 
% The fundamental reason that the result 
% The key 
% This result fundamentally follows from the fact that 
%         the quantum upload may generate entanglement among the user and the servers.
The paper \cite{SH19} showed that 
	symmetric QPIR is also possible with {the classical upload, quantum download, and shared entanglement,}
and derived that the symmetric QPIR capacity is $1$.

% Drawback: no-comm.  -> PIR with collusion
Furthermore, %the assumption of no communication among servers is a critical drawback of the multi-server PIR.
one critical weakness of multi-server PIR is the assumption of no communication among servers.
To relieve this assumption, PIR has also been studied in the case where some of the servers may communicate and collude to reveal the user's request.
{\em $\vT$-Private PIR} is 
% PIR with colluding servers is 
PIR %where
with stronger user secrecy, called {\em user $\vT$-secrecy} or {\em $\vT$-privacy}, in which
% with stronger user secrecy, called {\em user $\vT$-secrecy} or {\em $\vT$-privacy}, that 
the identity of the retrieved file is unknown to any collection of $\vT$ servers.
In the classical case,
	the paper \cite{SJ18} proved that the $\vT$-private PIR capacity is $(1-(\vT/\vN))/(1-(\vT/\vN)^{\vF})$ for any $1\leq \vT < \vN$.
Assuming shared randomness among servers, the paper \cite{WS17-2} derived that the symmetric $\vT$-private PIR capacity is $(\vN-\vT)/\vN$.
In the quantum case, the paper \cite{SH19-2} proved that 
the symmetric $(\vN-1)$-private QPIR capacity is $2/\vN$ for an even number of servers $\vN$, but it has not been proved for $1\leq \vT < \vN-1$.

\begin{table}[t]   
\renewcommand{\arraystretch}{1.6}
\begin{center}
%   \captionsetup{justification=centering, labelsep=newline}
\caption[caption]{Capacities of classical and quantum PIRs} \label{tab:1}
% \begin{tabular}{|c|c|c|}
{
\setlength\extrarowheight{0.0em}
\begin{tabular}{|M{7em}|M{9.5em}|M{10.5em}|N}
\cline{1-3}
                & Classical PIR Capacity & Quantum PIR Capacity  & \\
                \cline{1-3}
PIR             & $\displaystyle \frac{1-{\vN}^{-1}}{1-{\vN}^{-\vF}}$ \cite{SJ17} & $\frac{}{} 1$ ${}^{\ddagger}$ \cite{SH19} & \TBstrut \\[0.9em]
                \cline{1-3}
Symmetric PIR   & $\displaystyle {1-\frac{1}{\vN}}$   \cite{SJ17-2} ${}^{\dagger}$              & $\frac{}{} 1$ ${}^{\ddagger}$ \cite{SH19} & \TBstrut \\[0.6em]
                \cline{1-3}
$\vT$-Private PIR             & $\displaystyle \frac{1-\vT/\vN}{1-(\vT/\vN)^{\vF}}$ \cite{SJ18}& 
\multirow{2}{*}{
    \shortstack{\\ \\ \\
	$\min\px*{1, \displaystyle\frac{2(\vN-\vT)}{\vN}}$ ${}^{\ddagger}$\\[0.3em]{[This paper]}
%                    $\displaystyle\frac{2(\vN-\vT)}{\vN}$ 
%                    for $\vN/2< \vT $ ${}^{\ddagger}$} 
			}
}
                  &\TBstrut\\[0.8em]
                \cline{1-2}
$\vT$-Private symmetric PIR      & $\displaystyle \frac{\vN-\vT}{\vN}$   \cite{WS17-2} ${}^{\dagger}$              
            &   
                &\\[0.6em]
                \cline{1-3}               
\multicolumn{3}{l}{\footnotesize $\ast$ $\vN$, $\vF$: the numbers of servers and files, respectively.}\\[-0.6em]
\multicolumn{3}{l}{\footnotesize $\dagger$ Shared randomness among servers is necessary.}\\[-0.6em]
% \multicolumn{3}{l}{\scriptsize $\ddagger$ Files are coded by $(\vN,\vkc)$ maximum distance separable code.}\\[-0.8em]
\multicolumn{3}{l}{\footnotesize ${\ddagger}$ Capacities are derived with the strong converse bounds.}\\%[-1em]
\end{tabular}
}
\end{center}
\end{table}

\subsection{Contribution}

As a generalization of the QPIR capacities in \cite{SH19,SH19-2}, we derive the symmetric $\vT$-private QPIR capacity for any $\vT$ less than the number of servers $\vN$.
Similar to the papers \cite{SH19} and \cite{SH19-2}, 
we define the QPIR model as follows:
    a user retrieves one of $\vF$ files from $\vN$ non-communicating servers, each of which contains the whole classical file set;
    the servers share an entangled state before the protocol starts;
    the user uploads classical queries and downloads quantum systems; 
    and the user decodes the target file by quantum measurement.
The capacity is defined with four security parameters: error probability, server secrecy, user $\vT$-secrecy, and upload cost.
As a main result, we prove that the symmetric $\vT$-private QPIR capacity is $\min\{1,2(\vN-\vT)/\vN\}$ for $1\leq \vT < \vN$.
Especially, for $1\leq \vT \leq \vN/2$, the capacity is $1$ 
	{even if we require}
% 	regardless of requiring 
	the strongest security condition in which the protocol has zero-error, perfect user $\vT$-secrecy, and perfect server secrecy.
% 			,
%          or requiring the weakest security condition that the protocol has arbitrary error less than $1$, no user $\vT$-secrecy, and no server secrecy.
For the proof, we construct the capacity-achieving protocol by stabilizer formalism
	and present the tight upper bounds of the QPIR capacity for $1\le \vT \le \vN/2$ and $\vN/2 <  \vT < \vN$, respectively.

The derived quantum capacity is strictly greater than the classical symmetric $\vT$-private PIR capacity $(\vN-\vT)/\vN$ in \cite{WS17-2} (see Table \ref{tab:1}), and when more than half of the servers collude (i.e., $\vN/2 \le \vT$), the derived quantum capacity is exactly twice the classical capacity.
In addition, compared to the classical $\vT$-private PIR capacity $(1-\vT/\vN)/(1-(\vT/\vN)^{\vF})$ \cite{SJ18},
	our quantum capacity is greater when $\vT< \vN/2$ 
	or 
	$(\vN/\vT)^{\vF} >  2$, where the latter inequality is satisfied when the number of files $\vF$ is large enough.

Our result implies that 
	symmetric $\floor*{\vN/2}$-private QPIR can be constructed without sacrificing any communication efficiency
since the capacity is $1$ for $1\leq \vT \leq \vN/2$.
Moreover, QPIR with more servers may obtain stronger secrecy against collusion.
This result contrasts the result \cite{SH19} that 
	symmetric ($1$-private) QPIR has no benefit of increasing the number of servers
	because a two-server protocol achieves the capacity for $1$-private QPIR.

\subsection{Outline of protocol construction}

	Our protocol is constructed by the stabilizer formalism.
	Given a subspace $\VV$ of an even-dimensional vector space over a finite field $\Fq$, let $\VV^{\perp_J}$ be its orthogonal space with respect to the symplectic bilinear form.
	In the stabilizer formalism, 
	the stabilizer is described by a subspace $\VV$ such that $\VV\subset \VV^{\perp_J}$,
	and
	the state is prepared on the stabilized subspace.
	When the Weyl operator $\mathbf{\tilde{W}(s,t) \coloneqq X(s)Z(t)}$ with vectors $\mathbf{s,t}$ is applied,
	an appropriate quantum measurement in the decoding process outputs the outcome $\mathbf{(s,t)}+\VV^{\perp_J}$, 
	which is partial information of the Weyl operator.
% 	an appropriate quantum measurement outputs the outcome $(\mathbf{s,t})+\VV^{\perp_J}$, i.e., 
% 	some information of the Weyl operator is outputted.
	With this fact, 
	we design our QPIR protocol so that
	the servers share an entangled state on the stabilized subspace,
	the servers apply Weyl operators depending on the queries and files,
	and the user performs the measurement to obtain the target file.
% 	by the correlation between the target file and the servers' Weyl operators. 
	Here, for guaranteeing security, we choose the subspace $\VV$ and the queries to satisfy the following three conditions.
	\begin{enumerate}[1)]
		\item  The queries to any $\vT$ servers are independent of the user's request (for user secrecy).
		\item  When the Weyl operator applied by the servers is $\mathbf{\tilde{W}(s,t)}$ on the whole composite system,
		the target file information has a one-to-one correspondence with the value $\mathbf{(s,t)}+\VV^{\perp_J}$ (for correctness).
		\item  The information of other files is in $\VV^{\perp_J}$ (for server secrecy).
	\end{enumerate}
% 	Since the information in $\VV^{\perp_J}$ does not affect the measurement outcome $\mathbf{(s,t)}+\VV^{\perp_J}$
% 	and the final state is also the completely mixed state, the user obtains no information of the other files.
% % 		and 
% % 	In this case, the measurement outcome has only the information of $(s,t)$ which contains the target file.
The main difficulty of the protocol construction is to find an appropriate vector space $\VV$ which satisfies the properties 1), 2), and 3).
The problem reduces to the search of a symplectic matrix with a linear independence condition in row vectors and a symplectic orthogonality condition in column vectors.
We concretely construct a symplectic matrix satisfying those conditions. %in Section~\ref{sec:prelim}.
% 	by algebraic extension of the finite fields.

% 
% In addition, to implement the secrecy conditions in PIR, 
% 	a symplectic matrix with properties are necessary to be found, which generates a complicated linear algebra problem on finite fields.

Our QPIR protocol utilizes a multipartite entangled state whereas the QPIR protocols in \cite{SH19,SH19-2} are constructed from two well-known bipartite protocols, dense coding \cite{BW92} and quantum teleportation \cite{BBCJPW93}.
% 	which is defined by stabilizer formalism, 
The paper \cite{SH19} constructed a simple QPIR protocol by modifying the dense coding protocol,
and the paper \cite{SH19-2} constructed an $(\vN-1)$-private QPIR protocol by combining quantum teleportation and dense coding.
In contrast, our QPIR protocol utilizes a multipartite version of dense coding under the stabilizer formalism.
The proposed protocol includes the protocol in \cite{SH19} as an example of symmetric $1$-private QPIR protocols.

\subsection{Organization of paper}

The remainder of the paper is organized as follows.
Section~\ref{sec:quantum_basic} reviews the fundamentals of quantum information theory.
Section~\ref{sec:model} formally describes the QPIR protocol, defines the QPIR capacity, and presents the main result of the paper.
Section~\ref{sec:prelim} is the preliminary section for protocol construction.
	In this section, we define the stabilizer formalism,
	present a communication protocol for classical messages by stabilizer formalism,
	and
	give a fundamental lemma for protocol construction.  %on finite field linear algebra.
Section~\ref{sec:protocol} constructs the capacity-achieving symmetric $\vT$-private QPIR protocol.
The proposed protocol has no error, perfect user secrecy, and perfect server secrecy.
% Section~\ref{sec:analysis} analyzes the performance of the protocol.
Section~\ref{sec:converse} presents the converse bounds of the capacity result.
We present three upper bounds of the capacity depending on the number of colluding servers and the security parameters.
% , and presents a stabilizer state which generates $\floor{\vN/2}$-private QPIR protocol.
Section~\ref{sec:conclusion} is the conclusion of the paper.

\subsubsection*{Terms and notations}
\textit{
The matrix $I_{n}$ denotes the $n\times n$ identity matrix.
For a quantum system $\cH$, $I_{\cH}$ denotes identity matrix on $\cH$.
For a random variable $X$, $\pr_X [ f(X) ]$ is the probability that $X$ satisfies the condition $f(X)$.
%The quantum information measures used in the paper are reviewed in Appendix~\ref{sec:measures}.
% The $(\vN-1)$-colluded PIR is simply called {\em colluded PIR} because we only treat $(\vN-1)$-colluded PIR in the remainder of the paper.
% The operation $\oplus$ for two sets is the symmetric difference of the two sets: Given two sets $A,B$, $A\oplus B := A\backslash B \cup B\backslash A$.
% % For pure states on composite system, $|i,i\rangle := | i\rangle \otimes |i \rangle$.
}

\section{Fundamentals of Quantum Information Theory} \label{sec:quantum_basic}

%  I recommend having a brief introduction on their quantum notation (how they indicate Hilbert spaces, density operators, etc). 
% In general it would be useful to indicate on what system each density operator is acting (e.g., using ρA for a density operator on A).
% Especially in their converse argument, they do not need a new notation for partial traces.

In this section, we briefly introduces the fundamentals of quantum information theory.
More detailed introduction can be found at \cite{NC00, Hay17}.

A quantum system is a Hilbert space $\cH$.
	Throughout this paper, we only consider finite dimensional Hilbert spaces.
A quantum state is defined by a {\em density matrix}, which is a Hermitian matrix $\rho$ on $\cH$ such that 
\begin{align}
\rho \geq 0, \quad \Tr \rho  = 1.
\end{align}
The set of states on $\cH$ is written as $\mathcal{S}(\cH)$.
A state $\rho$ is called a {\em pure state} if $\rank \rho = 1$, which can also be described by a unit vector of $\cH$.
If a state $\rho$ is not a pure state, it is called a {\em mixed state}.
The state $\rho_{\mathrm{mix}} \coloneqq  I_{\cH} / \dim \cH$ is called the completely mixed state.
% 	and any pure state is written as a unit vector of $\cH$.
The composite system of two quantum systems $\cA$ and $\cB$ is given as the tensor product of the systems $\cA\otimes \cB$.
%If there is no confusion, we simply denoted $\cA\cB \coloneqq \cA\otimes \cB$.
For a state $\rho\in \mathcal{S}(\cA\otimes \cB)$, 
	the {\em reduced state} on $\cA$ is written as 
	\begin{align}
	\rho_{\cA} = \Tr_{\cB} \rho,
	\end{align}
where $\Tr_{\cB}$ is the partial trace with respect to the system $\cB$.
A state $\rho \in \mathcal{S}(\cA\otimes \cB)$ is called a {\em separable state} if 
	$\rho$ is written as 
	\begin{align}
	\rho = \sum_i p_i \rho_{\cA,i} \otimes \rho_{\cB,i},
	\end{align}
	for some distribution $p = \{p_i\}$ and states $\rho_{\cA,i} \in \mathcal{S}(\cA)$, $\rho_{\cB,i}\in \mathcal{S}(\cB)$.
A state $\rho \in \mathcal{S}(\cA\otimes \cB)$ is called an {\em entangled state} if it is not separable.
A quantum measurement is defined by a {\em positive-operator valued measure (POVM)}, which is the set of Hermitian matrices $\{M_{x}\}_{x\in\mathcal{X}}$ on $\cH$ such that 
\begin{align}
M_x \geq 0, \quad \sum_{x \in \mathcal{X}}  M_x = I_{\cH}.
\end{align}
When the elements of POVM are orthogonal projections, i.e., $M_x^2 = M_x$ and $M_x^\dagger = M_x$,
	we call the POVM a {\em projection-valued measure (PVM)}.
A quantum operation is described by a {\em trace-preserving completely positive (TP-CP) map} $\kappa$, which is a linear map such that 
	\begin{align}
	\Tr \kappa(\rho) &= 1 \quad \forall \rho \in \mathcal{S}(\cH), \\
	\kappa \otimes \iota_{\mathbb{C}^{d}} (\rho) &\geq 0 \quad \forall \rho \in \mathcal{S}(\cH\otimes \mathbb{C}^d), \ \forall d \geq 1,
	\end{align}
	where $\iota_{\mathbb{C}^{d}}$ is the identity map on $\mathbb{C}^d$.
An example of TP-CP maps is the unitary map, which is defined by $\kappa_U (\rho) = U \rho U^*$ for a unitary matrix $U$. 
% If a unitary map $\kappa_U$ is applied to the pure state $|\psi\rangle\langle \psi|$,

Next, we define the information measures used in this paper.
For a state $\rho$, %= \sum_x p_x |\psi_x\rangle \langle \psi_x|$, 
the von Neumann entropy is defined by $H(\rho) \coloneqq -\rho \log \rho $.
%For a composite system $\cA\otimes \cB$, we denote ${\cal AB} \coloneqq {\cal A} \otimes {\cal B}$.
%For a state $\rho$ on ${\cal A \otimes B \otimes C}$ and $\mathcal{X}\in\{{\cal A,B,C,AB,BC,CA,ABC}\}$,
%	we denote by $\rho_{\cal X}$ the reduced state on ${\cal X}$
%	and $H({\cal X})_{\rho} \coloneqq H(\rho_{\cal X})$.
%When $\rho \in \mathcal{S}(\cA)$, we denote $H(\cA)_{\rho}\coloneqq H(\rho)$ 
%and 
For $\rho \in \mathcal{S}(\cA \otimes \cB)$, we also denote 
\begin{align}
	H(\cA\cB)_{\rho} &\coloneqq H(\rho),\\
	H(\cA)_{\rho} &\coloneqq H(\rho_{\cA}), \\
	H(\cB)_{\rho} &\coloneqq H(\rho_{\cB}),
\end{align}
%	for emphasizing the quantum system
where $\rho_{\cA}$ and $\rho_{\cB}$ are the reduced states on $\cA$ and $\cB$, respectively.
For $\rho\in\mathcal{S}(\cA \otimes \cB \otimes \mathcal{C})$,
	the conditional entropy, mutual information, and conditional mutual information are defined by 
\begin{align}
H({\cal A|B})_{\rho} &\coloneqq H({\cal AB})_{\rho}- H({\cal B})_{\rho},\\
I({\cal A;B})_{\rho} &\coloneqq H({\cal A})_{\rho} - H({\cal A|B})_{\rho},\\
I({\cal A;B|C})_{\rho} &\coloneqq H({\cal A|C})_{\rho} - H({\cal A|BC})_{\rho}.
\end{align}
Let $X$ be a random variable with values in $\mathcal{X}$ and a probability distribution $p_X = \{ p_x \mid x \in \mathcal{X} \}$.
When a state $\rho(X)$ on ${\cal A}$ depends on the value of $X$,
	the joint state is written as $\tilde{\rho}(X) \coloneqq \sum_x p_x |x\rangle \langle x | \otimes \rho(x)$, which is called a classical-quantum state.
% For a quantity $C$,
% and denote $$.
For simplicity, we denote
\begin{align}
H(\cdot )_{\rho(X)} &\coloneqq H(\cdot)_{\tilde{\rho}(X)},\\
H(\cdot|\cdot)_{\rho(X)} &\coloneqq H(\cdot|\cdot)_{\tilde{\rho}(X)} ,\\
I(\cdot;\cdot)_{\rho(X)} &\coloneqq I(\cdot;\cdot)_{\tilde{\rho}(X)} ,\\
I(\cdot;\cdot|\cdot)_{\rho(X)} &\coloneqq I(\cdot;\cdot|\cdot)_{\tilde{\rho}(X)}.
\end{align}
For two distributions $p = \{p_x \mid x\in\mathcal{X} \}$ and $q = \{ q_x \mid x\in\mathcal{X} \}$,
	the classical relative entropy is defined as
\begin{align*}
	&D(p \| q) \coloneqq  
	\begin{cases}
	\displaystyle \sum_{ x\in\mathcal{X} }  p_x \log \frac{p_x}{q_x}	& \text{if }\supp(p) \subset \supp(q)\\
	\infty									& \text{otherwise}
	\end{cases},
	%\label{def:crel}
\end{align*}
	where $\supp(p) \coloneqq \{ x\in \mathcal{X} \mid p_x \neq 0\}$.
For two states $\rho$ and $\sigma$ on $\cH$, 
	the quantum relative entropy is defined as
\begin{align*}
	D(\rho \| \sigma) &\coloneqq  
	\begin{cases}
	\Tr \rho (\log \rho - \log \sigma) &\text{if }  \supp(\rho) \subset \supp(\sigma)\\
	\infty 		&\text{otherwise}
	\end{cases},
	%\label{def:qrel}
\end{align*}
	where $\supp(\rho) \coloneqq \{ |x\rangle \in \cH \mid \rho |x\rangle \neq 0 \}$.

\section{QPIR Model and Main Result} \label{sec:model}
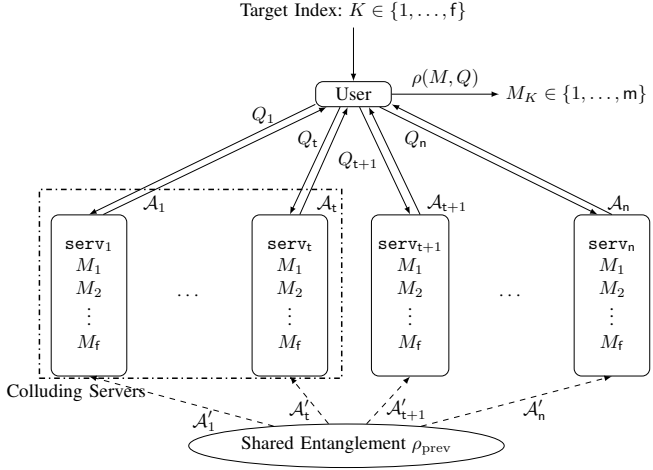
\begin{figure}[t]
\begin{center}
        \resizebox {1\linewidth} {!} {
\begin{tikzpicture}[node distance = 3.3cm, every text node part/.style={align=center}, auto]
    \node [block] (user) {User};
    \node [above=1cm of user] (sent) {Target Index: $K\in\{1,\ldots,\vF\}$};
    \node [block,minimum height = 3cm, below left=2cm and 3.5cm of user] (serv1) {$\mathtt{serv}_1$\\$M_1$\\$M_2$\\\vdots\\$M_{\vF}$};
    \node [right=0.8cm of serv1] (ten) {$\cdots$};
    \node [block,minimum height = 3cm, right=0.8cm of ten] (serv2) {$\mathtt{serv}_{\vT}$\\$M_1$\\$M_2$\\\vdots\\$M_{\vF}$};
    \node [block,minimum height = 3cm, right=0.8cm of serv2] (serv3) {$\mathtt{serv}_{\vT+1}$\\$M_1$\\$M_2$\\\vdots\\$M_{\vF}$};
    \node [right=0.8cm of serv3] (ten2) {$\cdots$};
    \node [block,minimum height = 3cm, right=0.8cm of ten2] (servn) {$\mathtt{serv}_{\vN}$\\$M_1$\\$M_2$\\\vdots\\$M_{\vF}$};
    
    \node [draw,ellipse,below right=1cm and 2cm of serv1] (shared) {Shared Entanglement $\rho_{\mathrm{prev}}$};
    
    \node [right=2cm of user] (receiv) {$M_K \in \{1,\ldots,\vM\}$};
    
    \node [thick, rec, above, above right= -5em and -9em of ten, minimum width=16em, minimum height=10em,dashdotted] (attack) {};
    \node [below left= -9em and -6em of attack] (text_collude) {\\\\\\\\\\\\\\\\Colluding Servers};
%     \tikzstyle{block} = [rectangle, draw, 
%     minimum width=4em, text centered, rounded corners, minimum height=1em]
    
    % Nraw edges
    \path [line] (sent) -- (user);
    
    \path [line] (user.195) --node[pos=0.1,left=2mm] {$Q_1$} (serv1.north);
    \path [line] (user.224) --node[pos=0.3,left] {$Q_{\vT}$} (serv2.north);
    \path [line] (user) --node[pos=0.5,left] {$Q_{\vT+1}$} (serv3.north);
    \path [line] (user) --node[pos=0.3,left=2mm] {$Q_{\vN}$} (servn.100);

    \path [line] (serv1.80)--node[pos=0.1,right=2mm] {$\cA_1$} (user);
    \path [line] (serv2.83) --node[pos=0.1,right=1mm] {$\cA_{\vT}$} (user);
    \path [line] (serv3.83) --node[pos=0.1,right=1mm] {$\cA_{\vT+1}$} (user.320);
    \path [line] (servn.north) --node[pos=0.1,right=2mm] {$\cA_{\vN}$} (user.345);
    
    \path [line] (user.east) -- node{$\rho(M,Q)$} (receiv);
    
    \path [line,dashed] (shared) --node[pos=0.1,left=6mm] {$\cA_{1}'$} (serv1.south);
    \path [line,dashed] (shared) --node[pos=0.3,left=0mm] {$\cA_{\vT}'$} (serv2.south);
    \path [line,dashed] (shared) --node[pos=0.3,right=0mm] {$\cA_{\vT+1}'$} (serv3.south);
    \path [line,dashed] (shared) --node[pos=0.33,right=6mm] {$\cA_{\vN}'$} (servn.south);
\end{tikzpicture}
}
\caption{Quantum private information retrieval protocol, where $\vT$ servers collude to know the target index $K$.
The user does not know which $\vT$ servers collude.}\label{fig:protocol_model}
\end{center}
\end{figure}

In this section, we formally state the definition of the QPIR protocol, which is illustrated in Fig.~\ref{fig:protocol_model}, and present the main result of the paper. 
Since the protocol is identically defined as in \cite{SH19,SH19-2}, we review the formal description of the QPIR protocol given in \cite{SH19,SH19-2}.
Then we define the security measures and the capacity of $\vT$-private QPIR, and state the main result.

\subsection{Formal definition of a QPIR protocol} \label{sec:foramal}

The task of $\vT$-private QPIR is described as follows.
%There are $\vN$ servers and a user.
The files $M_1,\ldots, M_{\vF}\in\{1,\ldots, \vM\}$ are uniformly and independently distributed.
Each of $\vN$ servers $\mathtt{serv}_1$, \ldots , $\mathtt{serv}_{\vN}$ contains
	a copy of all files $M \coloneqq (M_1,\ldots, M_{\vF})$.
The $\vN$ servers are assumed to share an entangled state.
A user chooses a file index $K \in\{1,\ldots, \vF\}$ uniformly and independently of $M$ 
	in order to retrieve the file $M_K$. 
The task of $\vT$-private QPIR is to construct a protocol that allows the user to retrieve $M_K$ without revealing $K$ to any collection of $\vT$ servers.
%The user wants to retrieve the $k$-th file $M_k$,
%	without revealing $k$ to any collection of $\vT$ servers.
% For the secrecy, the 
% For QPIR task, is 
%The QPIR task is achieved by a QPIR protocol, which is formally described by 
%	the shared entangled state, user encoder, server encoder, and decoder.
%To discuss the optimality of the class of QPIR protocols, 
%Since this paper discusses the optimality of QPIR protocols, 
In the following, we give the formal definition of the QPIR protocol. % before the concrete construction of the protocol.
The concrete construction of our QPIR protocol will be given in Section~\ref{sec:protocol}.

A QPIR protocol is formally defined as follows.
%\noindent\textbf{Shared entanglement}:\quad 
Let $\cA_1',\ldots, \cA_{\vN}'$ be $\vD'$-dimensional Hilbert spaces.
The state of the quantum system $\cA_1' \otimes \cdots \otimes \cA_{\vN}'$ is initialized as $\rho_{\mathrm{prev}}$ 
	and is distributed so that the $s$-th server $\mathtt{serv}_s$ contains $\cA_s'$. 
%\noindent\textbf{User encoder}:\quad 
The user %chooses a randomness $R_{\mathrm{user}}$ from a set $\mathcal{R}_{\mathrm{user}}$ and 
	randomly encodes the index $K$ to classical queries $Q_1,\ldots,Q_{\vN}$, i.e.,
% The user encodes $K$ by the user encoder:
\begin{align*}
\mathsf{Enc}_{\mathrm{user}} (K) = Q =  (Q_1,\ldots,Q_{\vN})  \in \mathcal{Q}_1\times\cdots\times \mathcal{Q}_{\vN},
\end{align*}
where $\mathcal{Q}_1,\ldots, \mathcal{Q}_{\vN}$ are finite sets.
% The user chooses a randomness $R_{\mathrm{user}}$ from a set $\mathcal{R}_{\mathrm{user}}$ and 
% encodes the index $K$ to queries $Q_1,\ldots,Q_{\vN}$, i.e.,
% % The user encodes $K$ by the user encoder:
% \begin{align*}
% \mathsf{Enc}_{\mathrm{user}} (K,R_{\mathrm{user}}) = Q =  (Q_1,\ldots,Q_{\vN})  \in \mathcal{Q}_1\times\cdots\times \mathcal{Q}_{\vN},
% \end{align*}
% where $\mathcal{Q}_1,\ldots, \mathcal{Q}_{\vN}$ are finite sets.
Then, the user sends $Q_s$ to the $s$-th server $\mathtt{serv}_s$ ($s=1,\ldots, \vN$).
%\noindent\textbf{Server encoder}:\quad
Let $\cA_1,\ldots, \cA_{\vN}$ be $\vD$-dimensional Hilbert spaces and $\cA \coloneqq\cA_1\otimes \cdots \otimes \cA_{\vN}$.
After receiving the query $Q_s$, the $s$-th server $\mathtt{serv}_s$ constructs a TP-CP map $\Lambda_s$ from $\cA_s'$ to $\cA_s$ by the server encoder $\mathsf{Enc}_{\mathrm{serv}_s}$ as
% encodes the query $Q_s$ 
\begin{align*}
\mathsf{Enc}_{\mathrm{serv}_s} (Q_s, M) =  \Lambda_s,
\end{align*}
applies $\Lambda_s$, and sends $\cA_s$ to the user.
Then, the state on $\cA_1\otimes \cdots \otimes \cA_{\vN}$ is written as
\begin{align*}
\rho(M,Q) \coloneqq \Lambda_1\otimes\cdots\otimes \Lambda_{\vN} (\rho_{\mathrm{prev}}) .
\end{align*}
%\noindent\textbf{Decoder}:\quad
The user decodes the received state by a decoder, which is defined as a POVM
$\mathsf{Dec}(K,Q) \coloneqq \{ {Y}_{KQ}(w) \mid  w\in\{1,\ldots,\vM\} \}$ on $\cA_1\otimes \cdots \otimes \cA_{\vN}$
depending on the variables $K$ and $Q$. 
%That is, the positive-semidefinite operators $Y_{KQ}(w)$ satisfy the condition $\sum_{w \in \{1,\ldots, \vM\}} Y_{KQ}(w)=I$. 
% , where the measurement outcome associated with $Y_{KQ}(w)$ is $w$.
% \sum_{m_k=0}^{\vM-1} {Y}_{KQ} (w) = {I}, \
The user outputs the measurement outcome $W$ as the retrieval result.

%Given the numbers of servers $\vN$ and files $\vF$,
%	a QPIR protocol is formally described by the four-tuple 
%$$\Psi_{\mathrm{QPIR}}^{(\vM)} \coloneqq (\rho_{\mathrm{prev}}, \mathsf{Enc}_{\mathrm{user}}, \mathsf{Enc}_{\mathrm{serv}}, \mathsf{Dec})$$
%of the shared entangled state, user encoder, server encoder, and decoder.
Given the numbers of servers $\vN$ and files $\vF$,
	a QPIR protocol of the file size $\vM$ is described by 
$$\Psi_{\mathrm{QPIR}}^{(\vM)} \coloneqq (\rho_{\mathrm{prev}}, \mathsf{Enc}_{\mathrm{user}}, \mathsf{Enc}_{\mathrm{serv}}, \mathsf{Dec})$$
of the shared entangled state, user encoder, server encoder, and decoder,
where $\mathsf{Enc}_{\mathrm{serv}} \coloneqq (\mathsf{Enc}_{\mathrm{serv}_{1}}, \ldots,\mathsf{Enc}_{\mathrm{serv}_{\vN}})$.
The upload cost, the download cost, and the rate of a QPIR protocol $\qprot$ are defined by 
\begin{align*}
U(\qprot) &\coloneqq \prod_{s=1}^{\vN} |\mathcal{Q}_s| , \\
D(\qprot) &\coloneqq \dim \bigotimes_{s=1}^{\vN} \cA_s =  \vD^\vN, \\
R(\qprot) &\coloneqq \frac{\log\vM}{\log D(\qprot)} = \frac{\log\vM}{\vN\log \vD} .
\end{align*}

%\subsection{Security parameters and capacity of $\vT$-private QPIR}
%

\subsection{Security measures of $\vT$-private QPIR protocol} \label{sec:securityme}

In this paper, we consider the security model that the servers and the user do not deviate from the protocol except for the collusion of servers.
%The collusion of servers ar
	In $\vT$-private QPIR for $\vT \in \{1,\ldots, \vN-1\}$, at most $\vT$ servers may collect their queries to reveal which file is requested by the user, and the user does not know which servers are colluding.
	We also consider the server secrecy in which the user only obtains the requested file but no information of the other files.

	We evaluate the security of a $\vT$-private QPIR protocol $\Phi_{\mathrm{QPIR}}^{(\vM)}$ by the following security measures.
Let
% $\mathcal{Q} \coloneqq \mathcal{Q}_1\times \cdots \times \mathcal{Q}_{\vN} $ be the set of queries,
$W\in\{1,\ldots,\vM\}$ be the protocol output, 
$M_k^c$ be the collection of all files except for $M_k$,
$\mathsf{S}_\vN$ be the symmetric group of $\{1,\ldots, \vN\}$, i.e., the set of all permutations on $\{1,\ldots, \vN \}$,
and $Q_{\pi,\vT} \coloneqq (Q_{\pi(1)},\ldots, Q_{\pi(\vT)})$ for $\pi\in\mathsf{S}_{\vN}$.
The error probability, server secrecy, and user $\vT$-secrecy are defined as
\begin{align*}
P_{\mathrm{err}}(\Psi_{\mathrm{QPIR}}^{(\vM)}) 
&\coloneqq 
    \max_{(\ast)}
%     \max_{ \substack{m=(m_1,\ldots,m_{\vN}),\\  q\in\mathcal{Q}_1\times \cdots \times \mathcal{Q}_{\vN} \\ k\in \{1,\ldots,\vF\} } } 
    \pr [\ww \neq m_k | M = m, Q=q, K=k]                  
% &=   \max_{ \substack{m=(m_1,\ldots,m_{\vN}),\\q\in\mathcal{Q}_1\times \cdots \times \mathcal{Q}_{\vN}  \\  k\in \{1,\ldots,\vF\} } } 1- \Tr \rho_{mq} Y_{k,q}(m_k) ,
    %\label{eq:error_probab2}
    \\
S_{\mathrm{serv}}(\Psi_{\mathrm{QPIR}}^{(\vM)}) &\coloneqq 
    \max_{(\ast)} 
    I(M_k^c; \cA|Q=q , K=k)_{\rho(M,q)}
    %\label{eq:serv_sec2}
    \\
S_{\mathrm{user}}^{(\vT)}(\Psi_{\mathrm{QPIR}}^{(\vM)}) &\coloneqq 
    \max_{\pi\in\mathsf{S}_{\vN}} I(K; Q_{\pi,\vT} ), 
    %\label{eq:user_sec2} 
\end{align*}
where 
the maximum $(\ast)$ is taken for all
$m=(m_1,\ldots,m_{\vF}),  q\in\mathcal{Q}_1\times \cdots \times \mathcal{Q}_{\vN},  k\in \{1,\ldots,\vF\}$ such that $\Pr[Q=q,K=k]\neq 0$.

The error probability $P_{\mathrm{err}}(\Psi_{\mathrm{QPIR}}^{(\vM)})$ is the worst-case probability that the protocol output $W$ is not the target file of the user.
%The server secrecy is the property that the servers' information of the non-targeted files are kept secret from the user.
%That is, 
The server secrecy $S_{\mathrm{serv}}(\Psi_{\mathrm{QPIR}}^{(\vM)})$ measures the independence between the non-targeted files and the quantum systems $\cA$ that the user obtains.
If $S_{\mathrm{serv}}(\Psi_{\mathrm{QPIR}}^{(\vM)}) = 0$, the user obtains no information of the non-targeted files.
The user $\vT$-secrecy $S_{\mathrm{user}}^{(\vT)}(\Psi_{\mathrm{QPIR}}^{(\vM)})$ is defined as the mutual information between the target index $K$ and the queries to any $\vT$ servers $Q_{\pi,\vT}$.
If $S_{\mathrm{user}}^{(\vT)}(\Psi_{\mathrm{QPIR}}^{(\vM)}) = 0$, 
	any $\vT$ servers obtain no information of $K$ even if they collect their queries.
	%$\mathtt{serv}_s$ only obtains $Q_s$ but not the other information.
% The three security figures are defined for the worst-case of all files $m$, queries $q$, and the target index $k$.
These security measures are defined for the worst-case of all files $m$, queries $q$, and the target index $k$.
A protocol $\Psi_{\mathrm{QPIR}}^{(\vM)}$ is a $\vT$-private QPIR protocol with perfect security if $P_{\mathrm{err}}(\Psi_{\mathrm{QPIR}}^{(\vM)}) = S_{\mathrm{user}}^{(\vT)}(\Psi_{\mathrm{QPIR}}^{(\vM)}) = 0$
and a symmetric $\vT$-private QPIR protocol with perfect security if
%$S_{\mathrm{serv}}(\Psi_{\mathrm{QPIR}}^{(\vM)}) = 0$ additionally.
$P_{\mathrm{err}}(\Psi_{\mathrm{QPIR}}^{(\vM)}) =S_{\mathrm{serv}}(\Psi_{\mathrm{QPIR}}^{(\vM)})=  S_{\mathrm{user}}^{(\vT)}(\Psi_{\mathrm{QPIR}}^{(\vM)}) = 0$.

\begin{remark} \label{rem:dsf1}
The server secrecy is also written as
\begin{align}
S_{\mathrm{serv}}(\Psi_{\mathrm{QPIR}}^{(\vM)}) 
    = \max_{(\ast)} 
    I(M_k^c; \cA|Q=q)_{\rho(M,q)}.
    \label{eq:newserv}
\end{align}
This equation follows from $I(M_k^c;\cA|Q=q)_{\rho(M,q)}=I(M_k^c;\cA|Q=q,K=k)_{\rho(M,q)}$ 
which is derived from the independence between $K$ and $(M_k^c, \cA)$ when $Q=q$ is fixed.
\end{remark}

\begin{remark}
When a QPIR protocol $\Psi_{\mathrm{QPIR}}^{(\vM)}$ satisfies $P_{\mathrm{err}}(\Psi_{\mathrm{QPIR}}^{(\vM)}) \leq \alpha$ and 
    $S_{\mathrm{serv}}(\Psi_{\mathrm{QPIR}}^{(\vM)}) \leq \beta$ for sufficiently small $\alpha,\beta \geq 0$,
the condition 
	$\Pr[K=k,Q=q] \neq 0$ 
	%$p_{K=k,Q=q} \neq 0$ 
	implies 
	$\Pr[K=i,Q=q] = 0$ for any $k\neq i \in \{1,\ldots, \vF\}$.
Otherwise, we derive a contradiction as follows.
If $\Pr[K=k,Q=q]\cdot \Pr[K=i,Q=q] \neq 0$ for some $k\neq i$, the server secrecy $S_{\mathrm{serv}}(\Psi_{\mathrm{QPIR}}^{(\vM)}) \leq \beta$  implies $I(M_i^c ; \cA | Q=q, K=i)_{\rho(M,q)}, I(M_k^c ; \cA | Q=q, K=k)_{\rho(M,q)} \leq \beta $.
However, we have the following contradiction 
% The definition of $S_{\mathrm{serv}}$ should be like (2) because if $I(M_i^c ; \cA | Q=q, K=i)  \leq \beta$ for $p_{iq}= 0$,
% for any $k\neq i \in \{1,\ldots, \vF\}$,
\begin{align*}
 &(1-\alpha) \log \vM  - h_2(\alpha) \stackrel{\mathclap{(a)}}{\leq}  I(M_k ; \cA |Q=q, K=k)_{\rho(M,q)}  \\
 & \stackrel{\mathclap{(b)}}{=} I(M_k ; \cA |Q=q)_{\rho(M,q)}  \leq I(M_i^c ; \cA | Q=q)_{\rho(M,q)} \\
 &\stackrel{\mathclap{(b)}}{=} I(M_i^c ; \cA | Q=q, K=i)_{\rho(M,q)}  \leq \beta, 
\end{align*}
where $(a)$ is 
	from Fano's inequality 
% 	will be derived in Lemma~\ref{lemm:upper_relate}
and two equalities with $(b)$ are from the independence of $K$ and $(M_1,\ldots,M_{\vF},\cA)$ when $Q=q$ is fixed.
Since $\beta$ can be chosen to be an arbitrary small number, these inequalities imply that the message size $\vM$ is also sufficiently close to zero, which is a contradiction.
\end{remark}

%\subsubsection{$\vT$-Private QPIR capacity}
\subsection{$\vT$-Private QPIR capacity}

When the numbers of servers $\vN$ and files $\vF$ are fixed, the $\vT$-private QPIR capacity is defined as the optimal rate of the QPIR protocols depending on the security and upload constraints.
% The $\vT$-private QPIR capacity is defined with the security and upload constraints.
For any $\alpha\in[0,1)$ and any $\beta$, $\gamma$, $\theta \in [0,\infty]$, 
the {\em asymptotic} and {\em exact security-constrained $\vT$-private QPIR capacities} are defined by
% \begin{strip}
\begin{align}%\\[-3em]
C_{\mathrm{asymp},\vT}^{\alpha,\beta,\gamma,\theta} \coloneqq 
\sup_{\eqref{con1}}
\liminf_{\ell\to\infty} R(\Psi_{\mathrm{QPIR}}^{(\vM_\ell)}), 
\label{eq:capa1}
\\
C_{\mathrm{exact},\vT}^{\alpha,\beta,\gamma,\theta} \coloneqq 
\sup_{\eqref{con2}}
\liminf_{\ell\to\infty} R(\Psi_{\mathrm{QPIR}}^{(\vM_\ell)}), 
\label{eq:capa2}
%C_{\mathrm{asymp},\vT}^{\alpha,\beta,\gamma,\theta} \coloneqq \!\!\!\!
%\sup_{\substack{\{\vM_\ell\}_{\ell=1}^{\infty},\\ \{\Psi_{\mathrm{QPIR}}^{(\vM_\ell)}\}_{\ell=1}^{\infty}}} 
%\bigg\{ 
%\liminf_{\ell\to\infty} R(\Psi_{\mathrm{QPIR}}^{(\vM_\ell)})  \ \bigg| \    
% &\limsup_{\ell\to\infty} P_{\mathrm{err}}(\Psi_{\mathrm{QPIR}}^{(\vM_\ell)}) \leq \alpha,    \enskip
%\limsup_{\ell\to\infty} S_{\mathrm{serv}}(\Psi_{\mathrm{QPIR}}^{(\vM_\ell)}) \leq \beta,    \nonumber \\[-1em]
%&\limsup_{\ell\to\infty} S_{\mathrm{user}}^{(\vT)}(\Psi_{\mathrm{QPIR}}^{(\vM_\ell)}) \leq \gamma, \enskip
%\limsup_{\ell\to\infty} \frac{\log U(\Psi_{\mathrm{QPIR}}^{(\vM_\ell)})}{\log D(\Psi_{\mathrm{QPIR}}^{(\vM_\ell)})} \leq  \theta
%\bigg\},
%\label{eq:capa1}
%\\
%C_{\mathrm{exact},\vT}^{\alpha,\beta,\gamma,\theta} \coloneqq \!\!\!\!
%\sup_{\substack{\{\vM_\ell\}_{\ell=1}^{\infty},\\ \{\Psi_{\mathrm{QPIR}}^{(\vM_\ell)}\}_{\ell=1}^{\infty}} }
%\bigg\{ 
%\liminf_{\ell\to\infty} R(\Psi_{\mathrm{QPIR}}^{(\vM_\ell)})  \ \bigg| \     
% &P_{\mathrm{err}}(\Psi_{\mathrm{QPIR}}^{(\vM_\ell)}) \leq \alpha, \enskip
% S_{\mathrm{serv}}(\Psi_{\mathrm{QPIR}}^{(\vM_\ell)}) \leq \beta,   \nonumber  \\[-1em]
% &S_{\mathrm{user}}^{(\vT)}(\Psi_{\mathrm{QPIR}}^{(\vM_\ell)}) \leq \gamma, \enskip
%\limsup_{\ell\to\infty} \frac{\log U(\Psi_{\mathrm{QPIR}}^{(\vM_\ell)})}{\log D(\Psi_{\mathrm{QPIR}}^{(\vM_\ell)})} \leq  \theta
%\bigg\},
%\label{eq:capa2}
%% \\[-2em]
\end{align}
% \end{strip}
% \!\!\!
where the supremum is taken for sequences $\{\vM_\ell\}_{\ell=1}^{\infty}$ such that $\lim_{\ell\to\infty} \vM_\ell = \infty$ 
and sequences $\{\Psi_{\mathrm{QPIR}}^{(\vM_\ell)}\}_{\ell=1}^{\infty}$ of QPIR protocols.
and sequences $\{\Psi_{\mathrm{QPIR}}^{(\vM_\ell)}\}_{\ell=1}^{\infty}$ of QPIR protocols
    to satisfy either \eqref{con1} or \eqref{con2} given by
\begin{align}    \label{con1} %\tag{$\ast$} 
\begin{split} 
&\limsup_{\ell\to\infty} P_{\mathrm{err}}(\Psi_{\mathrm{QPIR}}^{(\vM_\ell)}) \leq \alpha,  \\
& \limsup_{\ell\to\infty} S_{\mathrm{serv}}(\Psi_{\mathrm{QPIR}}^{(\vM_\ell)}) \leq \beta,    \\
&\limsup_{\ell\to\infty} S_{\mathrm{user}}^{(\vT)}(\Psi_{\mathrm{QPIR}}^{(\vM_\ell)}) \leq \gamma, \\
& \limsup_{\ell\to\infty} \frac{\log U(\Psi_{\mathrm{QPIR}}^{(\vM_\ell)})}{\log D(\Psi_{\mathrm{QPIR}}^{(\vM_\ell)})} \leq  \theta
,
\end{split}
\end{align}
and 
\begin{gather}  \label{con2} 
\begin{split}
& P_{\mathrm{err}}(\Psi_{\mathrm{QPIR}}^{(\vM_\ell)}) \leq \alpha, \\
& S_{\mathrm{serv}}(\Psi_{\mathrm{QPIR}}^{(\vM_\ell)}) \leq \beta,   \\
& S_{\mathrm{user}}^{(\vT)}(\Psi_{\mathrm{QPIR}}^{(\vM_\ell)}) \leq \gamma, \\
& \limsup_{\ell\to\infty} \frac{\log U(\Psi_{\mathrm{QPIR}}^{(\vM_\ell)})}{\log D(\Psi_{\mathrm{QPIR}}^{(\vM_\ell)})} \leq  \theta
%& \limsup_{\ell\to\infty} \frac{\log U(\Psi_{\mathrm{QPIR}}^{(\vM_\ell)})}{\log D(\Psi_{\mathrm{QPIR}}^{(\vM_\ell)})} \leq  \theta
.
\end{split}
\end{gather}

With this definition, we consider symmetric and non-symmetric capacities at the same time.
	The capacities $C_{\mathrm{asymp},\vT}^{0,\beta,0,\theta}$ and $C_{\mathrm{exact},\vT}^{0,\beta,0,\theta}$ are $\vT$-private QPIR capacities and 
	$C_{\mathrm{asymp},\vT}^{0,0,0,\theta}$ and $C_{\mathrm{exact},\vT}^{0,0,0,\theta}$ are symmetric $\vT$-private QPIR capacities with perfect security.
%The capacity for $\vT=1$ is the same as the capacity in \ref{SH19}.	
For any $\alpha,\beta,\gamma,\theta$ and $\vT \le \vT'$, we have the inequalities
%\begin{align}
$
C_{\mathrm{exact},\vT}^{0,0,0,0}
\le
C_{\mathrm{exact},\vT}^{\alpha,\beta,\gamma,\theta}
\le
C_{\mathrm{asymp},\vT}^{\alpha,\beta,\gamma,\theta}
$,
$
C_{\mathrm{asymp},\vT}^{\alpha,\beta,\gamma,\theta} \geq
	C_{\mathrm{asymp},\vT'}^{\alpha,\beta,\gamma,\theta}
$,
and
$
	C_{\mathrm{exact},\vT}^{\alpha,\beta,\gamma,\theta} \geq
	C_{\mathrm{exact},\vT'}^{\alpha,\beta,\gamma,\theta}
$
from definition.

\subsection{Main result} 

The following theorem is the main result of the paper.
\begin{theo} \label{theo:main}
The capacity of $\vT$-private QPIR with $\vN\geq2$ servers and $\vF\geq2$ files is
	derived for any  $\alpha\in [0, 1)$ and any $\beta,\gamma, \theta \in [0,\infty)$ as follows:
    \begin{align}
    C_{\mathrm{asymp},\vT}^{\alpha,\beta,\gamma,\theta} 
    &=
    C_{\mathrm{exact},\vT}^{\alpha,\beta,\gamma,\theta}
    = 
        1       & \text{ if $1\leq \vT\leq \frac{\vN}{2}$},\\
%     \end{align}
% for any $\alpha\in [0, 1)$ and any $\beta$, $\gamma$, $\theta \in [0,\infty]$, and
%     \begin{align}
    C_{\mathrm{asymp},\vT}^{0,\beta,0,\theta} 
    &=
    C_{\mathrm{exact},\vT}^{\alpha,0,0,\theta}
    =
        \frac{2(\vN-\vT)}{\vN} & \text{ if $\frac{\vN}{2} < \vT < \vN$}. \label{capacity2}
    \end{align}
% for any $\alpha\in [0, 1)$, $\beta\in[0,\infty)$, $\theta \in [0,\infty]$.
% for any $\alpha\in [0, 1)$ and $\beta, \gamma, \theta\geq 0$.
\end{theo}

Theorem~\ref{theo:main} includes the QPIR capacity without collusion \cite{SH19} as the case $\vT=1$.
In addition, our result implies that the remarkable result of capacity $1$ extends to the case $\vT\leq \vN/2$.
Furthermore, 
%For $\vT \leq \vN/2$, our result implies that the $\vT$-private QPIR capacity is also $1$.
	since the ($1$-private) multi-round QPIR capacity is $1$ \cite{SH19}, 
	we also obtain the following corollary.
\begin{coro}
	For $1\le \vT \le \vN/2$,
	the symmetric and non-symmetric $\vT$-private multi-round QPIR capacity is $1$.
\end{coro}
When $\vT \le \vN/2$,
	the capacity decreases but it is twice the symmetric $\vT$-private classical PIR capacity $(\vN-\vT)/\vN$ \cite{WS17-2} and 
	is still independent of the number of files $\vF$.
%Note that if $\vT \leq \vN/2$, the capacity is $1$ and does not depend on $\vN$ and $\vF$, 
%	which is similar result as the QPIR capacity without colluding servers \cite{SH19}.
%When $\vT > \vN/2$,
%	the capacity is twice the symmetric PIR capacity $(\vN-\vT)/\vN$ and 
%		is still independent of the number of files $\vF$.
%As discussed in the introduction and Table~\ref{tab:1},
%	the $\vT$-private QPIR capacity is greater than the classical PIR capacity.

In the remainder of the paper, we prove Theorem~\ref{theo:main} by two parts.
First, in Section~\ref{sec:protocol}, we 
	construct 
	a capacity-achieving symmetric $\vT$-private QPIR protocol by the stabilizer formalism.
To be precise, for $\vN/2\leq \vT < \vN$, we construct 
	%a symmetric $\vT$-private QPIR protocol with 
	a protocol with
	QPIR rate $R(\Psi_{\mathrm{QPIR}}^{(\vM)}) = 2(\vN-\vT)/\vN$ and 
	perfect security $P_{\mathrm{err}}(\Psi_{\mathrm{QPIR}}^{(\vM)}) = S_{\mathrm{serv}}(\Psi_{\mathrm{QPIR}}^{(\vM)}) = S_{\mathrm{user}}^{(\vT)}(\Psi_{\mathrm{QPIR}}^{(\vM)})=0$.
%For the protocol construction, we import a structure similar to a classical symmetric $\vT$-private PIR protocol with rate $(\vN-\vT)/\vN$ by \cite{WS17-2}.
%Our protocol achieves twice this rate by importing the structure by \cite{WS17-2} into both the computational basis and the Fourier basis under the stabilizer formalism.
For $1\leq \vT < \vN/2$,
	our $(\vN/2)$-private QPIR protocol also achieves $\vT$-private capacity since 
	the protocol achieves rate $1$ and the user $(\vN/2)$-secrecy guarantees the user $\vT$-secrecy.
Second, in Section~\ref{sec:converse}, we prove the converse bounds.
Furthermore, 
	we prove in Appendix~\ref{append:average} that the capacity result is the same
	even if we change the definition of the security measures 
	as the average measures for all files $m$, queries $q$, and target indexes $k$.

%	
%The capacity-achieving QPIR protocol is constructed  for $\vN/2\leq \vT < \vN$.
%The protocol obtains zero-error ($P_{\mathrm{err}}(\Psi_{\mathrm{QPIR}}^{(\vM)})=0$), perfect server secrecy ($S_{\mathrm{serv}}(\Psi_{\mathrm{QPIR}}^{(\vM)}) = 0$), and perfect user $\vT$-secrecy ($S_{\mathrm{user}}^{(\vT)}(\Psi_{\mathrm{QPIR}}^{(\vM)})=0$).
%Since 
%	our $(\vN/2)$-private QPIR protocol achieves the rate $1$ and 
%	and
%	the user $(\vN/2)$-secrecy guarantees the user $\vT$-secrecy for $1\leq \vT < \vN/2$,
%our $(\vN/2)$-private QPIR protocol is also the capacity-achieving protocol for $1\leq \vT < \vN/2$.
%% The converse bound of the capacity is identical to that of 
%The converse bounds are given in Section~\ref{sec:converse}.
%Furthermore, 
%	we prove in Appendix~\ref{append:average} that the capacity result is the same
%	even if we change the definition of the security measures 
%	as the average measures for all files $m$, queries $q$, and target indexes $k$.

\begin{remark}
In the definition of the protocol, we assumed the condition that the target index $K$ and the files $M_1,\ldots,M_{\vF}$ are chosen uniformly.
Indeed, this condition is necessary only for the proof of converse bounds.
Even if the distributions are arbitrary,
	the protocol in Section~\ref{sec:protocol} 
		guarantees that 
			any $\vT$-servers obtain no information about $K$
		and
			the user obtains no information of non-targeted files,
			except for the information obtained from the initial distributions of $K$ and $M_1,\ldots,M_{\vF}$.
% 
% The condition that the messages $M_1,\ldots,M_{\vN}$ are with uniform distribution is necessary only for the proof of converse bounds.
% Even if the distribution of the messages is arbitrary,
% 	the protocol in Section~\ref{sec:protocol2} achieves
% 	zero-error, perfect server secrecy, and perfect user $\vT$-secrecy.	
\end{remark}

\begin{remark}
The capacity
\eqref{capacity2}
% $C_{\mathrm{asymp},\vT}^{0,\beta,0,\theta} = 2(\vN-\vT)/\vN$ for $\vN/2 < \vT < \vN$ 
is 
derived for the case where any server secrecy $\beta \in [0,\infty)$ is allowed.
%the prefect user $\vT$-secrecy is required but 
However, one may notice that 
% even if $\beta$, $\theta \in [0,\infty)$,
for some parameters $(\vN,\vT,\vF)$,
the capacity 
\eqref{capacity2}
% $C_{\mathrm{asymp},\vT}^{0,\beta,0,\theta} = \frac{2(\vN-\vT)}{\vN}$ for $\frac{\vN}{2} < \vT < \vN$ 
is smaller than 
the capacity $(1-(\vT/\vN))/(1-(\vT/\vN)^{\vF})$ \cite{SJ18} of classical $\vT$-private PIR without server secrecy.
For instance, when $(\vN,\vT,\vF) = (4,3,2)$, the capacity \eqref{capacity2} is $0.5$ and the capacity in \cite{SJ18} is $0.57$. 
This follows from the fact that the capacity \eqref{capacity2} is derived 
    for finite $\beta$
% That is, the capacity \eqref{capacity2} is derived for the fixed $\beta$, 
but the capacity in \cite{SJ18} is derived for the case where $\beta$ is allowed to be infinite.
%     
% for the case where  
% the server secrecy constraint $\beta$is asymptotically negligible with respect to the file size $\vM_{\ell}$,
% but the capacity in \cite{SJ18} is derived for the case where the $\beta$ is allowed to be infinite.
\end{remark}

\section{Preliminaries for Protocol Construction}  \label{sec:prelim}

In this section, we give preliminaries for our protocol construction in Section~\ref{sec:protocol}.
Section~\ref{sec:stabilizer} introduces the stabilizer formalism 
	and Section~\ref{sec:comm_rp} presents a protocol for classical messages constructed defined from the stabilizer formalism.
Section~\ref{subsec:fund_lemm} gives a fundamental lemma for the construction of our QPIR protocol.

%In our QPIR protocol, the stabilizer formalism is an essential tool for classical communication.

\subsection{Stabilizer formalism over finite fields} \label{sec:stabilizer}

In this subsection, we introduce the stabilizer formalism for finite fields.
Stabilizer formalism gives an algebraic structure for quantum information processing.
We use this formalism for the construction of the QPIR protocol. 
Stabilizer formalism is often used for quantum error-correction.
% In addition to the introduction, 
At the end of this subsection (Remark~\ref{rem:qstabs}),
% 	Even if we do not consider error-correction, 
	we give a brief review of quantum stabilizer error-correcting codes with the notation introduced in this subsection.
More detailed introduction of the stabilizer formalism and stabilizer codes can be found at \cite{CRSS98,AK01, KKKS06, Haya2}.
% 	for binary case, 
% 	\cite{} over rings, 
% 	and \cite{} over the finite field.
% % We give the 

Let $\mathbb{F}_q$ be a finite field whose order is a prime power $q = p^r$ and
$\cH$ be a $q$-dimensional Hilbert space with a basis $\{ |j\rangle \mid  j\in \mathbb{F}_q \}$.
We define $\tr x \coloneqq \Tr T_x\in\mathbb{F}_p$ for $x\in\mathbb{F}_q$, where $T_x\in\mathbb{F}_p^{r\times r}$ denotes the matrix representation of the linear map $y\in\mathbb{F}_q \mapsto xy \in\mathbb{F}_q$ by identifying the finite field $\mathbb{F}_q$ with the vector space $\mathbb{F}_p^{r}$.
For $a,b\in\mathbb{F}_q$, we define two unitary matrices on $\cH$ 
\begin{gather*}
\mathsf{X}(a) \coloneqq \sum_{j\in\mathbb{F}_q} |j+a\rangle \langle j |, \quad 
\mathsf{Z}(b) \coloneqq \sum_{j\in\mathbb{F}_q} \omega^{\tr bj} |j\rangle \langle j |,
% \mathsf{W}(a,b) \coloneqq \mathsf{X}(a)\mathsf{Z}(b),
\end{gather*}
where $\omega \coloneqq \exp({2\pi i/p})$.
For $\mathbf{a} = (a_1,\ldots, a_n),\mathbf{b} = (b_1,\ldots, b_n) \in\mathbb{F}_q^n$,
	and $\mathbf{w} = (\mathbf{a},\mathbf{b}) \in \mathbb{F}_q^{2n}$, we define a unitary matrix on $\cH^{\otimes n}$
\begin{align*}
% \mathbf{X(a)} \coloneqq \mathsf{X}(a_1) \otimes \mathsf{X}(a_2)  \otimes \cdots \otimes \mathsf{X}(a_n),\\
% \mathbf{Z(b)} \coloneqq \mathsf{Z}(b_1) \otimes \mathsf{Z}(b_2)  \otimes \cdots \otimes \mathsf{Z}(b_n),\\
    &\mathbf{\tilde{W}(w)} = \mathbf{\tilde{W}(a,b)} \\
    &\coloneqq 
	\mathsf{X}(a_1) \mathsf{Z}(b_1) \otimes \mathsf{X}(a_2) \mathsf{Z}(b_2)   \otimes \cdots \otimes \mathsf{X}(a_n) \mathsf{Z}(b_n) 
% \mathbf{X(a)}\mathbf{Z(b)}
.
% \mathbf{W(a,b)} \coloneqq \tau^{\langle \mathbf{a}, \mathbf{b}\rangle } \mathbf{X(a)}\mathbf{Z(b)},
\end{align*}
The Heisenberg-Weyl group is defined as
\begin{align}
\mathrm{HW}_q^n \coloneqq \px*{c \mathbf{\tilde{W}(w)} \mid  \mathbf{w} \in \mathbb{F}_q^{2n},\  c \in \mathbb{C}  }.
\label{HWgrrdef}
\end{align}
For $\mathbf{x}, \mathbf{y} \in \mathbb{F}_q^{n}$,
	we denote $\langle \mathbf{x}, \mathbf{y} \rangle \coloneqq \tr \sum_{i=1}^{n} x_iy_i\in\mathbb{F}_p$
	and define a skew-symmetric matrix $J$ on $\mathbb{F}_q^{2n}$ by 
\begin{align*}
J = \begin{pmatrix}
    0 & -I_n \\ I_n & 0
    \end{pmatrix}.
\end{align*}
Since $\mathsf{X}(a)\mathsf{Z}(b) = \omega^{-\tr ab} \mathsf{Z}(b)\mathsf{X}(a)$,
	for any $(\mathbf{a}, \mathbf{b}), (\mathbf{c}, \mathbf{d}) \in \mathbb{F}_q^{2n}$,
	we have 
\begin{align}
\mathbf{\tilde{W}(a,b)}\mathbf{\tilde{W}(c,d)} &= 
            \omega^{\langle\mathbf{(a,b)}, J\mathbf{(c,d)} \rangle} 
            \mathbf{\tilde{W}(c,d)} \mathbf{\tilde{W}(a,b)}, \label{eq:commutative}\\
\mathbf{\tilde{W}(a,b)}\mathbf{\tilde{W}(c,d)} &= 
%             \omega^{\langle\mathbf{b}, \mathbf{c}\rangle} 
            \omega^{\langle\mathbf{b}, \mathbf{c} \rangle} \mathbf{\tilde{W}(a+c, b+d)}.
             \label{eq:sum}
\end{align}

A commutative subgroup of $\mathrm{HW}_q^n$ not containing $cI_{\cH^{\otimes n}}$ for any $c\neq 0$ is called a {\em stabilizer}.
A subspace $\VV$ of $\mathbb{F}_q^{2n}$ is called {\em self-orthogonal} with respect to the bilinear form $\langle \cdot, J \cdot \rangle$ if
$$\VV\subset \VV^{\perp_J} \coloneqq \{ \mathbf{w}\in \mathbb{F}_q^{2n} \mid  \langle \mathbf{v} , J \mathbf{w} \rangle = 0 \text{ for any } \mathbf{v}\in \VV \}.$$
% Let $\VV$ be a self-orthogonal $d$-dimensional subspace of $\mathbb{F}_q^{2n}$.
We can define a stabilizer from any self-orthogonal subspace of $\mathbb{F}_q^{2n}$ by the following proposition.
% We have the following lemma.
\begin{prop} \label{prop:stab}
Let $\VV$ be a self-orthogonal subspace of $\mathbb{F}_q^{2n}$.
% Let $\VV$ be a $d$-dimensional subspace of $\mathbb{F}_q^{2n}$ which is self-orthogonal with respect to the bilinear form $\langle \cdot, J \cdot \rangle$, 
% i.e., 
% $$\VV\subset \VV^{\perp_J} \coloneqq \{ \mathbf{w}\in \mathbb{F}_q^{2n} \mid  \langle \mathbf{v} , J \mathbf{w} \rangle = 0 \text{ for any } \mathbf{v}\in \VV \}.$$
There exists $\{c_{\mathbf{v}}\in \mathbb{C}\mid \mathbf{v} \in \VV\}$ such that 
\begin{align}
S(\VV)  \coloneqq \{ \mathbf{W(v)}  \coloneqq c_{\mathbf{v}} \mathbf{\tilde{W}(v)} \mid  \mathbf{v} \in \VV \} \subset \mathrm{HW}_q^n
	\label{eq:123stab}
\end{align}
	is a stabilizer. 
\end{prop}
For the completeness, we give the proof of Proposition~\ref{prop:stab} in Appendix~\ref{append:stab}.

\begin{prop}[{\cite[(8.22), (8.24), Lemma 8.7]{Haya2}}]	\label{prop:2stab}
Let $\VV$ be a self-orthogonal $d$-dimensional subspace of $\mathbb{F}_q^{2n}$ and $S(\VV)$ be a stabilizer defined from $\VV$. % \eqref{eq:123stab}.
For the quotient space $\mathbb{F}_q^{2n} / \VV^{\perp_J}$, we denote the elements by $[\mathbf{w}] = \mathbf{w} + \VV^{\perp_J} \in \mathbb{F}_q^{2n} / \VV^{\perp_J}$.
Then, we obtain the following statements.

\begin{enumerate}
\item All elements $\mathbf{W(v)} \in S(\VV)$ are simultaneously and uniquely decomposed as 
\begin{align}
\mathbf{W(v)} = \sum_{[\mathbf{w}]\in\mathbb{F}_q^{2n} / \VV^{\perp_J} } \omega^{ \langle \mathbf{v}, J\mathbf{w}\rangle } P_{\mathbf{[w]}}^{\VV}
\qquad(\forall\mathbf{v}\in\VV)
	\label{eq:edededefcm}
\end{align}
with orthogonal projections $\{P_{[\mathbf{w}]}^{\VV}\}$ such that 
	\begin{align}
	P_{[\mathbf{w}]}^{\VV} P_{[\mathbf{w}']}^{\VV} &= 0 \text{ for any } [\mathbf{w}]\neq [\mathbf{w}'], \\
	\sum_{[\mathbf{w}]\in\mathbb{F}_q^{2n} / \VV^{\perp_J} } P_{[\mathbf{w}]}^{\VV} &= I_{\mathcal{H}^{\otimes n}}.
	\label{eq:svcxvijlkfdecomp}
	\end{align}
\item Let $\cH_{[\mathbf{w}]}^{\VV} \coloneqq \Ima P_{[\mathbf{w}]}^{\VV}$.
For any $\mathbf{w},\mathbf{w'}\in \mathbb{F}_q^{2n}$, we have the relation
    \begin{align}
    \mathbf{W(w)} \cH_\mathbf{[w']}^{\VV} = \cH_\mathbf{[w+w']}^{\VV}.
    \label{eq:hw_mov}
    \end{align}
\item For any $[\mathbf{w}]\in\mathbb{F}_q^{2n} / \VV^{\perp_J}$, 
	\begin{align}
	\dim \cH_{[\mathbf{w}]}^{\VV} = q^{n-d} . 
	\label{eq:sdfefdfsdimx}
	\end{align}
\end{enumerate}
\end{prop}
For the completeness, we give the proof of Proposition~\ref{prop:2stab} in Appendix~\ref{append:2stabb}.
As a corollary of Proposition~\ref{prop:2stab}, we obtain the following decomposition theorem.
We use this decomposition in our protocol construction.
\begin{coro} \label{coro:defsfvckl}
% By 1) and 3), 
	The quantum system $\cH^{\otimes n}$ is decomposed as 
\begin{align}
\cH^{\otimes n} = \bigotimes_{[\mathbf{w}]\in \mathbb{F}_q^{2n} / \VV^{\perp_J}} \cH_{\mathbf{[w]}}^{\VV} = \mathcal{W} \otimes \mathbb{C}^{q^{n-d}},
\label{eq:decompose}
\end{align}
where the system $\mathcal{W}$ is the $q^d$-dimensional subspace with the basis $\{ |[\mathbf{w}]\rangle  \mid [\mathbf{w}]\in \mathbb{F}_q^{2n} / \VV^{\perp_J}  \}$ such that 
$\cH_{[\mathbf{w}]}^{\VV} = |[\mathbf{w}] \rangle \otimes  \mathbb{C}^{q^{n-d}} \coloneqq \{ |[\mathbf{w} ] \rangle \otimes | v \rangle \mid |v\rangle \in\mathbb{C}^{q^{n-d}}  \}$.
With this decomposition, 
	\begin{align}
	\mathbf{W(w)} |[\mathbf{w'}] \rangle \otimes  \mathbb{C}^{q^{n-d}} = |[\mathbf{w}+\mathbf{w'}] \rangle \otimes  \mathbb{C}^{q^{n-d}}.
	\label{eq:sdfe123adec}
	\end{align}
\end{coro}
\begin{IEEEproof}
Eq.~\eqref{eq:decompose} follows from \eqref{eq:svcxvijlkfdecomp} and \eqref{eq:sdfefdfsdimx}
	and 
	Eq.~\eqref{eq:sdfe123adec} follows directly from the relation \eqref{eq:hw_mov}.
\end{IEEEproof}

We also have the following lemma.
\begin{lemm}   \label{lemm:identity_mov}
For any $\mathbf{w}, \mathbf{w'}\in\mathbb{F}_q^{2n}$, we have
\begin{align*}
&\mathbf{W(w')} \paren*{|[\mathbf{w}] \rangle\langle [\mathbf{w}] | \otimes  I_{q^{n-d}}  } \mathbf{W(w')}^{\ast}\\
&= |[\mathbf{w}+\mathbf{w'}] \rangle \langle [\mathbf{w}+\mathbf{w'}] | \otimes  I_{q^{n-d}}.
\end{align*}
\end{lemm}
\begin{IEEEproof}
Let $X \coloneqq \mathbf{W(w')} \paren*{|[\mathbf{w}] \rangle\langle [\mathbf{w}] | \otimes  I_{q^{n-d}}  } \mathbf{W(w')}^{\ast}$.
Since $X^2 = X$ and $X^{\ast} = X$, the matrix $X$ is an orthogonal projection.
Since $|[\mathbf{w}+\mathbf{w'}] \rangle \otimes  \mathbb{C}^{q^{n-d}} $ is an invariant subspace of $X$ and $\rank X = \dim |[\mathbf{w}+\mathbf{w'}] \rangle \otimes  \mathbb{C}^{q^{n-d}}  = q^{n-d}$, the matrix $X$ is the orthogonal projection onto $|[\mathbf{w}+\mathbf{w'}] \rangle  \otimes  \mathbb{C}^{q^{n-d}}$, which implies the lemma.
\end{IEEEproof}

% In Appendix~\ref{sec:comm_rp}, 
% 	we give a communication protocol for classical messages using Lemma~\ref{lemm:identity_mov}.
% }
% % Our QPIR protocol is a 	
% % The protocol in Appendix~\ref{sec:comm_rp} can be considered as a simplified 

\begin{remark} \label{rem:qstabs}
% For any element $x \in \cH_{\mathbf{[0]}}^{\VV}$ and $\mathbf{v} \in \VV$, we have 
% 	$\mathbf{W(v)} x = x$.
In terms of quantum stabilizer code, 
% 	the state on $\mathbb{C}^{q^{n-d}}$ is encoded to 
	the space $\cH_{\mathbf{[0]}}^{\VV} = |[\mathbf{0}] \rangle \otimes  \mathbb{C}^{q^{n-d}}$ is called the {\em code space}, which is 
% 	he space $\cH_{\mathbf{[0]}}^{\VV}$ 
% 	is
	the stabilized space by the action of the group $S(\VV)$. 
	In other words, from \eqref{eq:edededefcm}, the code space $\cH_{\mathbf{[0]}}^{\VV}$ is the intersection of eigenspaces of $S(\VV)$ with eigenvalue $1$.
In quantum stabilizer code, a message state is prepared in the code space $\cH_{\mathbf{[0]}}^{\VV}$.  
If an error $\mathbf{W(e)}$ is applied, 
	the encoded state on $\cH_{\mathbf{[0]}}^{\VV}$ is changed to a state on $\cH_{\mathbf{[e]}}^{\VV}$ by \eqref{eq:hw_mov}.
Then, the error correction is preformed 
	by obtaining the identity of the subspace $\cH_{\mathbf{[e]}}^{\VV}$ by
		the measurement $\{ P_{[\mathbf{e}]}^{\VV} \mid  [\mathbf{e}] \in \mathbb{F}_q^{2n} / \VV^{\perp_J}\}$ on $\cH^{\otimes n}$
		and performing the recovery operation 
		$\mathbf{W(-r)}$ for some $\mathbf{r} \in [\mathbf{e}]$, which maps 
		the state from $\cH_{\mathbf{[e]}}^{\VV}$ to $\cH_{\mathbf{[0]}}^{\VV}$.
This error correction is performed correctly if 
	$\mathbf{e-r} \in \VV$
	since the combined operation of the error and the correction is $\mathbf{W(-r) W(e)} = \mathbf{W(e-r)}$ 
		and the code space $\cH_{\mathbf{[0]}}^{\VV}$ is invariant with respect to the operation $\mathbf{W(e-r)}$
		if $\mathbf{e-r} \in \VV$.
However, if $\mathbf{e-r}  \in \VV^{\perp_J} \setminus \VV$,
	the error correction may be incorrect.
See \cite{CRSS98,AK01, KKKS06, Haya2} for details.
% Note that this error correction is not always correctly,
% 	since $\mathbf{W(-w') W(w)} = \mathbf{W(w-w')}$ 
% 		$\mathbf{w-w'} \in \mathbb{V}^{\perp}$
% 		if $\mathbf{w-w'} \in \mathbb{V}^{\perp}$
% 	
\end{remark}

\begin{remark}	\label{remark:rhomix}
Lemma~\ref{lemm:identity_mov} is equivalent to considering the state on $\mathbb{C}^{q^{n-d}}$ as completely mixed state $\rho_{\mathrm{mix}} = I_{q^{n-d}}/q^{n-d}$.
% Since $\rho = \rho_{\mathrm{mix}}$  in Lemma~\ref{lemm:identity_mov}, 
% 	the state on $\mathbb{C}^{q^{n-d}}$ is unchanged after applying any $\mathbf{W(w')}$. 
% However, 
If the state $\rho$ on $\mathbb{C}^{q^{n-d}}$ is not the completely mixed state,
% If it is not the completely mixed state,
	there always exists an operation $\mathbf{W(w')}$ %for $\mathbf{w'} \in \FF_q^{2n}$
		such that $\rho$ is changed to another state $\rho'_{\mathbf{w'}}$ as 
$$\mathbf{W(w')} \paren*{|[\mathbf{w}] \rangle\langle [\mathbf{w}] | \otimes  \rho  } \mathbf{W(w')}^{\ast}= |[\mathbf{w}+\mathbf{w'}] \rangle \langle [\mathbf{w}+\mathbf{w'}] | \otimes  \rho'_{\mathbf{w'}}.$$
For example, %if the state on $\mathbb{C}^{q^{n-d}}$ is $\rho$, which is not the completely mixed state,
	 we have $\rho\neq \rho'_{\mathbf{w'}}$ for $[\mathbf{w}] = [\mathbf{0}]$ and some $\mathbf{w'} \in \VV^{\perp_J} \setminus \VV$.
% 	\begin{align}
% 	\mathbf{W(w')} \paren*{|[\mathbf{w}] \rangle\langle [\mathbf{w}] | \otimes  \rho  } \mathbf{W(w')}^{\ast}
% 	&= \mathbf{W(w+w')} \paren*{|[\mathbf{0}] \rangle\langle [\mathbf{0}] | \otimes  \rho  } \mathbf{W(w+w')}^{\ast} \\
% 	&= |[\mathbf{w}+\mathbf{w'}] \rangle \langle [\mathbf{w}+\mathbf{w'}] | \otimes  \rho'_{\mathbf{w'}}.
% 	\end{align}
% 	 
% If logical quantum state is the completely mixed state, 
% 	it does not changed by any discrete Weyl operation.
% However, 
\end{remark}

	\begin{figure} 
	\begin{center}
        \resizebox {1\linewidth} {!} {
\begin{tikzpicture}[node distance = 3.3cm, every text node part/.style={align=center}, auto]
    \node [block] (user) {{$|{[(\mathbf{a,b})]}\rangle \langle {[(\mathbf{a,b})]} | \otimes \rho_{\mathrm{mix}}$} on {$\cH^{\otimes n} = \mathcal{W}\otimes \mathbb{C}^{q^{n-d}}$}};
    \node [above left=-0.3em and -3em of user.north west] (userlabel) {$\mathtt{receiver}$};
    
%     \node [left=1cm of user] (sent) {${K}\in\{1,\ldots,\vF\}$};
    \node [block,minimum height = 0cm, below left=2cm and -1cm of user] (serv1) {Apply\\{$\mathsf{X}({a_1})\mathsf{Z}({b_1})$}};
% 	\node [above left=-0.3em and -1.5em of serv1.north west]  {$\mathtt{p}_1$};
	
    \node [block,minimum height = 0cm, right of=serv1] (serv2) {Apply\\{$\mathsf{X}({a_2})\mathsf{Z}({b_2})$}};
%     \node [above left=-0.3em and -1.5em of serv2.north west]  {$\mathtt{p}_2$};
    
    \node [right=1.8cm of serv2] (ten) {$\cdots$};
    \node [block,minimum height = 0cm, right=1.8cm of ten] (servn) {Apply\\{$\mathsf{X}({a_n})\mathsf{Z}({b_n})$}}; %\\$M_1$\\$M_2$\\\vdots\\$M_{\vF}$};
%     \node [above right=-0.3em and -1em of servn.north east]  {$\mathtt{p}_\vN$};

    \node [draw,ellipse,below right=1cm and 0cm of serv1] (shared) {{$|{[\mathbf{0}]}\rangle \langle {[\mathbf{0}]}| \otimes \rho_{\mathrm{mix}}$} on {$\cH^{\otimes n} = \mathcal{W}\otimes \mathbb{C}^{q^{n-d}}$}};
    
    \node [right=1.5cm of user] (receiv) {${[(\mathbf{a,b})]}$}; % = {(\mathbf{a,b})+\VV^{\perp}}$};

    \path [line] (serv1.80)--node[pos=0.1,right=2mm] {{$\cH$}} (user);
    \path [line] (serv2.83) --node[pos=0.1,right=1mm] {{$\cH$}} (user.287);
    \path [line] (servn.north) --node[pos=0.1,left=4mm] {{$\cH$}} (user.346);
    
    \path [line] (user.east) -- node{} (receiv);
    
    \path [line,dashed] (shared) --node[pos=0.3,right=6mm] {{$\cH$}} (serv1.south);
    \path [line,dashed] (shared) --node[pos=0.3,right=2mm] {{$\cH$}} (serv2.south);
    \path [line,dashed] (shared) --node[pos=0.33,left=6mm] {{$\cH$}} (servn.south);
\end{tikzpicture}
    }
\caption{Protocol~\ref{prot:cc}.} \label{fig:simplestab}
\end{center}
\end{figure}
% Lemma~\ref{lemm:identity_mov}

\subsection{Communication protocol for classical message by stabilizer formalism} \label{sec:comm_rp}

% Lemma \ref{lemm:identity_mov}

% With the notations defined in the previous subsection,
In this subsection, we propose a communication protocol for classical messages from $n$ players to a receiver. 
The protocol is constructed by the stabilizer formalism.
	We will construct our QPIR protocol in Section~\ref{sec:protocol} by modifying the protocol in this subsection.
% The protocol in this section helps understanding the QPIR protocol in Section~\ref{sec:protocol}
% 	since the protocol in this section is simple and has a similar communication structure 
% 	to the QPIR protocol in Section~\ref{sec:protocol}. 

% 	we use the similar communication structure to Protocol~\ref{prot:cc}.
% However, 
% For fulfilling the QPIR task, 
% 	we will choose a suitable self-orthogonal subspace $\VV$ 
% 		and design query structures and server encoders in Section~\ref{sec:protocol}.

% We denote the completely mixed state by $\rho_{\mathrm{mix}} = (1/q^{n-d})\cdot I_{q^{n-d}}$.
% Let $\VV$ be a self-orthogonal subspace defined in Section~\ref{sec:stabilizer} and 
% 	
% 	$\cH^{\otimes n} = \mathcal{W}\otimes \mathbb{C}^{q^{n-d}}$ be the quantum system defined in \eqref{eq:decompose}.
In the following protocol,
	$n$ players encode $(a_1,b_1)$, \ldots, $(a_n,b_n) \in \mathbb{F}_q^{2}$
	and 
	the receiver decodes
	$$\mathbf{[(a,b)]} = [(a_1,\ldots,a_n, b_1,\ldots,b_n)] \in \mathbb{F}_q^{2n}/\VV^{\perp_J},$$
	where $\VV$ is a self-orthogonal subspace of $\mathbb{F}_q^{2n}$.
The protocol is depicted in Fig.~\ref{fig:simplestab}.

\begin{prot} \label{prot:cc}
Let $\VV$ be a self-orthogonal $d$-dimensional subspace of $\mathbb{F}_q^{2n}$ and $S(\VV)$ be a stabilizer associated with $\VV$. 
By Corollary~\ref{coro:defsfvckl},
	we decompose $\cH^{\otimes n}$ as 
	\begin{align}
	\cH^{\otimes n} = \bigotimes_{[\mathbf{w}]\in \mathbb{F}_q^{2n} / \VV^{\perp_J}} \cH_{\mathbf{[w]}}^{\VV} = \mathcal{W} \otimes \mathbb{C}^{q^{n-d}}.
% 	\cH^{\otimes n} = \mathcal{W} \otimes \mathbb{C}^{q^{n-d}}.
	\end{align}

% The following protocol consists of a receiver and 
% 	$n$ players, namely, player 1, \ldots, player $n$.
\begin{enumerate}
%\item {\textbf{[Shared entanglement]}}
\item {\textbf{[Distribution of entangled state]}}
	The state of $\cH^{\otimes n} = \mathcal{W}\otimes \mathbb{C}^{q^{n-d}}$ is initialized as $|[\mathbf{0}] \rangle\langle [\mathbf{0}] | \otimes \rho_{\mathrm{mix}}$,
% 	The $n$ players share a state 
	where $\rho_{\mathrm{mix}}$ is the completely mixed state on $\mathbb{C}^{q^{n-d}}$, i.e., $\rho_{\mathrm{mix}} = I_{q^{n-d}}/q^{n-d} $.
% 	The systems $\cH$
	The $n$ subsystems of $\cH^{\otimes n}$ are distributed to the $n$ players, respectively.

\item {\textbf{[Message encoding]}} 
	For each $s\in\{1,\ldots, n\}$, the player $s$ applies $\mathsf{X}(a_s) \mathsf{Z}(b_s)$ to the distributed system $\cH$ and sends the system $\cH$ to the receiver.
	
\item {\textbf{[Message decoding]}} 
	The receiver applies the PVM 
	$\mathbf{M}^{\VV} = \{ P_{[\mathbf{w}]}^{\VV} \mid  [\mathbf{w}] \in \mathbb{F}_{\qq}^{2n} / \VV^{\perp_J}\}$
on $\cH^{\otimes n}$,
where $[\mathbf{w}]$ is the measurement outcome 
{associated with} $P_{[\mathbf{w}]}^{\VV}$. \QEDB
% The measurement outcome of the user is denoted by $[\mathbf{w}_{\mathrm{out}}]$.
% In the expansion $\mathbf{w}_{\mathrm{out}} = \sum_{i=1}^{2\vN} c_i \mathbf{v}_i$,  
% the user outputs $(c_{2\vT+1},c_{2\vT+2},\ldots, c_{2\vN})\in\mathbb{F}_{\qq}^{2(\vN-\vT)}$.
\end{enumerate}
\end{prot}

% In the above protocol,
% the receiver obtains 
% 	$[(a_1,\ldots,a_n, b_1,\ldots,b_n)] \in \mathbb{F}_q^{2n}/\VV^{\perp_J}$
% 	as the measurement outcome
% since Lemma \ref{lemm:identity_mov} implies that the receiver receives the state 
% 	$|[(a_1,\ldots,a_n, b_1,\ldots,b_n)] \rangle\langle [(a_1,\ldots,a_n, b_1,\ldots,b_n)] | \otimes \rho_{\mathrm{mix}}$.
	In the above protocol, the receiver receives the state 
	$|[(\mathbf{a,b})] \rangle\langle [(\mathbf{a,b})] | \otimes \rho_{\mathrm{mix}}$ by Lemma~\ref{lemm:identity_mov}.
Thus, the receiver obtains 
	$$\mathbf{[(a,b)]} = [(a_1,\ldots,a_n, b_1,\ldots,b_n)] \in \mathbb{F}_q^{2n}/\VV^{\perp_J}$$
	as the measurement outcome.
% 	since Lemma~\ref{lemm:identity_mov} implies that the receiver receives the state 
% 	$|[(\mathbf{a,b})] \rangle\langle [(\mathbf{a,b})] | \otimes \rho_{\mathrm{mix}}$.
Note that the receiver can retrieve no more information than $[(\mathbf{a},\mathbf{b})]$ from the state $|[(\mathbf{a,b})] \rangle\langle [(\mathbf{a,b})] | \otimes \rho_{\mathrm{mix}}$.
	However, if the initial state of $\mathbb{C}^{q^{n-d}}$ is not $\rho_{\mathrm{mix}}$ in Step 1,
% 	the initial state of $\mathbb{C}^{q^{n-d}}$ is set to $\rho_{\mathrm{mix}}$.
	the receiver may obtain more information about $(\mathbf{a,b})$ than $[(\mathbf{a},\mathbf{b})]$
	since the final state on $\mathbb{C}^{q^{n-d}}$ may depend on $(\mathbf{a,b})$. 
	See Remark~\ref{remark:rhomix} for more detail.
% In Section~\ref{sec:protocol}, we import this protocol in our QPIR protocol as a method to transmit the classical file information from the servers to the user.

Protocol~\ref{prot:cc} is a generalization of the two-sum communication protocol \cite[Protocol III.1]{SH19-2}.
In the two-sum communication protocol,
% 	the stabilizer is $[0,0]$
	two players share a maximally entangled state
		and have two-bit classical messages $(a,b),(c,d) \in \mathbb{F}_2^2$, respectively.
	Each player applies $\mathsf{X}(a) \mathsf{Z}(b)$ and $\mathsf{X}(c) \mathsf{Z}(d)$ on each entangled system
		and sends the system to the receiver.
Finally, the receiver obtains the sum $(a+c,b+d)$.
% Since the maximally entangled state is 
This protocol is a special case of Protocol~\ref{prot:cc}
	since the maximally entangled state is the state $|[\mathbf{0}] \rangle\langle [\mathbf{0}] | \otimes \rho_{\mathrm{mix}}$ by choosing $\VV = \spann \{ (1,1,0,0)^{\top}, (0,0,1,1)^{\top} \}$ with $(n,d) = (2,2)$.

% 				$\mathcal{W} \otimes \mathbb{C}^{q^{n-d}} = $ ).

In Section~\ref{sec:protocol},
	we will construct our QPIR protocol by modifying Protocol~\ref{prot:cc}.
% 	we use the similar communication structure to Protocol~\ref{prot:cc}.
% However, 
For fulfilling the QPIR task, 
	we will choose a suitable self-orthogonal subspace $\VV$ 
		and design the query structure in Section~\ref{sec:protocol}.

\subsection{Fundamental lemma for protocol construction} \label{subsec:fund_lemm}

In this subsection, we prepare a fundamental lemma for the QPIR protocol construction.
Our QPIR protocol will use the self-orthogonal subspace $\VV$ defined from the lemma in this subsection.
% The paper {\cite[Appendix A]{CH17}} proved the statement only for $a_{ij}= \alpha_{i+j-2}$.
In the statement of the lemma, we use an extension of a finite field \cite{MacWilliam, finite}.
% For a set $A$, 
% 	the algebraic extension $\mathbb{F}_{q'}(A)$ of $\mathbb{F}_{q'}$ 
% 	is the smallest field containing that includes $\mathbb{F}_{q'}$ and $A$. 
    Let $\mathbb{F}$ be a subfield of $\mathbb{G}$, i.e., $\mathbb{G}$ is an extension field of $\mathbb{F}$.
    For $\alpha\in\mathbb{G}$,
        the smallest field containing $\mathbb{F}$ and $\alpha$ is denoted by $\mathbb{F}(\alpha)$.
%    When $\alpha$ is a root of some polynomial over a field $\mathbb{F}$,
%	an {\em algebraic extension} $\mathbb{F}(\alpha)$ is 
%	the smallest field containing $\mathbb{F}$ and $\alpha$.}
%Any algebraic extension of a finite field $\mathbb{F}_q$ is another finite field.
%See \cite{MacWilliam} and \cite{finite} for details.
We denote by $\mathbb{F}(\alpha_1,\ldots, \alpha_n)$ the field defined recursively by the relation $\mathbb{F}(\alpha_1,\ldots, \alpha_k) = [\mathbb{F}(\alpha_1, \ldots, \alpha_{k-1})]( \alpha_k)$.
% $\mathbb{F}_q$ be the algebraic extension $\mathbb{F}_{q'}(\alpha_1,\ldots, \alpha_{k-2})$

%The following lemma is fundamental to guarantee the secrecy in our QPIR protocol.
The lemma is given as follows.

\begin{lemm} \label{lemm:fund}
Let $n,t$ be positive integers such that $n/2 \leq t < n$.
Let $q'$ be an arbitrary prime power, $\mathbb{F}_{q'}$ be the finite field of order $q'$, and 
	$\mathbb{F}_q$ be 
        an extension field of $\mathbb{F}_{q'}$ such that 
        $\mathbb{F}_q = \mathbb{F}_{q'}(\alpha_1,\ldots, \alpha_{n+2t-2})$,
        where $\alpha_i \not \in \mathbb{F}_{q'}(\alpha_{1},\ldots, \alpha_{i-1})$ for any $i$.
	%$\mathbb{F}_q$ be the algebraic extension $\mathbb{F}_{q'}(\alpha_1,\ldots, \alpha_{n+2t-2})$,
There exist $2t$ linearly independent vectors $\mathbf{v}_1,\ldots, \mathbf{v}_{2t}\in\mathbb{F}_q^{2n}$ satisfying the following conditions.
% there exists a $2n \times 2m$ matrix $D$ over a finite field $\mathbb{F}_q$  of order $q$
% such that
    \begin{description}
    \item[$\mathrm{(a)}$] 
    Let $\mathbf{w}_1, \ldots, \mathbf{w}_{2n}$ be the row vectors of the matrix $D = (\mathbf{v}_1,\ldots, \mathbf{v}_{2t})\in\mathbb{F}_q^{2n\times 2t}$.
    Then, 
    $\mathbf{w}_{\pi(1)},\ldots, \mathbf{w}_{\pi(t)}$, $\mathbf{w}_{\pi(1)+n}, \ldots,  \mathbf{w}_{\pi(t)+n}$ are linearly independent for any 
%     injective map $\pi : \{1,\ldots,m\}\to \{1,\ldots,n\}$, 
    permutation $\pi$ in $\mathsf{S}_{n}$.
    \item[$\mathrm{(b)}$] $\langle \mathbf{v}_i,J \mathbf{v}_j\rangle  = 0$ for any $i\in\{1,\ldots, 2n-2t\}$ and any $j\in\{1,\ldots, 2t\}$.
    \end{description}
\end{lemm}
The proof of Proposition~\ref{lemm:fund} is given in Appendix \ref{append:fund}.

Many studies in classical information theory have already studied 
	the matrices $D\in\mathbb{F}_q^{n\times t}$ whose arbitrary $t$ ($\leq n$) row vectors are linearly independent,
	which is similar to condition $\mathrm{(a)}$ of Lemma \ref{lemm:fund}.
For instance, matrices of this kind have been studied as a generator matrix of the maximum distance separable (MDS) codes \cite{Singleton64}
	and have been widely used in the construction of secure communication protocols, e.g., classical private information retrievals \cite{WS17,BU18, SJ18}, wiretap channel II \cite{OW84}, and secure network coding \cite{CY11, CH17}.

\section{Construction of QPIR protocol with colluding servers}        \label{sec:protocol}

In this section, 
% by combining Protocol \ref{prot:cc} and Lemma \ref{lemm:fund},
we construct the capacity-achieving QPIR protocol for $\vN\geq 2$ servers, $\vF\geq 2$ files,
and $\vN/2 \leq \vT < \vN$ colluding servers.
For collusion of $1\leq \vT < \vN/2$ servers, our protocol for $\vT= \vN/2$ is the capacity-achieving protocol.

\subsection{Construction of protocol} \label{sec:protocol2}

\begin{figure}
\begin{center}
        \resizebox {1\linewidth} {!} {
\begin{tikzpicture}[node distance = 3.3cm, every text node part/.style={align=center}, auto]
    \node [block] (user) {{$|{[\mathbf{q}^\top \mathbf{m}]}\rangle \langle {[\mathbf{q}^\top \mathbf{m}]}| \otimes \rho_{\mathrm{mix}}$} on {$\cA = \mathcal{W} \otimes \mathbb{C}^{q^{n-d}}$}};
    \node [above left=-0.3em and -3em of user.north west] (userlabel) {$\mathtt{user}$};
    
    \node [left=1cm of user] (sent) {${k}\in\{1,\ldots,\vF\}$};
    \node [block,minimum height = 0cm, below left=2.5cm and 0cm of user] (serv1) {Apply\\{$\mathsf{X}({\mathbf{q}_{1X}^\top \mathbf{m}} )\mathsf{Z}({\mathbf{q}_{1Z}^\top \mathbf{m}})$}};
    \node [above left=-0.3em and -3.5em of serv1.north west]  {$\mathtt{serv}_1$};
    
    \node [block,minimum height = 0cm, right=2.5em of serv1] (serv2) {Apply\\{$\mathsf{X}({\mathbf{q}_{2X}^\top \mathbf{m}})\mathsf{Z}({\mathbf{q}_{2Z}^\top \mathbf{m}})$}};
    \node [above left=-0.3em and -3.5em of serv2.north west]  {$\mathtt{serv}_2$};
    
    \node [right=1.8cm of serv2] (ten) {$\cdots$};
    \node [block,minimum height = 0cm, right=4em of ten] (servn) {Apply\\{$\mathsf{X}({\mathbf{q}_{\vN X}^\top \mathbf{m}})\mathsf{Z}({\mathbf{q}_{\vN Z}^\top \mathbf{m}})$}}; %\\$M_1$\\$M_2$\\\vdots\\$M_{\vF}$};
    \node [above right=-0.3em and -3.5em of servn.north east]  {$\mathtt{serv}_\vN$};
	
    \node [draw,ellipse,below right=1cm and 0cm of serv1] (shared) {{$|{[\mathbf{0}]}\rangle \langle {[\mathbf{0}]}| \otimes \rho_{\mathrm{mix}}$} on {$\cA = \mathcal{W} \otimes \mathbb{C}^{q^{n-d}}$}};
	
    \node [right=1.5cm of user] (receiv) {\\[0.7em]${[{\mathbf{q}^\top \mathbf{m}}]} = {{\mathbf{q}^\top \mathbf{m}} + \VV^\perp}$\\$\in \mathbb{F}_q^{2n}/\VV^{\perp}\simeq \mathbb{F}_q^{2(\vN-\vT)}$};
%     \node [right=2cm of user] (receiv) {${[{\mathbf{q}^\top \mathbf{m}}]} \in \mathbb{F}_q^{2n}/\VV^{\perp}\simeq \mathbb{F}_q^{2(\vN-\vT)}$};

    \path [line] (serv1.75)--node[pos=0.1,right=2mm] {{$\cA_1$}} (user.192);
    \path [line] (serv2.70) --node[pos=0.1,right=1mm] {{$\cA_2$}} (user.280);
    \path [line] (servn.70) --node[pos=0.2,right=4mm] {{$\cA_{\vN}$}} (user.352);
	
    \path [line] (user.east) -- node{} (receiv);
	
    \path [line,dashed] (shared) --node[pos=0.3,right=6mm] {{$\cA_1$}} (serv1.south);
    \path [line,dashed] (shared) --node[pos=0.3,right=2mm] {{$\cA_2$}} (serv2.south);
    \path [line,dashed] (shared) --node[pos=0.33,left=6mm, above=1mm] {{$\cA_{\vN}$}} (servn.south);
    
    \path [line] (sent) -- (user);
    
    \path [line] (user.190) --node[pos=0.3,left=2mm] {${(\mathbf{q}_{1X}, \mathbf{q}_{1Z})}$} (serv1.110);
    \path [line] (user) --node[pos=0.3,left] {${(\mathbf{q}_{2X}, \mathbf{q}_{2Z})}$} (serv2.north);
    \path [line] (user.351) --node[pos=0.3,left=2mm] {${(\mathbf{q}_{\vN X}, \mathbf{q}_{\vN Z})}$} (servn.100);
\end{tikzpicture}
}
\caption{$\vT$-Private QPIR protocol. 
		$\mathbf{m} = (\mathbf{m}_1^\top,\ldots, \mathbf{m}_k^\top)^\top$
		and $\mathbf{q} = (\mathbf{q}_{1X}^\top, \ldots, \mathbf{q}_{\vN X}^\top, \mathbf{q}_{1Z}^\top, \ldots, \mathbf{q}_{\vN Z}^\top )^{\top}$ are the collections of files and queries, respectively.}
\label{fig:tcolludpro}
\end{center}
\end{figure}
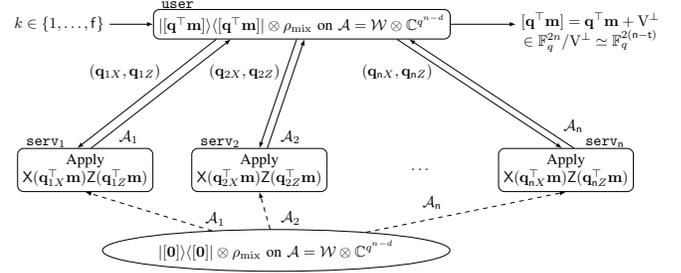

%The protocol presented in this section 
We construct our QPIR protocol by modifying Protocol~\ref{prot:cc}, which is defined with a self-orthogonal subspace $\VV$.
Similar to Protocol~\ref{prot:cc}, in our QPIR protocol,
	$\vN$ servers encode $(a_1,b_1)$, \ldots, $(a_\vN,b_\vN) \in \mathbb{F}_q^{2}$
	and 
	a user receives
	$\mathbf{[(a,b)]} = [(a_1,\ldots,a_\vN, b_1,\ldots,b_\vN)] \in \mathbb{F}_q^{2\vN}/\VV^{\perp_J}$
	but no more information of $\mathbf{(a,b)}$.
For guaranteeing the correctness and the secrecies of our protocol,
	we will choose a self-orthogonal subspace $\VV$ by Lemma~\ref{lemm:fund} and design the query structure so that 
	\begin{enumerate}
	\item the user generates queries depending on the file index $k$, while any $\vT$ queries are independent of $k$,
	\item the $s$-th server encodes $(a_i,b_i)$ depending on the query and the files, and 
	\item the user's received message $\mathbf{[(a,b)]}= \mathbf{(a,b)}+\VV^{\perp}$ is identical to the $k$-th file.
	\end{enumerate}
	%\begin{enumerate}
	%\item the user generates queries $\mathbf{q}_s$ ($s=\{1,\ldots,\vN \})$ depending on the file index $k$, while the collection of queries to any $\vT$ servers is independent of $k$,
	%\item the $s$-th server encodes $(a_i,b_i)$ depending on query $\mathbf{q}_s$ and the files $\mathbf{m} = (\mathbf{m}_1^\top,\ldots, \mathbf{m}_k^\top)^\top$, and 
	%\item the user's received message $\mathbf{[(a,b)]}= \mathbf{(a,b)}+\VV^{\perp}$ is identical to the target file $\mathbf{m}_k$.
	%\end{enumerate}

	%\begin{itemize}
	%\item server encoder: each server encodes query and the messages as $(a_i,b_i)$.
	%\item the server secrecy is guaranteed, since the user only receives $\mathbf{[(a,b)]}$. 
	%\end{itemize}

% With the above choice of self-orthogonal space $\VV$ and notations, our QPIR protocol is described as follows.
Our QPIR protocol is described as follows.
\begin{prot} \label{prot:1}
Let  $\vN\geq 2$, $\vF\geq 2$, and $\vN/2 \leq \vT < \vN$.
The files are $2(\vN-\vT)$-dimensional vectors $\mathbf{m}_1,\ldots, \mathbf{m}_{\vF} \in \mathbb{F}_{\qq}^{2(\vN-\vT)}$.
Each of $\vN$ servers contains a copy of all files $\mathbf{m} \coloneqq (\mathbf{m}_1^\top,\ldots, \mathbf{m}_k^\top)^\top \in \mathbb{F}_{\qq}^{2(\vN-\vT)\vF}$.
The index of the user's target file is $k$, i.e., the user retrieves $\mathbf{m}_k$.

% The QPIR protocol is described as follows.
We choose a self-orthogonal vector space $\VV$ and introduce some notations as follows.
Let $q'$ be an arbitrary prime power, $\mathbb{F}_{q'}$ be the finite field of order $q'$, and 
	$\mathbb{F}_q$ be an extension field of $\mathbb{F}_{q'}$ such that $\mathbb{F}_q = \mathbb{F}_{q'}(\alpha_1,\ldots, \alpha_{n+2t-2})$,
	where $\alpha_i \not \in \mathbb{F}_{q'}(\alpha_{1},\ldots, \alpha_{i-1})$ for any $i$.
We choose a basis $\mathbf{v}_1,\ldots, \mathbf{v}_{2\vN}$ of $\mathbb{F}_{\qq}^{2\vN}$ 
	such that the first $2\vT$ vectors $\mathbf{v}_1,\ldots, \mathbf{v}_{2\vT}$ satisfy the conditions of Lemma~\ref{lemm:fund}.
% a matrix $V_1 \in\mathbb{F}_{\qq}^{2\vN\times 2\vT}$ 
% The columns vectors of $V_1$ is denoted by $\mathbf{v}_1,\ldots, \mathbf{v}_{2\vT}$.
Let $\VV\coloneqq \spann \{\mathbf{v}_1    , \mathbf{v}_2 , \ldots , \mathbf{v}_{2\vN-2\vT}\}\subset\mathbb{F}_{\qq}^{2\vN}$.
 Then, 
 from condition $\mathrm{(b)}$ of Lemma~\ref{lemm:fund},
 the subspace $\VV$
  is self-orthogonal with respect to $\langle \cdot, J \cdot \rangle$,
%   i.e., $\VV\subset \VV^{\perp_J}$,
%   we have the relation 
  $\VV^{\perp_J} = \spann\{ \mathbf{v}_1, \ldots, \mathbf{v}_{2\vT}\}$
%   and $\mathbb{F}_q^{2n}/\VV^{\perp_J}  = \spann\{[\mathbf{v}_{2\vT+1}], \ldots, [\mathbf{v}_{2\vN}]\}$.
  and the quotient space $\mathbb{F}_q^{2n}/\VV^{\perp_J}$ is written as 
  $$\mathbb{F}_q^{2n}/\VV^{\perp_J}  = \{ [\mathbf{w}] \coloneqq \mathbf{w} + \VV^{\perp_J}  \mid \mathbf{w} \in \spann\{\mathbf{v}_{2\vT+1}, \ldots, \mathbf{v}_{2\vN}\} \}.$$
  Let 
\begin{align}
D_1 &\coloneqq \begin{pmatrix}
        \mathbf{v}_1    & \mathbf{v}_2 & \cdots & \mathbf{v}_{2\vT}
        \end{pmatrix}\in \mathbb{F}_{\qq}^{2\vN\times 2\vT},\\
D_2 &\coloneqq \begin{pmatrix}
        \mathbf{v}_{2\vT+1}    & \mathbf{v}_{2\vT+2} & \cdots & \mathbf{v}_{2\vN}
        \end{pmatrix}\in \mathbb{F}_{\qq}^{2\vN\times 2(\vN-\vT)}.
%         ,\\
% D_0 &\coloneqq 
% 		( D_1 , D_2 ) \in \mathbb{F}_{\qq}^{2\vN\times 2\vN}.
% 		\begin{pmatrix}
%         \mathbf{v}_1    & \mathbf{v}_2 & \cdots & \mathbf{v}_{2\vN}
%         \end{pmatrix} \in \mathbb{F}_{\qq}^{2\vN\times 2\vN},\\       
\end{align}
  We assume that the vectors $\mathbf{v}_1    , \ldots , \mathbf{v}_{2\vN}$ are publicly known to the user and all servers.
% Each server contains the entire file set which is described as follows.
% % Suppose there exists $\vN$ servers and
% % each server contains 
% The files $\mathbf{m}_1,\ldots, \mathbf{m}_{\vF}$ are the elements of $\mathbb{F}_{\qq}^{2(\vN-\vT)}$.
With this choice of the self-orthogonal space $\VV$ and vectors $\mathbf{v}_1    , \ldots , \mathbf{v}_{2\vN}$, the protocol works as follows.

\begin{enumerate}
\item {\textbf{[Distribution of Entangled State]}}
%\item {\textbf{[Shared entanglement]}}
Let $\cA_1, \ldots , \cA_{\vN}$ be $q$-dimensional Hilbert spaces.
From Corollary~\ref{coro:defsfvckl}, the quantum system $\cA \coloneqq \cA_1\otimes \cdots \otimes \cA_{\vN}$ is decomposed as 
	$\cA = \mathcal{W}\otimes \mathbb{C}^{q^{2\vT-\vN}}$,
	where $\mathcal{W} = \spann\{ |[\mathbf{w}]\rangle  \mid [\mathbf{w}]\in \mathbb{F}_q^{2n} / \VV^{\perp_J}  \}$.
The state of $\cA$ is initialized as 
	$|[\mathbf{0}]\rangle \langle [\mathbf{0}]| \otimes \rho_{\mathrm{mix}}$
	%	, where $\rho_{\mathrm{mix}} =  I_{q^{2\vT-\vN}}/ q^{2\vT-\vN} $,
and is distributed so that the $s$-th server has $\cA_s$ for $s=1,2,\ldots, \vN$.

\item {\textbf{[Query]}} 
%\item {\textbf{[User encoder]}} 
The user randomly chooses a matrix $R$ in $\mathbb{F}_q^{2\vT\times 2(\vN-\vT)\vF}$ with the uniform distribution.
Depending on $k$, let $E_k \coloneqq (\delta_{i ,j - 2(\vN-\vT)(k-1) } )_{i,j}\in\mathbb{F}_{\qq}^{2(\vN-\vT)\times 2(\vN-\vT)\vF}$, 
where $\delta_{x,y} = 1 $ if $x=y$ and $\delta_{x,y} =0$ if $x\neq y$.
That is, $E_k$ is the block matrix whose $k$-th block is the identity matrix $I\in\mathbb{F}_q^{2(\vN-\vT)\times 2(\vN-\vT)}$ and all other blocks are zero.
Let
\begin{align*}
\mathbf{q} &= 
(\mathbf{q}_{1X}^\top,
\ldots,
\mathbf{q}_{\vN X}^\top,
\mathbf{q}_{1Z}^\top,
\ldots,
\mathbf{q}_{\vN Z}^\top
)^{\top}
\\
&\coloneqq
	%(D_1, D_2) \begin{pmatrix} R \\ E_k \end{pmatrix}
	%= 
	D_1 R + D_2 E_k
	\in \mathbb{F}_{q}^{ 2\vN \times 2(\vN-\vT)\vF}.
\end{align*}
The user sends the query $\mathbf{q}_s = (\mathbf{q}_{sX}, \mathbf{q}_{sZ}) \in \mathbb{F}_{\qq}^{2(\vN-\vT) \vF} \times \mathbb{F}_{\qq}^{2(\vN-\vT) \vF}$ to the $s$-th server for $s=1,2, \ldots, \vN$.

% (R_1, \ldots, R_{2\vT}) V_1^\top
%  + ( \mathbf{e}_{2(\vN-\vT) (k-1)+1}, \mathbf{e}_{2(\vN-\vT) (k-1)+2}, \ldots, \mathbf{e}_{2(\vN-\vT) k} ) V_2^{\top} ,
% where 
% $\mathbf{e}_j\in\mathbb{F}_{\qq}^{2(\vN-\vT)\vF}$ ($j=1,\ldots, 2(\vN-\vT)\vF$) is the unit vector whose $j$-th element is $1$.
% The user sends
% \begin{align}
% Q_s &\coloneqq (Q_{sX}, Q_{sZ}) \in \mathbb{F}_{\qq}^{d \vF} \times \mathbb{F}_{\qq}^{d \vF}
% \end{align}
% to $s$-th server ($s=1,2, \ldots, \vN$).

%\item {\textbf{[Server encoder]}} 
\item {\textbf{[Download]}}
For each $s=1,2,\ldots, \vN$,
the $s$-th server applies the unitary operation $\mathsf{X}( \mathbf{q}_{sX}^\top \mathbf{m} )\mathsf{Z} (\mathbf{q}_{sZ}^\top \mathbf{m} )$ to $\cA_s$ and sends $\cA_s$ to the user.

%\item {\textbf{[Decoder]}} 
\item {\textbf{[Recovery]}} 
	The user applies the {PVM}
	$\mathbf{M}^{\VV} = \{ P_{[\mathbf{w}]}^{\VV} \mid  [\mathbf{w}] \in \mathbb{F}_{\qq}^{2\vN} / \VV^{\perp_J}\}$
	on $\cA$,
where $[\mathbf{w}]$ is the measurement outcome 
{associated with} $P_{[\mathbf{w}]}^{\VV}$.
The measurement outcome of the user is denoted by $[\mathbf{w}_{\mathrm{out}}]$.
In the expansion $\mathbf{w}_{\mathrm{out}} = \sum_{i=1}^{2\vN} c_i \mathbf{v}_i$,  
the user outputs $(c_{2\vT+1},c_{2\vT+2},\ldots, c_{2\vN})\in\mathbb{F}_{\qq}^{2(\vN-\vT)}$.
\QEDB

% In the expansion $\mathbf{w'} = \sum_{i=1}^{2\vN} c_i \mathbf{v}_i$ of a vector $\mathbf{w'}\in[\mathbf{w}_{\mathrm{out}}]$,  
% the coefficients $(c_{2\vT+1},c_{2\vT+2},\ldots, c_{2\vN})$ is uniquely determined for any element $\mathbf{w'} \in [\mathbf{w}_{\mathrm{out}}]$, and the user returns $(c_{2\vT+1},c_{2\vT+2},\ldots, c_{2\vN})\in\mathbb{F}_{\qq}^{2(\vN-\vT)}$.
\end{enumerate}
\end{prot}

\subsection{Analysis of Protocol~\ref{prot:1}} \label{sec:analysis}

In this section, we analyze the performance of Protocol \ref{prot:1}.

\subsubsection{Costs and QPIR rate}

The file size is $\vM = |\mathbb{F}_{\qq}^{2(\vN-\vT)}|=  q^{2(\vN-\vT)}$,
the download cost is $D(\qprot) = \dim \bigotimes_{s=1}^{\vN} \cA_s = q^{\vN}$,
and the upload cost is $U(\qprot) = |\mathbb{F}_{\qq}^{2(\vN-\vT)\vF \times 2\vN}| = q^{4\vN\vF(\vN-\vT)}$.
Therefore, the QPIR rate is $R(\qprot) = \log \vM/ \log D(\qprot) = 2(\vN-\vT)/\vN$.

\subsubsection{Error probability}
We show that the user obtains $\mathbf{m}_k$ without error.
Let 
\begin{align}
\mathbf{w}' &\coloneqq \mathbf{q}^{\top} \mathbf{m} \\
    &= 
% (\mathbf{q}_{1X},
% \ldots,
% \mathbf{q}_{\vN X},
% \mathbf{q}_{1Z},
% \ldots,
% \mathbf{q}_{\vN Z}
% )^{\top} \mathbf{m} 
% =
(\mathbf{q}_{1X}^{\top} \mathbf{m} ,
\ldots,
\mathbf{q}_{\vN X}^{\top} \mathbf{m} ,
\mathbf{q}_{1Z}^{\top} \mathbf{m} ,
\ldots,
\mathbf{q}_{\vN Z}^{\top} \mathbf{m} 
)^{\top}
\in \mathbb{F}_q^{2\vN}.
\end{align}
% \begin{align}
% \mathbb{F}_{\qq}^{2\vN}
% \ni
% % \mathbf{w}_{\mathrm{out}} =
%     \mathbf{w}' 
%             &=
%             (\mathbf{q}_{1X} ,\ldots ,
%               \mathbf{q}_{\vN X} ,
%               \mathbf{q}_{1Z} ,
%               \ldots,
%               \mathbf{q}_{\vN Z} )^{\top} \mathbf{m} \\
% %         &=
% %         V
% %         (R_1,\ldots,R_{2\vN-d}, \mathbf{e}_{d(k-1)+1}, \mathbf{e}_{d(k-1)+2}, \ldots, \mathbf{e}_{d k} )^{\top}
% %          W\\
%         &= D_1 R \mathbf{m}  + D_2 E_k \mathbf{m}
%          \\
%         &= 
%         D_1 R \mathbf{m} 
%          + 
%         \sum_{i=1}^{2\vN-2\vT}
%          m_{k,i} \mathbf{v}_{2\vT+i} .
% \end{align}
The state after the servers' encoding is 
\begin{align}
 &\mathbf{W} ( \mathbf{w}' ) \paren*{|[\mathbf{0}]\rangle \langle [\mathbf{0}]| \otimes \rho_{\mathrm{mix}} } \mathbf{W}  ( \mathbf{w'} )^{\ast} \\
 &= |[\mathbf{w'}]\rangle \langle [\mathbf{w'}]| \otimes \rho_{\mathrm{mix}}
,
\label{eqref:after_encoding}
\end{align}
where the equality follows from Lemma~\ref{lemm:identity_mov}.
Thus, the measurement outcome $[\mathbf{w}_{\mathrm{out}} ]$ is $[\mathbf{w}']$. 
Note that we have 
\begin{align}
\mathbb{F}_{\qq}^{2\vN}
\ni
% \mathbf{w}_{\mathrm{out}} =
    \mathbf{w}' 
            &=
            \mathbf{q}^{\top} \mathbf{m} \\
%         &=
%         V
%         (R_1,\ldots,R_{2\vN-d}, \mathbf{e}_{d(k-1)+1}, \mathbf{e}_{d(k-1)+2}, \ldots, \mathbf{e}_{d k} )^{\top}
%          W\\
        &= D_1 R \mathbf{m}  + D_2 E_k \mathbf{m}
         \\
        &= 
        D_1 R \mathbf{m} 
         + 
        \sum_{i=1}^{2\vN-2\vT}
         m_{k,i} \mathbf{v}_{2\vT+i} 
         \label{eq:fsdfffedccvwjio}
%         \\
%         &= 
%         \sum_{i=1}^{2\vT} R_{i}^{\top} M \mathbf{v}_i + 
%         \sum_{i=1}^{2\vN-2\vT}
%          M_{k,i} \mathbf{v}_{2\vT+i} .
\end{align}
and the first term $D_1 R\mathbf{m}$ of \eqref{eq:fsdfffedccvwjio} is a vector in $\VV^{\perp_J}$,
which implies 
% Thus, we have 
\begin{align}
 &[\mathbf{w}_{\mathrm{out}}] 
= [\mathbf{w}'] 
= \mathbf{w}'  + \VV^{\perp} \\
% = \mathbf{w}' + \VV^{\perp}
% = \mathbf{w}_{\mathrm{out}} + \VV^{\perp}
% =
&=\sum_{i=1}^{2\vN-2\vT}
         m_{k,i} \mathbf{v}_{2\vT+i}  + \VV^{\perp}
= \bigg[ \sum_{i=1}^{2\vN-2\vT}
         m_{k,i} \mathbf{v}_{2\vT+i} \bigg].
\end{align}
% and the protocol output $(c_{2\vT+1},c_{2\vT+2},\ldots, c_{2\vN})\in\mathbb{F}_{\qq}^{2(\vN-\vT)}$ is independent of the representative of $[\mathbf{w}_{\mathrm{out}}]$,
Thus, the user obtains $(c_{2\vT+1},c_{2\vT+2},\ldots, c_{2\vN}) = (m_{k,1}, \ldots,  m_{k,2(\vN-\vT)}) = \mathbf{m}_k$.

% Notice that the protocol output $(c_{2\vT+1},c_{2\vT+2},\ldots, c_{2\vN})\in\mathbb{F}_{\qq}^{2(\vN-\vT)}$ is independent of the representative of $[\mathbf{w}_{\mathrm{out}}]$,

\subsubsection{Server secrecy}
The protocol has perfect server secrecy because 
from \eqref{eqref:after_encoding},
the state after the servers' encoding is 
$|[\mathbf{w'}]\rangle \langle [\mathbf{w'}]| \otimes \rho_{\mathrm{mix}}$, which is independent of the non-retrieved files.

\begin{remark}
The server secrecy is not perfect 
	if the prior entangled state is 
	$|[\mathbf{0}]\rangle \langle [\mathbf{0}]| \otimes \rho$ with non-completely mixed state $\rho$.
% 	if the state on $\mathbb{C}^{q^{2\vT-\vN}}$ of the prior entangled state is not completely mixed state, i.e., 
% 	it is $|[\mathbf{0}]\rangle \langle [\mathbf{0}]| \otimes \rho$ for some non completely mixed state $\rho$.
% 	then the resultant state is 
As remarked in Remark~\ref{remark:rhomix}, if $|[\mathbf{0}]\rangle \langle [\mathbf{0}]| \otimes \rho$ is the initial entangled state, 
	the state $\rho$ may be changed depending on the servers' operation $\mathbf{W(w')}$, i.e.,
	\begin{align}
	\mathbf{W(w')} \paren*{|[\mathbf{0}] \rangle\langle [\mathbf{0}] | \otimes  \rho  } \mathbf{W(w')}^{\ast}
	= |[\mathbf{w'}] \rangle \langle [\mathbf{w'}] | \otimes  \rho'_{\mathbf{w'}}
	\end{align}
	for some state $\rho'_{\mathbf{w'}}$.
Thus, the user may obtain some information of $\mathbf{w}'$ from the state $\rho'_{\mathbf{w'}}$, i.e., some information of the non-targeted files is leaked.
\end{remark}

\subsubsection{User secrecy}
To discuss the user secrecy of Protocol \ref{prot:1}, we introduce the following notations.
% Let $\mathsf{S}_\vN$ be the symmetric group of $\{1,\ldots, \vN\}$, i.e., the set of all permutations on $\{1,\ldots, \vN \}$.
We denote $\mathbf{v}_i = (v_{1,i},\ldots, v_{2\vN,i})^{\top}\in\mathbb{F}_{\qq}^{2\vN}$ for $i=1,\ldots, 2\vN$.
For any permutation $\pi$ in $\mathsf{S}_\vN$, we denote 
\begin{align*}
\mathbf{v}_{i,\pi} &\coloneqq 
\begin{pmatrix}
v_{\pi(1),i}\\ \vdots \\ v_{\pi(\vT),i}\\ v_{\vN+\pi(1),i} \\ \vdots \\ v_{\vN+\pi(\vT),i}
\end{pmatrix}
\in\mathbb{F}_{\qq}^{2\vT}
,
\\
    D_{1,\pi}
    &\coloneqq
    (\mathbf{v}_{1,\pi}, \ldots , \mathbf{v}_{2\vT,\pi})
    \in\mathbb{F}_{\qq}^{2\vT \times 2\vT},
    \\
    D_{2,\pi}
    &\coloneqq
    (\mathbf{v}_{2\vT+1,\pi}, \ldots , \mathbf{v}_{2\vN,\pi})
    \in\mathbb{F}_{\qq}^{2\vT \times 2(\vN-\vT)}.
\end{align*}

The user $\vT$-secrecy is proved as follows.
% We assume that the servers know any information of the protocol but not the user's private randomness $R$.
Let $\pi$ be an arbitrary permutation in $\mathsf{S}_{\vN}$.
The queries to the $\pi(1)$-th server, \ldots, $\pi(\vT)$-th server are written as
\begin{align}
&(\mathbf{q}_{\pi(1)X}, \ldots, \mathbf{q}_{\pi(\vT) X}, \mathbf{q}_{\pi(1)Z}, \ldots, \mathbf{q}_{\pi(\vT) Z})^{\top}\\
&=
  D_{1,\pi} R
 +
  D_{2,\pi} E_k
 \in\mathbb{F}_{\qq}^{  2\vT \times 2(\vN-\vT)\vF}.
\end{align}
Since condition $\mathrm{(a)}$ of Lemma \ref{lemm:fund} implies $\rank D_{1,\pi} = 2\vT$, i.e., $D_{1,\pi}$ is invertible,
when %$V_{1,\pi}$, $V_{2,\pi}$, $E_k$ are fixed but 
$R$ is uniformly random in $\mathbb{F}_{\qq}^{2\vT\times 2(\vN-\vT)\vF}$,
the distribution of $(\mathbf{q}_{\pi(1)X}, \ldots, \mathbf{q}_{\pi(\vT) X}, \mathbf{q}_{\pi(1)Z}, \ldots, \mathbf{q}_{\pi(\vT) Z})^{\top}$
is the uniform distribution on $\mathbb{F}_{\qq}^{ 2\vT\times 2(\vN-\vT)\vF}$.
Therefore, 
the colluding servers obtain no information of the target file index $k$
since the matrix $R$ is unknown to the colluding servers and is uniformly random in $\mathbb{F}_{\qq}^{ 2\vT \times 2(\vN-\vT)\vF}$.

\section{Converse Bounds} \label{sec:converse}

The converse bounds of Theorem \ref{theo:main} are written for any  $\alpha\in [0, 1)$ and any $\beta,\gamma, \theta \in [0,\infty)$ as
% \begin{align}
% \tilde{C}_{\mathrm{asymp},\vT}^{\alpha,\beta,\gamma,\theta} &\leq 1 & \text{ if $1\leq \vT\leq \frac{\vN}{2}$},     \label{ineq:conv1} \\ 
% \tilde{C}_{\mathrm{exact},\vT}^{\alpha,0,0,\theta} &\leq \frac{2(\vN-\vT)}{\vN} & \text{ if $\frac{\vN}{2} < \vT < \vN$}, \label{ineq:conv2}\\
% \tilde{C}_{\mathrm{asymp},\vT}^{0,\beta,0,\theta} &\leq \frac{2(\vN-\vT)}{\vN} & \text{ if $\frac{\vN}{2} < \vT < \vN$}. \label{ineq:conv3}
% \end{align} 
\begin{align}
C_{\mathrm{asymp},\vT}^{\alpha,\beta,\gamma,\theta} &\leq 1 & \text{if $1\leq \vT\leq \frac{\vN}{2}$},  \label{ineq:conv1} \\
% \end{align}
% for any $\alpha\in [0, 1)$ and any $\beta, \gamma, \theta \in [0,\infty]$, and
% \begin{align}
C_{\mathrm{exact},\vT}^{\alpha,0,0,\theta}  &\leq \frac{2(\vN-\vT)}{\vN} & \text{if $\frac{\vN}{2} < \vT < \vN$},
	\label{ineq:conv2}
    \\
C_{\mathrm{asymp},\vT}^{0,\beta,0,\theta}         &\leq \frac{2(\vN-\vT)}{\vN} & \text{if $\frac{\vN}{2} < \vT < \vN$}.
	\label{ineq:conv3}
\end{align}
The bounds \eqref{ineq:conv1} and \eqref{ineq:conv2} are proved similar to the converse proofs of \cite{SH19} and \cite{SH19-2}, respectively.
Therefore, we give the details of the two proofs in Appendix~\ref{append:converse} and in this section, we give the proof of \eqref{ineq:conv3}.

\begin{figure}
\begin{center}
\subfloat[Downloading step of QPIR protocol. The user shares $Q_{\vT}$ with colluding servers and $Q_{\vT^c}$ with non-colluding servers.]{
\centering
\begin{tikzpicture}[node distance = 3.3cm, every text node part/.style={align=center}, auto]
    \node [block] (user) {User};
    \node [block,minimum height = 0.1cm, below left=2cm and -0.5cm of user] (serv1) {Colluding servers\\($\vT$ servers)};
%     \node [block,minimum height = 3cm, right of=serv1] (serv2) {$\mathtt{serv}_2$\\$W_1$\\$W_2$\\\vdots\\$W_{\vF}$};
%     \node [right=1.8cm of serv2] (ten) {$\cdots$};
    \node [block,minimum height = 0.1cm, below right=2cm and -0.5cm of user] (servn) {Non-colluding servers\\($\vN-\vT$ servers)};

    \path [line] (serv1.80)--node[left=2mm] {{$\cA_{\vT}$}} (user);
    \path [line] (servn.110)--node[right=2mm] {{$\cA_{\vT^c}$}} (user);
    
    \path [line,-,dashed] (serv1.180) edge[bend left=90] node[left=2mm] {$Q_{\vT}$} (user.180);
    \path [line,-,dashed] (servn.0) edge[bend right=90] node[right=2mm] {{$Q_{\vT^c}$}} (user.0);
    
%     \node [draw,ellipse,below right=1cm and 2cm of serv1] (shared) {Shared Entanglement $\rho_{\mathrm{prev}}$};
   
    \path [line,-,dashed] (serv1.240) edge [bend right=30] node[block,below] {{Shared Entanglement}} (servn.240);
    
%     \node [below right = 1cm and 5cm of user] {$\mathbf{\rightarrow}$};
%     \node [below = 5cm of user] {$\mathbf{\downarrow}$};
\end{tikzpicture}
} \\[3em]
\subfloat[Entanglement-assisted communication of classical message with shared randomness $Q_{\vT^c}$.
	Note that the user know $Q$ but not which query $Q_{\vT^c}$ the non-colluding servers contain.]{
\centering
\begin{tikzpicture}[node distance = 3.3cm, every text node part/.style={align=center}, auto]
    \node [block] (user) {User};
    \node [block,minimum height = 0.1cm, below =3cm of user] (servn) {Non-colluding servers\\($\vN-\vT$ servers)};
      
    \path [line] (servn)--node[right=1mm] {{$\cA_{\vT^c}$}} (user);
    
%     \node [draw,ellipse,below right=1cm and 2cm of serv1] (shared) {Shared Entanglement $\rho_{\mathrm{prev}}$};
   
    \path [line,-,dashed] (user.190) edge [bend right=40] node[block,left=1mm] (sr) {Shared\\Entanglement} (servn.170);
    
    \path [line,-,dashed] (servn.10) edge[bend right=40] node[right=2mm] {{$Q_{\vT^c}$}} (user.350);
%     \path [line,<->,dashed] (serv1.270) edge [bend right=50] node[below] {{Shared Entanglement}} (servn.240);
%     \node [below right = 1cm and 5cm of user] {$\mathbf{\rightarrow}$};
%     \node [below = 5cm of user] {$\mathbf{\downarrow}$};
\end{tikzpicture}
}
\caption{Proof idea of converse bound. 
	By the secrecy conditions, the downloading step (a) can be considered as (b).
Here, we denote $\cA_{\vT} \coloneqq \bigotimes_{s=1}^{\vT} \cA_s$, $\cA_{\vT^c} \coloneqq \bigotimes_{s=\vT+1}^{\vN} \cA_s$, $Q_{\vT} \coloneqq (Q_1,\ldots, Q_{\vT})$, and $Q_{\vT^c} \coloneqq (Q_{\vT+1},\ldots,Q_{\vN})$ for $\pi \in \mathsf{S}$.
% Downloading Step of Protocol. 
% By the secrecy conditions, the system $\bigotimes_{s=1}^{\vT}\cA_{\pi(s)}$ of the colluding servers is independent of the file information.
% Therefore, the
}   \label{fig:ds}
\end{center}
\end{figure}

The proof idea of the converse bounds \eqref{ineq:conv2}, \eqref{ineq:conv3} is illustrated in Fig.~\ref{fig:ds} and is explained as follows.
By the secrecy conditions, 
	the state on $\bigotimes_{s=1}^{\vT}\cA_{\pi(s)}$ from the colluding servers is independent of the file information, which will be precisely stated in Lemma \ref{L1}.
With this fact, 
%Therefore, 
	the state on $\bigotimes_{s=1}^{\vT}\cA_{\pi(s)}$ can be considered as shared entanglement between the user and the non-colluding servers.
That is,
the downloading step of the protocol (Fig.~\ref{fig:ds}-(a)) can be considered as the entanglement-assisted communication of a classical message (Fig.~\ref{fig:ds}-(b)). 
Since the capacity of the entanglement-assisted classical communication \cite{BSST99} for the identity channel is two times the dimension of the transmitted quantum systems, the PIR capacity is upper bounded by $2(\vN-\vT)/\vN$ and the tightness of this bound is guaranteed by the QPIR protocol in Section~\ref{sec:protocol}.
The bound \eqref{ineq:conv1} is satisfied because the retrieved file size cannot exceed the dimension of downloaded quantum systems, but we give the formal proof in Appendix~\ref{append:converse}.

\subsection{Proof of converse bound \eqref{ineq:conv3}}

% Now, we prove the 
First, we prepare the following lemma for the converse bound \eqref{ineq:conv3}.
\begin{lemm} \label{lemm:diffffd}
Let $\Psi_{\mathrm{QPIR}}^{(\vM)}$ be a $\vT$-private QPIR protocol such that   
% $P_{\mathrm{err}}(\Psi_{\mathrm{QPIR}}^{(\vM)}) \le 1/2$,
	\begin{align}
	S_{\mathrm{serv}}(\Psi_{\mathrm{QPIR}}^{(\vM)})	&\le\beta,	\\
	S_{\mathrm{user}}^{(\vT)}(\Psi_{\mathrm{QPIR}}^{(\vM)}) &\le \gamma,	\\
	P_{\mathrm{err}}(\Psi_{\mathrm{QPIR}}^{(\vM)})  &\leq \min\{1/2,1 - 10\sqrt{2\vF\gamma} \}.
%	\\
%	P_{\mathrm{err}}(\Psi_{\mathrm{QPIR}}^{(\vM)})  &\leq 1 - 10\sqrt{2\vF\gamma}.
	\end{align}
Then, the protocol $\Psi_{\mathrm{QPIR}}^{(\vM)}$ satisfies
	\begin{align}
	&\log \vM  \nonumber
	\le 
	\frac{2 (\vN-\vT) \log \vD 
	 + f(P_{\mathrm{err}}(\Psi_{\mathrm{QPIR}}^{(\vM)}), \beta, \gamma, \vF) }
	 {1-P_{\mathrm{err}}(\Psi_{\mathrm{QPIR}}^{(\vM)})- 10\sqrt{2\vF\gamma}   },
	 \label{H6}
	\end{align}
	where 
	$f(\alpha, \beta, \gamma, \vF) \coloneqq
	\beta
	 +\eta_0(2\sqrt{2\vF\gamma})
	 + 2h_2(2\sqrt{2\vF\gamma})
	 + h_2\paren*{\alpha }$,
	$h_2(x)$ is the binary entropy function $h_2(x)\coloneqq -x\log x -(1-x)\log (1-x)$,
	and 
	\begin{align}
	\eta_0 (x) \coloneqq  
		\begin{cases}
		1/e & \text{if }  1/e < x,\\
		-x \log x  & \text{if } 0 < x < 1/e.
		\end{cases}
% 		\label{def:eta}
	\end{align}
\end{lemm}
Lemma~\ref{lemm:diffffd} will be proved in Section~\ref{lemmProffd}.

Now, we prove the converse bound \eqref{ineq:conv3}.
% {\noindent{\textbf{Step 3}:}\quad
% Finally, we prove the converse bound \eqref{ineq:conv3}.}
Let 
	$\{\Psi_{\mathrm{QPIR}}^{(\vM_\ell)}\}_{\ell=1}^{\infty}$
be any sequence of QPIR protocols 
	such that 
$$(\alpha_\ell, \beta_\ell, \gamma_\ell) \coloneqq (P_{\mathrm{err}}(\Psi_{\mathrm{QPIR}}^{(\vM_{\ell})}),S_{\mathrm{serv}}(\Psi_{\mathrm{QPIR}}^{(\vM_\ell)}) , {S}_{\mathrm{user}}^{(\vT)}(\qprot{\vM_{\ell}}) )$$
satisfies
\begin{align}
 \limsup_{\ell\to\infty} \alpha_{\ell} &= 0,\\
 \limsup_{\ell\to\infty} \gamma_{\ell} & = 0,\\
% \limsup_{\ell \to \infty} 
 \limsup_{\ell\to\infty} \beta_{\ell} 
	&=  \beta.
\end{align}
Let $\vD_\ell$ be the dimension of $\cA_s$ ($\forall s\in\{1,\ldots,\vN\}$) for the protocol $\Psi_{\mathrm{QPIR}}^{(\vM_\ell)}$.
Then, for any sufficiently large $\ell$ such that $\alpha_\ell\leq \min\{1/2,1 - 10\sqrt{2\vF\gamma_\ell} \}$,
	Lemma~\ref{lemm:diffffd} gives 
% 	the inequality $(d)$ can be rewritten as
\begin{align*}
  \log \vM_{\ell}  
  & \le 
%  \frac{ 2 (\vN-\vT) \log \vD_{\ell} + \beta_{\ell} + \eta_0(2\sqrt{2\vF\gamma})+ 2h_2(2\sqrt{2\vF\gamma_\ell}) + h_2\paren*{P_{\mathrm{err},k}(\Psi_{\mathrm{QPIR}}^{(\vM)})} }{1- P_{\mathrm{err},k}(\Psi_{\mathrm{QPIR}}^{(\vM)}) - 10\sqrt{2\vF\gamma_\ell} }\\
%   & \stackrel{\mathclap{(e)}}{\le}
  \frac{ 2 (\vN-\vT) \log \vD_{\ell} + 
    f(\alpha_\ell, \beta_\ell, \gamma_\ell, \vF) 
    %\beta_{\ell} + \eta_0(2\sqrt{2\vF\gamma_\ell}) +2h_2(2\sqrt{2\vF\gamma_\ell}) + h_2\paren*{\alpha_{\ell}} 
    }{1- \alpha_{\ell} - 10\sqrt{2\vF\gamma_\ell} } 
 .
\end{align*}
% where $(e)$ follows from $P_{\mathrm{err},k}(\Psi_{\mathrm{QPIR}}^{(\vM_\ell)})\le \alpha_\ell = P_{\mathrm{err}}(\Psi_{\mathrm{QPIR}}^{(\vM_\ell)}) < 1/2$ for  $P_{\mathrm{err}}(\Psi_{\mathrm{QPIR}}^{(\vM_\ell)})$.
Hence, 
the asymptotic QPIR rate satisfies
% we have
\begin{align*}
    \lim_{\ell \to \infty} R(\Psi_{\mathrm{QPIR}}^{(\vM_\ell)}) 
    &= \lim_{\ell \to \infty}\frac{\log \vM_{\ell}}{\vN \log \vD_{\ell}}\\
    &\le \lim_{\ell \to \infty} 
		\frac{ 2 (\vN-\vT) \log \vD_{\ell} + 
            f(\alpha_\ell, \beta_\ell, \gamma_\ell, \vF) 
            %\beta_{\ell} + \eta_0(2\sqrt{2\vF\gamma_\ell})+ 2h_2(2\sqrt{2\vF\gamma_\ell}) + h_2\paren*{\alpha_{\ell}} 
            }{(1- \alpha_{\ell} - 10\sqrt{2\vF\gamma_\ell})\vN \log \vD_{\ell}  }\\
% 		\frac{2(\vN-\vT)}{\vN (1- \alpha_{\ell} - 16\sqrt{2\vF\gamma_\ell})} + 
% 	\frac{\beta_{\ell} + 4h_2(2\sqrt{2\vF\gamma_\ell}) + h_2\paren*{\alpha_\ell} }{ \vN \log \vD_{\ell} }
	&= \frac{2(\vN-\vT)}{\vN},  
\end{align*}
where the last equality follows from the relation $(\alpha_{\ell}, \beta_{\ell}, \gamma_\ell, \vD_{\ell}) \to (0,\beta,0,\infty)$ as $\ell \to \infty$.
Thus, we obtain the converse bound \eqref{ineq:conv3}.

\subsection{Proof of Lemma~\ref{lemm:diffffd}} \label{lemmProffd}

% In the remainder of this section, we present the proof of \eqref{ineq:conv3}.
In this section, we prove Lemma~\ref{lemm:diffffd}.
Without losing generality, we fix an arbitrary $\pi\in\mathsf{S}_{\vN}$ in the following proofs.
We use the notation 
\begin{align}
Q_{\vT} &\coloneqq (Q_{\pi(1)},\ldots, Q_{\pi(\vT)}),\\
Q_{\vT^c} &\coloneqq (Q_{\pi(\vT+1)},\ldots, Q_{\pi(\vN)}),\\
\cA_{\vT} &\coloneqq \bigotimes_{s=1}^{\vT} \cA_{\pi(s)},\\
\cA_{\vT^c} &\coloneqq \bigotimes_{s=\vT+1}^{\vN} \cA_{\pi(s)}.
\end{align}
% Let $\vD_{\vT} \coloneqq \vD^\vT $ and $\vD^{\vN-\vT} \coloneqq \vD^{\vN-\vT}$ be the dimensions of $ \cA_{\vT}$ and $\cA_{\vT^c}$, respectively.
We also denote by $\rho(M,Q_{\vT})$ the state on $\cA_{\vT}$ of the $\vT$-colluding servers after the servers' encoding.

To show Lemma~\ref{lemm:diffffd}, we prepare the following lemma.
\begin{lemm}\label{L1}
If a QPIR protocol $\Psi_{\mathrm{QPIR}}^{(\vM)}$ satisfies
$S_{\mathrm{serv}}(\Psi_{\mathrm{QPIR}}^{(\vM)}) \le \beta$ and ${S}_{\mathrm{user}}^{(\vT)}(\Psi_{\mathrm{QPIR}}^{(\vM)}) \le \gamma$,
then
for any $k \in \{1,\ldots, \vF\}$,
we have the relation 
\begin{align}
I(M_k;\cA_{\vT}|Q_{\vT},K=k)_{\rho(M,Q_{\vT})} \le  \beta
 + g(\vM,\gamma) 
\end{align}
after the servers' encoding, where 
\begin{align*}
% f(\vM,\alpha) &\coloneqq \alpha \log \vM + h_2(\alpha), \\
  g(\vM,\gamma) \coloneqq 10\sqrt{2\vF\gamma} \log \vM + \eta_0(2\sqrt{2\vF\gamma}) + 2h_2(2\sqrt{2\vF\gamma}).
% 
% g(\vM,\gamma) &\coloneqq \min \px*{ 16\sqrt{2\vF\gamma} \log \vM + 4h(2\sqrt{2\vF\gamma}),\
% \ 
% 4\log \vM  } .
%\label{eq:gdef}
\end{align*}
In particular, when $\gamma=0$, 
we have $I(M_k;\cA_{\vT}|Q_{\vT}, K=k)_{\rho(M,Q_{\vT})} \le  \beta$ for any $k \in \{1,\ldots, \vF\}$. 
\end{lemm}
Lemma~\ref{L1} is proved in Appendix~\ref{append:upper}.
We also prepare the following proposition for the classical and quantum relative entropies defined in Section~\ref{sec:quantum_basic}. %\eqref{def:crel} and \eqref{def:qrel}.
\begin{prop}[{\cite[(3.18)]{Hay17} }] \label{prop:ineqda}
The information-processing inequality for quantum relative entropy with respect to the two-valued measurement $\{Y,I-Y\}$ is written as
% $p = \{p_i\}_i = \{
\begin{align*}
D(\rho \| \sigma)  &\geq D(P_{\rho}\|P_{\sigma}) \\
    &= -h_2(P_{\rho}(1)) - P_{\rho}(1) \log P_{\sigma}(1) - P_{\rho}(2) \log P_{\sigma}(2),
%\label{eq:info_pr}
\end{align*} 
where $P_{\rho} = \{P_{\rho}(1), P_{\rho}(2)\} = \{\Tr \rho Y, \Tr \rho (I-Y) \}$, $P_{\sigma} = \{P_{\sigma}(1), P_{\sigma}(2)\} = \{\Tr \sigma Y,  \Tr \sigma (I-Y)\}$.
\end{prop}
% どうやって示すのか本文で簡単に説明してください。
% }

Now, we prove Lemma~\ref{lemm:diffffd} by four steps.

\noindent{\textbf{Step 1}:}\quad
% For the proof of \eqref{H4},
First, we prepare the following notation.
Fix $K=k$ arbitrarily.
Let $\rho(m_k,m_k^c,q)$ be the quantum state on the composite system $\bigotimes_{s=1}^{\vN} \cA_s$,
where $m_k$ is the file to be retrieved, $m_k^c$ is the collection of non-retrieved $\vF-1$ files,
 and $q$ is the collection of queries.
Note that 
	in view of Fig.~\ref{fig:ds}-(b), 
	the target file $m_k$ corresponds to the classical message and 
	the query $q$ determines the decoding algorithm, 
	but $m_k^c$ is redundant information since it is independent of $m_k$ and $q$.
Therefore, in the following,
    we only consider the states averaged with respect to $m_k^c$ as
\begin{align}
    \rho(m_k,q) &= \frac{1}{\vM^{\vF-1}} \sum_{m_k^c} \rho(m_k,m_k^c, q) ,
    \\
    \rho(q) &= \frac{1}{\vM} \sum_{m_k=1}^{\vM} \rho(m_k, q)
    .
\end{align}
Considering the entire system $\cA$ as a bipartite system $\cA_{\vT}\otimes \cA_{\vT^c}$,
let $\rho(m_k,q_{\vT})$ and $\rho(q_{\vT}) $ be the reduced densities of $\rho(m_k,q)$ and $\rho(q)$ on $ \cA_{\vT}$, respectively.
Remind that the decoding POVM is denoted by $\{ Y_{k,q}(w) \}_{w\in\{1,\ldots,\vM\}}$, which depends on $k$ and $q$.
Then, we define the states 
\begin{align}
\tilde{\rho}(M_k|q)&\coloneqq \sum_{m_k=1}^{\vM} \frac{1}{\vM} | m_k \rangle \langle m_k| \otimes \rho(m_k,q),\\
\tilde{\rho}(q_\vT)&\coloneqq \sum_{m_k=1}^{\vM} \frac{1}{\vM} |  m_k \rangle \langle m_k| \otimes \rho(q_{\vT}) \otimes I/\vD^{\vN-\vT} \\
		&= \frac{I}{\vM}\otimes \rho(q_{\vT}) \otimes \frac{I}{\vD^{\vN-\vT}}
\end{align}
and the operator $Y_{k,q}\coloneqq  \sum_{m_k=1}^{\vM} | m_k \rangle \langle m_k | \otimes Y_{k,q}(m_k)$.

\noindent{\textbf{Step 2}:}\quad
In this step, we derive 
\begin{align}
&{(1-P_{\mathrm{err},k}(\Psi_{\mathrm{QPIR}}^{(\vM)}) )\log \vM} \nonumber \\
& \le 
\mathbb{E}_{Q}  D( \tilde{\rho}(M_k|Q) \| \tilde{\rho}(Q_\vT) ) {+ h_2\paren*{P_{\mathrm{err},k}(\Psi_{\mathrm{QPIR}}^{(\vM)})} }
\label{H4}
\end{align}
for any $k\in\{1,\ldots,\vF\}$,
where 
	\begin{align}
% 	\tilde{\rho}(M_k) &\coloneqq  \Pr[Q=q]\tilde{\rho}(M_k|q),\\
% 	\tilde{\rho} &\coloneqq  \Pr[Q_\vT=q_\vT]\tilde{\rho}(q_\vT),\\
	P_{\mathrm{err},k}(\Psi_{\mathrm{QPIR}}^{(\vM)}) 
		&\coloneqq \pr_{W,M_k} [\ww \neq M_k| K=k].
	\end{align}
We prove \eqref{H4} by applying Proposition~\ref{prop:ineqda} to the states $\tilde{\rho}(M_k|q)$ and $\tilde{\rho}(q_\vT)$ 
	and the measurement $\{Y_{k,q},I-Y_{k,q}\}$ as follows.
% 	When we apply Proposition~\ref{prop:ineqda} to
% 	the states $\tilde{\rho}(M_k|q)$ and $\tilde{\rho}(q_\vT)$
% 	and the measurement $\{Y_{k,q},I-Y_{k,q}\}$,
	In this application of Proposition~\ref{prop:ineqda},
	the distributions $P_{\tilde{\rho}(M_k|q)}$ and $P_{\tilde{\rho}(q_\vT)}$ are written as 
	\begin{align}
	P_{\tilde{\rho}(M_k|q)} &= \{P_{\tilde{\rho}(M_k|q)}(1), P_{\tilde{\rho}(M_k|q)}(2)\} \nonumber \\
        &= \{P_{\mathrm{err},k,q}(\Psi_{\mathrm{QPIR}}^{(\vM)}), 1- P_{\mathrm{err},k,q}(\Psi_{\mathrm{QPIR}}^{(\vM)})\}, \label{whyQ}\\
	P_{\tilde{\rho}(q_\vT)} &= \{P_{\tilde{\rho}(q_\vT)}(1), P_{\tilde{\rho}(q_\vT)}(2)\} = \px*{\frac{1}{\vM},  1-\frac{1}{\vM} }, \nonumber 
	\end{align}
where
$$P_{\mathrm{err},k,q}(\Psi_{\mathrm{QPIR}}^{(\vM)}) \coloneqq \pr_{W,M_k} [\ww \neq M_k| K=k, Q=q	].$$
% Here, $P_{\tilde{\rho}(q_\vT)}$ is independent of $k$ and $q$.
Thus, applying Proposition~\ref{prop:ineqda}, we have
% From Eq.~\eqref{eq:info_pr}, we have
	\begin{align}
	&(1-P_{\mathrm{err},k,q}(\Psi_{\mathrm{QPIR}}^{(\vM)}) )\log \vM  \nonumber \\
		&= - P_{\tilde{\rho}(M_k|q)}(2)\log P_{\tilde{\rho}(q_\vT)}(2)	\nonumber \\
		&\stackrel{\mathclap{{(a)}}}{\le} - P_{\tilde{\rho}(M_k|q)}(2)\log P_{\tilde{\rho}(q_\vT)}(2) - P_{\tilde{\rho}(M_k|q)}(1)\log P_{\tilde{\rho}(q_\vT)}(1) \nonumber \\
% 		D(\tilde{\rho}(M_k|q)\|\tilde{\rho}(q_\vT))+h_2(P_{\tilde{\rho}(M_k|q)}(1))	\\
		&\le D\paren*{\tilde{\rho}(M_k|q) \|  \tilde{\rho}(q_\vT)} +h_2\paren*{P_{\tilde{\rho}(M_k|q)}(1)}	\nonumber \\
		&= D(\tilde{\rho}(M_k|q) \| \tilde{\rho}(q_\vT)) + h_2\paren*{P_{\mathrm{err},k,q}(\Psi_{\mathrm{QPIR}}^{(\vM)})},
		\label{eq:whyqq}
	\end{align}
where ${(a)}$ is from $P_{\tilde{\rho}(M_k|q)}(1)\log P_{\tilde{\rho}(q_\vT)}(1) \leq 0$.
Taking the expectation with respect to $Q$ and using the concavity of $h_2$, we have
\begin{align}
&{(1-P_{\mathrm{err},k}(\Psi_{\mathrm{QPIR}}^{(\vM)}) )\log \vM} \nonumber \\
& = \mathbb{E}_{Q} {(1-P_{\mathrm{err},k,Q}(\Psi_{\mathrm{QPIR}}^{(\vM)}) )\log \vM} \label{EQQ}\\
& \le 
\mathbb{E}_{Q}  D( \tilde{\rho}(M_k|Q) \| \tilde{\rho}(Q_\vT) ) + \mathbb{E}_{Q} h_2\paren*{P_{\mathrm{err},k,Q}(\Psi_{\mathrm{QPIR}}^{(\vM)})} \nonumber \\
& \le
\mathbb{E}_{Q}  D( \tilde{\rho}(M_k|Q) \| \tilde{\rho}(Q_\vT) ) + h_2\paren*{P_{\mathrm{err},k}(\Psi_{\mathrm{QPIR}}^{(\vM)})} ,
\label{SS1}
\end{align}
which is the desired inequality \eqref{H4}.
% where 
% 	\begin{align}
% % 	\tilde{\rho}(M_k) &\coloneqq  \Pr[Q=q]\tilde{\rho}(M_k|q),\\
% % 	\tilde{\rho} &\coloneqq  \Pr[Q_\vT=q_\vT]\tilde{\rho}(q_\vT),\\
% 	P_{\mathrm{err},k}(\Psi_{\mathrm{QPIR}}^{(\vM)}) 
% 		&\coloneqq \mathbb{E}_Q P_{\mathrm{err},k,Q}(\Psi_{\mathrm{QPIR}}^{(\vM)})
% 		= \pr_{W,M_k} [\ww \neq M_k| K=k].
% 	\end{align}
% 
% \begin{align}
% \mathbb{E}_Q P_{\mathrm{err},k,Q}(\Psi_{\mathrm{QPIR}}^{(\vM)})
% 		= 
% 	P_{\mathrm{err},k}(\Psi_{\mathrm{QPIR}}^{(\vM)}) 
% \end{align}
% 

\noindent{\textbf{Step 3}:}\quad
Next, we derive
\begin{align}
\mathbb{E}_{Q} 
 &D(\tilde{\rho}(M_k|Q) \| \tilde{\rho}(Q_\vT) ) \nonumber \\
 &\leq 2(\vN-\vT)\log \vD + I(M_k;\cA_{\vT}|Q_{\vT},K=k)_{\rho(M,Q_{\vT})} \label{H33}
\end{align}
as follows.
The inequality \eqref{H33} is derived by 
\begin{align*}
&\mathbb{E}_{Q} D(\tilde{\rho}(M_k|Q) \| \tilde{\rho}(Q_\vT) ) \\
&=
\mathbb{E}_{Q} 
\frac{1}{\vM} \sum_{m_k=1}^{\vM}   
D\paren*{ \rho(m_k,Q) \bigg\| \rho(Q_{\vT})\otimes \frac{I}{\vD^{\vN-\vT}}}  \\
&
\stackrel{\mathclap{(b)}}{=} 
\mathbb{E}_{Q} 
\frac{1}{\vM} \sum_{m_k=1}^{\vM}
D\paren*{ \rho(m_k,Q)\bigg\| \rho(m_k,Q_{\vT})\otimes \frac{I}{\vD^{\vN-\vT}}} \\
&\quad +
\mathbb{E}_{Q} 
\frac{1}{\vM} \sum_{m_k=1}^{\vM}
D( \rho(m_k,Q_{\vT})\| \rho(Q_{\vT}))  \\
&
=
\mathbb{E}_{Q} 
\frac{1}{\vM} \sum_{m_k=1}^{\vM}
D\paren*{ \rho(m_k,Q)\bigg\| \rho(m_k,Q_{\vT})\otimes \frac{I}{\vD^{\vN-\vT}} }
    \\
    &\quad +
  I(M_k;\cA_{\vT}|Q_{\vT},K=k)_{\rho(M,Q_{\vT}) } 
\\
&
\stackrel{\mathclap{(c)}}{\le} 
2(\vN-\vT)\log \vD + I(M_k;\cA_{\vT}|Q_{\vT},K=k)_{\rho(M,Q_{\vT})} . %\label{H3} %_{\rho_{M_k,Z}}
\end{align*}
The equation $(b)$ can be shown as follows.
\begin{align*}
 & { D\paren*{ \rho(m_k,Q) \bigg\| \rho(Q_{\vT})\otimes \frac{I}{\vD^{\vN-\vT}}}  } \\
 &=
\Tr  \rho(m_k,Q)\paren*{\log \rho(m_k,Q) - \log  \paren*{\rho(Q_{\vT})\otimes \frac{I}{\vD^{\vN-\vT}}} }\\
 &=
\Tr  \rho(m_k,Q)
    \Big(\log \rho(m_k,Q)- \log  \Big(\rho(m_k,Q_{\vT})\otimes \frac{I}{\vD^{\vN-\vT}}\Big) 
            \\
	& \quad +\log \Big( \rho(m_k,Q_{\vT}) \otimes \frac{I}{\vD^{\vN-\vT}} \Big)-\log  \Big( \rho(Q_{\vT})\otimes \frac{I}{\vD^{\vN-\vT}} \Big) \Big) \\
 &=
D\paren*{ \rho(m_k,Q)\bigg\|  \rho(m_k,Q_{\vT})\otimes \frac{I}{\vD^{\vN-\vT}}} \\
    &\quad  + D( \rho(m_k,Q_{\vT})\| \rho(Q_{\vT})) .
\end{align*}
The inequality $(c)$ can be shown as follows.
We diagonalize $\rho(m_k,Q)=\sum_i p_i |\phi_i\rangle \langle \phi_i|$
and denote by $\rho_i$ the reduced density of $|\phi_i\rangle \langle \phi_i| $ on $ \cA_{\vT}$.
Then, 
\begin{align*}
 & D\paren*{ \rho(m_k,Q) \bigg\| \rho(m_k,Q_{\vT})\otimes \frac{I}{\vD^{\vN-\vT}} } \\
 &\le 
	\sum_i p_i  D\paren*{ |\phi_i\rangle \langle \phi_i| \bigg\| \rho_i\otimes \frac{I}{\vD^{\vN-\vT}} }    
	\\
 &=\log \vD^{\vN-\vT}+\sum_i p_i H(\rho_i)  
 \le 2 \log \vD^{\vN-\vT},
\end{align*}
where the last inequality is proved from 
$H(\rho_i) = H(\rho_i') \leq \log \vD^{\vN-\vT}$ for the reduced density $\rho_i'$ of $|\phi_i\rangle\langle\phi_i|$ on $\cA_{\vT^c}$.

\noindent{\textbf{Step 4}:}\quad
Lastly, we prove \eqref{H6} of Lemma~\ref{lemm:diffffd}.
Combining \eqref{H4}, \eqref{H33}, and Lemma~\ref{L1}, we have
\begin{align*}
%   &(1-P_{\mathrm{err}}(\Psi_{\mathrm{QPIR}}^{(\vM)}) ) \log \vM  \\
%   &\stackrel{(c)}{\le} 
  &(1-P_{\mathrm{err},k}(\Psi_{\mathrm{QPIR}}^{(\vM)}) ) \log \vM   \\
  &\le 
		2  (\vN-\vT) \log \vD +
		I(M_k;\cA_{\vT}|Q_{\vT},K=k)_{\rho(M,Q_{\vT})}\\
		&\quad + h_2\paren*{P_{\mathrm{err},k}(\Psi_{\mathrm{QPIR}}^{(\vM)})  }\\
 &\le  2 (\vN-\vT) \log \vD 
		+ \beta +  g(\vM,\gamma) 
		+ h_2\paren*{P_{\mathrm{err},k}(\Psi_{\mathrm{QPIR}}^{(\vM)})  }	\\
 &=		2 (\vN-\vT) \log \vD 
	 +\beta 
	 +10\sqrt{2\vF\gamma} \log \vM + \eta_0(2\sqrt{2\vF\gamma}) \\
% 	 +\min \px*{ 16\sqrt{2\vF\gamma} \log \vM + 4h_2(2\sqrt{2\vF\gamma}),\ 4\log \vM  }	
	 &\quad + 2h_2(2\sqrt{2\vF\gamma})+ h_2\paren*{P_{\mathrm{err},k}(\Psi_{\mathrm{QPIR}}^{(\vM)})  }  	\\
 &= 2 (\vN-\vT) \log \vD  + f(P_{\mathrm{err},k}(\Psi_{\mathrm{QPIR}}^{(\vM)}), \beta, \gamma, \vF)
\end{align*}
% 	and $(d)$ follows because $\gamma$ is taken to satisfy $g(\vM,\gamma) = 16\sqrt{2\vF\gamma} \log \vM + 4h_2(2\sqrt{2\vF\gamma})$. 
% Taking the expectation with respect to $k$ and using the concavity of $h$, the inequality \eqref{H5} can be written independently of $k$ as
	Then, rewriting the above inequality, we obtain Lemma~\ref{lemm:diffffd} as 
% 		the desired inequality \eqref{H6}.
% 	\begin{align}
% 	&(1-P_{\mathrm{err},k}(\Psi_{\mathrm{QPIR}}^{(\vM)})- 10\sqrt{2\vF\gamma}  ) \log \vM  \nonumber\\
% 	&\stackrel{\mathclap{(d)}}{\le} 2 (\vN-\vT) \log \vD 
% 	 +\beta 
% 	 +\eta_0(2\sqrt{2\vF\gamma})
% 	 + 2h_2(2\sqrt{2\vF\gamma})
% 	 + h_2\paren*{P_{\mathrm{err},k}(\Psi_{\mathrm{QPIR}}^{(\vM)})  } .
% 	 \label{H7}
% 	\end{align}
\begin{align*}
  \log \vM
  & \le 
 \frac{ 2 (\vN-\vT) \log \vD  + f(P_{\mathrm{err},k}(\Psi_{\mathrm{QPIR}}^{(\vM)}), \beta, \gamma, \vF)}
 {1- P_{\mathrm{err},k}(\Psi_{\mathrm{QPIR}}^{(\vM)}) - 10\sqrt{2\vF\gamma} }\\
  & \stackrel{\mathclap{(d)}}{\le}
  \frac{ 2 (\vN-\vT) \log \vD +
  f(P_{\mathrm{err}}(\Psi_{\mathrm{QPIR}}^{(\vM)}), \beta, \gamma, \vF)} {{1- P_{\mathrm{err}}(\Psi_{\mathrm{QPIR}}^{(\vM)})} - 10\sqrt{2\vF\gamma} } 
 ,
\end{align*}
where $(d)$ holds because $P_{\mathrm{err},k}(\Psi_{\mathrm{QPIR}}^{(\vM)}) \le P_{\mathrm{err}}(\Psi_{\mathrm{QPIR}}^{(\vM)}) < 1/2$ and the binary entropy $h_2(p)$ is inceasing for $p\leq 1/2$.

\begin{remark}
In Step 2, we condition on $Q=q$ and then take expectation with respect to $Q$.
The reason why we condition on $Q=q$ is that the states and the decoder are determined depending on the value of $Q$.
Thus, to derive \eqref{whyQ} which relates the error probability and Proposition~\ref{prop:ineqda}, we need to condition on $Q=q$.
On the other hand, we need to take expectation on $Q$ in \eqref{SS1} because 
	we need to recover $Q_{\vT}$ as a random variable to apply the user secrecy condition $I(K; Q_{\vT}) \le \gamma$.
To be precise, 
	we use the user secrecy condition $I(K; Q_{\vT}) \le \gamma$ in the proof of Lemma~\ref{L1}
	and 
	we apply Lemma~\ref{L1} in Step~4.
%Thus, $Q_{\vT}$ should be a random variable but not a fixed value.
\end{remark}

\section{Conclusion} \label{sec:conclusion}

We have discussed symmetric $\vT$-private quantum private information retrieval. %, where the identity of the target file  is secret to the colluding $\vT$ servers.
We have considered two secrecy conditions, the user $\vT$-secrecy and the server secrecy.
The user $\vT$-secrecy is the secrecy in which any collection of $\vT$ queries contains no information of the user's request,
    and the server secrecy is the secrecy in which the user obtains no information of other files than the target file.
We have derived two capacities of $\vT$-private QPIR, the exact and asymptotic security-constrained capacities.
When the number of colluding servers $\vT$ is less than or equal to half of the number of servers $\vN$,
    the capacities are exactly $1$ whether considering the security conditions or not and 
    if $\vT>\vN/2$,
    the capacities are $2(\vN-\vT)/\vN$.
%     decreases by $2/\vN$ as the $\vT$ increases by one.
For the proof of the capacities, we have constructed a symmetric $\vT$-private QPIR protocol with perfect security conditions by the stabilizer formalism.
We have also derived the converse bounds, which complete the optimality of our protocol.

There are many open problems related to $\vT$-private QPIR.
$\vT$-Private QPIR with adversarial models needs to be discussed since we only considered the most trivial security model in which the user and the servers follow the protocol and do not deviate from the protocol. 
$\vT$-Private QPIR capacity should also be discussed for the multi-round case and for the model without shared entanglement.

\section*{Acknowledgments}
The authors are grateful to Yuuya Yoshida for providing a concise proof of Proposition \ref{prop:al_ex} and helpful comments.
% SS is supported by Lotte Foundation and JSPS Grant-in-Aid for JSPS Fellows No. JP20J11484.
% MH is supported in part by a JSPS Grant-in-Aids for 
% 	Scientific Research (A) No.17H01280 and for Scientific Research (B) No.16KT0017, 
% 	and Kayamori Foundation of Information Science Advancement.

\appendices

\section{QPIR Capacity with Average Security Measures} \label{append:average}

In Section~\ref{sec:securityme},
	we defined the security measures as the worst-case definition.
In this appendix, 
	we show that the capacity does not changes even if we change the definition of the security measures as for the average case.

Define the average security measures as
	\begin{align}
\tilde{P}_{\mathrm{err}}(\Psi_{\mathrm{QPIR}}^{(\vM)}) 
	&\coloneqq 
    \pr_{\ww} [\ww \neq M_K | M Q K]                  \\
\tilde{S}_{\mathrm{serv}}(\Psi_{\mathrm{QPIR}}^{(\vM)}) &\coloneqq 
    I(M_K^c; \cA|QK)_{\rho(M,Q)}
    \\
\tilde{S}_{\mathrm{user}}^{(\vT)}(\Psi_{\mathrm{QPIR}}^{(\vM)}) &\coloneqq 
	\mathbb{E}_{\pi\in\mathsf{S}_{\vN}} I(K; Q_{\pi,\vT} ), 
\end{align}
and QPIR capacities
$\tilde{C}_{\mathrm{exact},\vT}^{\alpha,\beta,\gamma,\theta}$ and $\tilde{C}_{\mathrm{asymp},\vT}^{\alpha,\beta,\gamma,\theta}$ 
	are defined the same as 
	\eqref{eq:capa1} and \eqref{eq:capa2} except that
		the security measures
		${P}_{\mathrm{err}}(\Psi_{\mathrm{QPIR}}^{(\vM)}), 
		{S}_{\mathrm{serv}}(\Psi_{\mathrm{QPIR}}^{(\vM)}),
		{S}_{\mathrm{user}}^{(\vT)}(\Psi_{\mathrm{QPIR}}^{(\vM)})$ 
		are replaced by 
	$\tilde{P}_{\mathrm{err}}(\Psi_{\mathrm{QPIR}}^{(\vM)})$, 
		$\tilde{S}_{\mathrm{serv}}(\Psi_{\mathrm{QPIR}}^{(\vM)})$,
		$\tilde{S}_{\mathrm{user}}^{(\vT)}(\Psi_{\mathrm{QPIR}}^{(\vM)})$.
Then, similar to Theorem~\ref{theo:main}, the average capacity is derived as 
	\begin{align}
    \tilde{C}_{\mathrm{asymp},\vT}^{\alpha,\beta,\gamma,\theta} 
    &=
    \tilde{C}_{\mathrm{exact},\vT}^{\alpha,\beta,\gamma,\theta}
    = 
        1       & \text{ if $1\leq \vT\leq \frac{\vN}{2}$},
        \label{111EQ}
        \\
    \tilde{C}_{\mathrm{asymp},\vT}^{0,\beta,0,\theta} 
    &=
    \tilde{C}_{\mathrm{exact},\vT}^{\alpha,0,0,\theta}
    =
        \frac{2(\vN-\vT)}{\vN} & \text{ if $\frac{\vN}{2} < \vT < \vN$}.
        \label{111EQ2}
    \end{align}
For the achievability proof of \eqref{111EQ} and \eqref{111EQ2}, the QPIR protocol in Section~\ref{sec:protocol} achieves the capacity.

The converse bounds are also proved similar to the case of the worst-case security.
The converse bounds are written for any  $\alpha\in [0, 1)$ and any $\beta,\gamma, \theta \in [0,\infty)$ as
\begin{align}
\tilde{C}_{\mathrm{asymp},\vT}^{\alpha,\beta,\gamma,\theta} &\leq 1 & \text{if $1\leq \vT\leq \frac{\vN}{2}$},  \label{cb1}\\
\tilde{C}_{\mathrm{exact},\vT}^{\alpha,0,0,\theta}  &\leq \frac{2(\vN-\vT)}{\vN} & \text{if $\frac{\vN}{2} < \vT < \vN$}, \label{cb2}\\
\tilde{C}_{\mathrm{asymp},\vT}^{0,\beta,0,\theta}         &\leq \frac{2(\vN-\vT)}{\vN} & \text{if $\frac{\vN}{2} < \vT < \vN$}.
\label{cb3}
\end{align}

First, the converse bound \eqref{cb3} is proved by the same steps as Section~\ref{sec:converse} except for the following part.
In Section~\ref{sec:converse},
	Eq.~\eqref{eq:whyqq} is written as 
	\begin{align*}
	&(1-P_{\mathrm{err},k,q}(\Psi_{\mathrm{QPIR}}^{(\vM)}) )\log \vM \\
	&\le D(\tilde{\rho}(M_k|q) \| \tilde{\rho}(q_\vT)) + h_2\paren*{P_{\mathrm{err},k,q}(\Psi_{\mathrm{QPIR}}^{(\vM)})},
	\end{align*}
	and by taking the expectation of \eqref{eq:whyqq} with respect to $Q$, we obtain \eqref{EQQ}.
Similarly, we take the expectation of \eqref{eq:whyqq} with respect to $K,Q$ and then, we obtain
	\begin{align*}
	&{(1-\tilde{P}_{\mathrm{err}}(\Psi_{\mathrm{QPIR}}^{(\vM)}) )\log \vM} \\
    &\le
	\mathbb{E}_{K,Q}  D( \tilde{\rho}(M_K|Q) \| \tilde{\rho}(Q_\vT)  | K) + h_2\paren*{\tilde{P}_{\mathrm{err}}(\Psi_{\mathrm{QPIR}}^{(\vM)})} .
	\end{align*}
Then, by the similar steps as Section~\ref{sec:converse}, the converse bound \eqref{cb3} is proved.

For the converse bounds \eqref{cb1} and \eqref{cb2},
	in the last steps of the proofs in Appendix~\ref{append:converse},  
	we replace \eqref{EQQ1} and \eqref{EQQ2} by
	\begin{align}
	1- \tilde{P}_{\mathrm{err}}(\Psi_{\mathrm{QPIR}}^{(\vM_\ell)})  
	\leq 1 - \inf_{k,q} P_{\mathrm{err},k,q}(\Psi_{\mathrm{QPIR}}^{(\vM_\ell)}).
	\end{align}
Then, the converse bounds \eqref{cb1} and \eqref{cb2} are proved in the same way.

\section{Proof of Proposition~\ref{prop:stab}} \label{append:stab}

We concretely construct the subgroup $S(\VV)$ as follows.
From \eqref{eq:commutative}, all elements of $S(\VV)$ are commutative regardless of the choice of $c_{\mathbf{v}}$.
Since $I \in S(\VV)$, we set $\mathbf{W(0)} = I$, i.e., $c_{\mathbf{0}} = 1$.
Then, it is enough to choose $c_{\mathbf{v}}$ so that $S(\VV)$ satisfies the closure for the multiplication.

We choose $\{c_{\mathbf{v}}\in \mathbb{C}\mid \mathbf{v} \in \VV\}$ as follows.
For a fixed basis $\mathbf{v}_1,\ldots, \mathbf{v}_d$ of $\VV$,
	we choose $c_{\mathbf{v}_i}$ as follows:
	if $p>2$, choose $c_{\mathbf{v}_i}$ as a $p$-th root of unity, i.e., $c_{\mathbf{v}_i} = \omega^k$ for some integer $k$;
	if $p=2$, 
		choose $c_{\mathbf{v}_i}$ as 
		$$c_{\mathbf{v}_i} = 
		\pm \sqrt{-1}^{\langle \mathbf{b}_i, \mathbf{a}_i\rangle}
		=
		\begin{cases}
		\pm 1 	& \text{if $\langle \mathbf{b}_i, \mathbf{a}_i\rangle = 0$},\\
		\pm \sqrt{-1} & \text{if $\langle \mathbf{b}_i, \mathbf{a}_i\rangle = 1$},\\
		\end{cases}
			$$
		where $\mathbf{a}_i, \mathbf{b}_i$ are vectors in $\mathbb{F}_q^n$ such that $(\mathbf{a}_i, \mathbf{b}_i) = \mathbf{v}_i$.
For any $\mathbf{v} = \sum_{i} a_i  \mathbf{v}_i \in \VV$, we choose $c_{\mathbf{v}}$ 
% 	define $\mathbf{W(v)}$ uniquely as
	by the relation
	\begin{align}
	\mathbf{W(v)} &= \mathbf{W}(\mathbf{v}_1)^{a_1} \cdots \mathbf{W}(\mathbf{v}_d)^{a_d}.
% 		&= c_{\mathbf{v}_1} \cdots  c_{\mathbf{v}_n} \mathbf{\tilde{W}}(\mathbf{v}_1) \cdots \mathbf{\tilde{W}}(\mathbf{v}_n)	\\
% 		&= c_{\mathbf{v}_1} \cdots  c_{\mathbf{v}_n} \sqrt{-1}^{ \sum_{i,j: i\neq j} \tr  \mathbf{v}_{i,Z} \mathbf{v}_{j,X} }
% 			\mathbf{\tilde{W}(v)}.
	\end{align}

Next, we prove the closure for the multiplication in $S(\VV)$ by the above choice of $c_{\mathbf{v}}$.
For any basis element $\mathbf{v}_i$, we have 
	\begin{align}
	\mathbf{W}(\mathbf{v}_i)^p 
					&= c_{\mathbf{v}_i}^p \mathbf{\tilde{W}(v}_i)^p 
					\stackrel{\mathclap{(a)}}{=} c_{\mathbf{v}_i}^p \omega^{p(p-1)\langle \mathbf{b}_i, \mathbf{a}_i\rangle /2} \mathbf{\tilde{W}}(p\mathbf{v}_i) \nonumber \\
					&= c_{\mathbf{v}_i}^p \omega^{p(p-1)\langle \mathbf{b}_i, \mathbf{a}_i\rangle /2} I
					= I ,
					\label{eq:IdsfIIIII}
	\end{align}
	where $(a)$ follows from \eqref{eq:sum}.
Then, we can confirm %\eqref{eq:cond111ffsd},
	the closure for the multiplication as
\begin{align}
% \mathbf{W(v)}^p &= I,\\
    &\mathbf{W(v) W(v') } \nonumber \\
    &= \mathbf{W}(\mathbf{v}_1)^{a_1} \cdots \mathbf{W}(\mathbf{v}_d)^{a_d}
						\mathbf{W}(\mathbf{v}_1)^{a_1'} \cdots \mathbf{W}(\mathbf{v}_d)^{a_d'} \nonumber\\
					&= \mathbf{W}(\mathbf{v}_1)^{a_1+a_1'} \cdots \mathbf{W}(\mathbf{v}_d)^{a_d+a_d'} \label{eqsf:sfocvv}\\
					&= \mathbf{W}(\mathbf{v}_1)^{a_1+a_1' \!\!\!\!\!\mod p} \cdots \mathbf{W}(\mathbf{v}_d)^{a_d+a_d'\!\!\!\!\!\mod p} \label{eqsf:sfocvv2}\\
					&= \mathbf{W(v+v')} \nonumber
\end{align}
for any $\mathbf{v}, \mathbf{v'}\in \VV$,
where the equality \eqref{eqsf:sfocvv} is from the commutative property of $S(\VV)$ and 
the equality \eqref{eqsf:sfocvv2} is from \eqref{eq:IdsfIIIII}.
Thus, $S(\VV)$ is a commutative subgroup of $\mathrm{HW}_q^n$ not containing $cI$ for any $c\neq 0$, i.e., a stabilizer.

Alternatively, if $p> 2$, the set $S(\VV)$ is a stabilizer 
	by choosing $c_{(\mathbf{a,b})} = (\omega^{(p+1)/2})^{\langle \mathbf{a}, \mathbf{b} \rangle}$ 
	since 
	$$\mathbf{W(a,b)W(c,d)} = (\omega^{(p+1)/2})^{\langle \mathbf{(a,b)} , J\mathbf{(c,d)} \rangle} \mathbf{W(a+c,b+d)}$$ for any $\mathbf{(a,b),(c,d)}\in\mathbb{F}_q^{n}$
	and $\VV$ is self-orthogonal.

\section{Proof of Proposition~\ref{prop:2stab}} \label{append:2stabb}

% \begin{proof}
Let $\VV$ be a self-orthogonal $d$-dimensional subspace of $\mathbb{F}_q^{2n}$ and $S(\VV)$ be a stabilizer defined by \eqref{eq:123stab}.
Notice the following facts.
\begin{enumerate}[\text{Fact} 1)]
\item $\mathbf{W(v)}\mathbf{W(v')} = \mathbf{W(v+v')}$ for any $\mathbf{v},\mathbf{v'} \in \VV$
	by the closure for the multiplication of $S(\VV)$, 
\item All eigenvalues of $\mathbf{W}(\mathbf{v})$ are in $\{\omega^k \mid k\in\mathbb{F}_p \}$, since $(\mathbf{W}(\mathbf{v}) )^p =\mathbf{W}(p\mathbf{v}) = \mathbf{W}(\mathbf{0}) = I_{q^n}$ for any $\mathbf{v} \in \VV$. 
\item All elements of $S(\VV)$ are simultaneously diagonalized, since $S(\VV)$ is a commutative group.
\end{enumerate}

First, we prove 1) of the proposition.
By Facts 2 and 3, we have the simultaneous decomposition of all elements $\mathbf{W(v)} \in S(\VV)$ as 
\begin{align}
\mathbf{W}(\mathbf{v}) =  \sum_{f:\VV\to\mathbb{F}_p} \omega^{f(\mathbf{v})} P_f^{\VV}
\qquad(\forall\mathbf{v}\in\VV) ,
\label{eq:hw_f0}
\end{align}
where the summation is taken for all maps $f$ from $\VV$ to $\mathbb{F}_p$
and $\{P_f^{\VV}\}$ are orthogonal projections (including the zero matrix) such that 
	\begin{align}
	P_f^{\VV} P_{f'}^{\VV} &= 0 \text{ for any } f\neq f',\\
	\sum_{f \in \VV^{*}} &= I_{\cH^{\otimes n} }.
	\end{align}
Let $\VV^*$ be the space of linear maps from $\VV$ to $\mathbb{F}_p$.
Since Fact 1 implies 
	$\omega^{f(\mathbf{v})+f(\mathbf{v}')} P_{f}^{\VV} = 
	 \omega^{f(\mathbf{v} + \mathbf{v}')} P_{f}^{\VV}$ for any $\mathbf{v} \in \VV$ and $f : \VV \to \mathbb{F}_p$,  
	we have $P_f^{\VV} = 0$ for any $f \not \in \VV^*$.
	Thus, \eqref{eq:hw_f0} is written as 
\begin{align}
\mathbf{W}(\mathbf{v}) =  \sum_{f\in\VV^*} \omega^{f(\mathbf{v})} P_f^{\VV}
\qquad(\forall\mathbf{v}\in\VV) .
\label{eq:hw_f}
\end{align}
% 
% By the above three facts, we have the simultaneous spectral decomposition of all elements $\mathbf{W(v)} \in S(\VV)$ as 
% Let $\VV^*$ be the space of linear maps from $\VV$ to $\mathbb{F}_p$.
% % 	all elements $\mathbf{W(v)} \in S(\VV)$ are simultaneously diagonalized with the eigenvalues are $\omega^{f(\mathbf{v})}$ for $f \in \VV^*$.
% % We denote by $\cH_{f}^{\VV}$ the simultaneous eigenspace of $S(\VV)$ corresponding to the eigenvalues $\omega^{f(\mathbf{v})}$.
% % That is, with the projection $P_{f}^{\VV}$ onto $\cH_{f}^{\VV}$, we have the relation
% By the above three facts, we have the simultaneous spectral decomposition of all elements $\mathbf{W(v)} \in S(\VV)$ as 
% \begin{align}
% \mathbf{W}(\mathbf{v}) =  \sum_{f\in\VV^*} \omega^{f(\mathbf{v})} P_f^{\VV}
% \qquad(\forall\mathbf{v}\in\VV) .
% \label{eq:hw_f}
% \end{align}
Furthermore, the space $\VV^*$ is isomorphic to $\mathbb{F}_q^{2n} / \VV^{\perp_J}$ by the following identification:
	we identify $f \in \VV^*$ and $[\mathbf{w}] \coloneqq \mathbf{w} + \VV^{\perp_J} \in \mathbb{F}_q^{2n} / \VV^{\perp_J}$
	if $f(\mathbf{v}) = \langle \mathbf{v}, J \mathbf{w} \rangle$ for any $\mathbf{v}\in\VV$.
Therefore, 
	we denote $P_{[\mathbf{w}]}^{\VV} \coloneqq P_{f}^{\VV}$ if $f $ and $[\mathbf{w}]$ are identical
and
Eq.~\eqref{eq:hw_f} is written as 
\begin{align}
\mathbf{W(v)} = \sum_{[\mathbf{w}]\in\mathbb{F}_q^{2n} / \VV^{\perp_J} } \omega^{ \langle \mathbf{v}, J\mathbf{w}\rangle } P_{\mathbf{[w]}}^{\VV}
\qquad(\forall\mathbf{v}\in\VV),
\label{eq:sdffollwos}
\end{align}
which implies 1) of the proposition.
The uniqueness of the decomposition \eqref{eq:sdffollwos} is from the uniqueness of the eigendecomposition.
% \end{proof}

Next, we prove 2) of the proposition.
Let $\cH_{\mathbf{[w]}}^{\VV} \coloneqq \Ima P_{\mathbf{[w]}}^{\VV}$.
For any $\mathbf{v}\in \VV$, we have 
   \begin{align} 
   \mathbf{W(v) W(w)} \cH_\mathbf{[w']}^{\VV}
   &\stackrel{\mathclap{(a)}}{=} \omega^{\langle \mathbf{v} , J\mathbf{w} \rangle }  \mathbf{W(w)W(v)}  \cH_\mathbf{[w']}^{\VV} \label{eq:123qqd1}\\
   &\stackrel{\mathclap{(b)}}{=} \omega^{\langle \mathbf{v} , J\mathbf{(w+w')} \rangle}  \mathbf{W(w)} \cH_\mathbf{[w']}^{\VV}, \label{eq:123qqd2}
   \end{align}
where $(a)$ is from 
	$$\mathbf{W(v) W(w)} = \omega^{\langle \mathbf{v} , J\mathbf{w} \rangle }  \mathbf{W(w)W(v)},$$ 
	which follows from \eqref{eq:commutative},
	and $(b)$ is from 
	$$\mathbf{W(v)}  \cH_\mathbf{[w']}^{\VV} = \omega^{\langle \mathbf{v} , J\mathbf{w}' \rangle}  \cH_\mathbf{[w']}^{\VV} ,$$
	which follows from \eqref{eq:sdffollwos}.
Since \eqref{eq:123qqd2} implies that $\mathbf{W(v)}$ maps $\mathbf{W(w)} \cH_\mathbf{[w']}^{\VV}$ to $\omega^{\langle \mathbf{v} , J\mathbf{(w+w')} \rangle}\mathbf{W(w)} \cH_\mathbf{[w']}^{\VV}$, 
	we have $\mathbf{W(w)} \cH_\mathbf{[w']}^{\VV} \subseteq \cH_\mathbf{[w+ w']}^{\VV}$ from \eqref{eq:sdffollwos}.
Conversely, we also have $\mathbf{W(-w)} \cH_\mathbf{[w+ w']}^{\VV} \subseteq \cH_\mathbf{[w']}^{\VV}$.
%, since $\dim \mathbf{W(w)} \cH_\mathbf{[w']}^{\VV} = \dim \cH_\mathbf{[w+ w']}^{\VV}$, 
Thus, we have $\dim \cH_\mathbf{[w']}^{\VV} = \dim \cH_\mathbf{[w+ w']}^{\VV}$ and therefore, obtain the desired relation $\mathbf{W(w)} \cH_\mathbf{[w']}^{\VV} = \cH_\mathbf{[w+w']}^{\VV}$.
   
Lastly, we prove 3) of the proposition.
By 2) of the proposition,  
	all spaces $\cH_{[\mathbf{w}]}^{\VV}$ have the same dimension.
Therefore, we have 
	$$\dim \cH_{[\mathbf{w}]}^{\VV} 
		= \frac{\dim \cH^{\otimes n}}{|\mathbb{F}_q^{2n} / \VV^{\perp_J}|} 
		= \frac{\dim \cH^{\otimes n}}{|\VV|} 
%		= \frac{q^{n}}{q^{d}} 
		= q^{n-d}.$$

%\section{Proof of Proposition~\ref{prop:al_ex}} \label{append:al_ex}

\section{Proof of Lemma~\ref{lemm:fund}} \label{append:fund}

For the proof of Lemma~\ref{lemm:fund}, we prepare the following proposition.
\begin{prop} \label{prop:al_ex}
Let $\mathbb{F}_{q'}$ be the finite field of order $q'$ and 
% Let $\mathbb{F}_{q'}$ be the finite field of prime order $p$ and 
$\mathbb{F}_q$ be an extension field of $\mathbb{F}_{q'}$ such that $\mathbb{F}_{q'}(\alpha_1,\ldots, \alpha_{k-2})$,
where $\alpha_i \not \in \mathbb{F}_{q'}(\alpha_{1},\ldots, \alpha_{i-1})$ for any $i$. %\in\{1,\ldots, n+2m-2\}$.
Given two positive integers $r < k$,
we define a matrix $A = (a_{ij}) \in \mathbb{F}_q^{(k-r) \times r}$ such that
\begin{align}
a_{11} &= 1, \\ 
a_{ij} &\in \mathbb{F}_{q'}(\alpha_1,\ldots, \alpha_{i+j-2} ) \setminus \mathbb{F}_{q'}(\alpha_1,\ldots, \alpha_{i+j-3} )
    %\\ &\qquad {\text{}}.
    \label{def:A}
\end{align}
if $(i,j) \neq (1,1)$.
Then, any $r$ row vectors of 
\begin{align}
\bar{A} \coloneqq
\begin{pmatrix}
 A \\ I_r
\end{pmatrix}
\in\mathbb{F}_q^{k\times r}
\label{def:A234}
\end{align}
are linearly independent. 
In particular, when $k= 2r$, the square matrix $A\in\mathbb{F}_q^{r\times r}$ is invertible.
\end{prop}

\begin{IEEEproof}[{Proof of Proposition~\ref{prop:al_ex}}]
Before we give the proof, we introduce the following notation:
For an $n\times m$ matrix $M = (m_{ij})$, $S\subset \{1,\ldots, n\}$ and $T\subset \{1,\ldots, m\}$,
define a submatrix $M(S,T) \coloneqq (m_{ij})_{i\in S, j \in T}$.

Let $S\subset \{1,\ldots, k-r\}$ and $T\subset \{1,\ldots,r\}$ be subsets such that $|S|+ |T|= r$.
Choose $r$ row vectors of $\bar{A}$ as %from $S,T$ such that the chosen row vectors form the matrix 
\begin{align*}
\bar{A}( S\cup (k-r+T), \{1,\ldots, r\} ) = 
\begin{pmatrix}
 A(S,\{1,\ldots,r\})  \\ I_r(T,\{1,\ldots,r\}) 
\end{pmatrix}.
\end{align*}
% $\bar{A}( S\cup (k-r+T), \{1,\ldots, k\} )$.
% If $|T| = r$, we have $\bar{A}( S\cup (k-r+T), \{1,\ldots, k\} ) = I_r$, and therefore, the chosen $r$ row vectors are linearly independent.
% Now, consider the case $|T| < r$.
% Row vectors of 
The row vectors of $\bar{A}( S\cup (k-r+T), \{1,\ldots, r\} )$ are linearly independent
if and only if $A(S,T^c)\in\mathbb{F}_q^{|S|\times|S|}$ is invertible, where $T^c \coloneqq \{1,\ldots,r\} \setminus T $.
Therefore, we show in the following that the determinant of $A(S,T^c)$ is nonzero.

% Since the only element of $A(S,T^c)$ such that 
From the definition of $A$ in \eqref{def:A}, 
the $(|S|,|S|)$ element $a_{\max S, \max T^c}$ of $A(S,T^c)$ is in
$\mathbb{F}_{q'}(\alpha_1,\ldots, \alpha_{\max S + \max T^c -2} ) \setminus \mathbb{F}_{q'}(\alpha_1,\ldots, \alpha_{\max S + \max T^c -3})$
but the other $|S|^2-1$ elements are in $\mathbb{F}_{q'}(\alpha_1,\ldots, \alpha_{\max S + \max T^c -3})$.
Thus, by the cofactor expansion of the determinant, i.e., 
% 	the determinant of a matrix $M$ is given as
	$\det M = \sum_{j} (-1)^{i+j} m_{i,j} M_{i,j}$ for a matrix $M=(m_{ij})$ and its $i,j$ minor $M_{i,j}$,
	we have
	\begin{align*}
	&\det A(S,T^c) \\
% 	= \sum_{
	&= a_{\max S, \max T^c} \cdot \det A(S , T^c)_{|S|,|S|} + x\\
	&= a_{\max S, \max T^c} \cdot \det A(S \setminus \{\max S\}, T^c\setminus \{ \max T^c \} ) + x
	\end{align*}
with some $x\in \mathbb{F}_{q'}(\alpha_1,\ldots, \alpha_{\max S + \max T^c -3})$.
If 
    $$\det A(S \setminus \{\max S\}, T^c\setminus \{ \max T^c \} ) \neq 0,$$ 
    then $\det A(S,T^c) \neq 0$.
   Thus, by induction,
   we have $\det A(S,T^c) \neq 0$ since $\det A(\{\min S\}, \{ \min T^c \} ) = a_{\min S,\min T^c} \neq 0$.
\end{IEEEproof}
%The proof of Proposition~\ref{prop:al_ex} is given in Appendix \ref{append:al_ex}.

\begin{remark} \label{remark:sfedvckjio}
Proposition~\ref{prop:al_ex} is a slight generalization of the construction {\cite[Appendix A]{CH17}}, which proposed the same construction only for $a_{ij}= \alpha_{i+j-2}$ in \eqref{def:A}.
\end{remark}

Now, we prove Lemma~\ref{lemm:fund}.
Let $S \in \mathbb{F}_q^{2n\times 2n}$ be a symplectic matrix, i.e., $S^{\top} J S = J$, 
and $\mathbf{s}_i \in\mathbb{F}_q^{2n}$ be the $i$-th column vector of $S$.
Then, the following $\mathbf{v}_1,\ldots, \mathbf{v}_{2t}\in\mathbb{F}_q^{2\vN}$ satisfy condition $\mathrm{(b)}$:
\begin{align*}
(\mathbf{v}_1, \ldots, \mathbf{v}_{2n-2t} ) &\coloneqq ( \mathbf{s}_{2t-n+1}, \ldots, \mathbf{s}_{n} ) \\
(\mathbf{v}_{2n-2t+1}, \ldots, \mathbf{v}_{2t} ) &\coloneqq 
( \mathbf{s}_{1}, \ldots, \mathbf{s}_{2t-n}, \mathbf{s}_{n+1}, \ldots, \mathbf{s}_{2t}) .
\end{align*}
Therefore, in the following, we prove that there exists a symplectic matrix $S =(\mathbf{s}_{1}, \ldots,  \mathbf{s}_{2n}) $ such that the row vectors of $S' \coloneqq (\mathbf{s}_{1}, \ldots,  \mathbf{s}_{2t})$ satisfy condition $\mathrm{(a)}$.

First, we construct a symplectic matrix as follows.
% Let $\mathbb{F}_{q'}$ be the finite field of order $q'$ and 
% $\mathbb{F}_q$ be the algebraic extension $\mathbb{F}_{q'}(\alpha_1,\ldots, \alpha_{n+2m-2})$,
% where $\alpha_i \not \in \mathbb{F}_{q'}(\alpha_{1},\ldots, \alpha_{i-1})$ for any $i$. %\in\{1,\ldots, n+2m-2\}$.
For convenience, let $\alpha_0 \coloneqq 1$.
Define two square symmetric matrices $A = (a_{ij}), B= (b_{ij})\in\mathbb{F}_q^{n\times n}$ as 
\begin{align*}
    a_{ij} &= \alpha_{i+j-2}, \quad  b_{ij} = \alpha_{i+j-2 + (2t-n)},
\end{align*}
i.e.,
\begin{align*}
    A &= 
    \begin{pmatrix}
    \alpha_0         &  \alpha_1  &  \cdots      &  \alpha_{n-1} \\
    \alpha_1   &  \alpha_2  &  \cdots      &  \alpha_{n-2}   \\
    \vdots    &  \vdots   &  \ddots      &  \vdots \\
    \alpha_{n-1} & \alpha_{n}   &  \cdots  &  \alpha_{2n-2}    
    \end{pmatrix}
    ,
    \\
    B &= 
    \begin{pmatrix}
    \alpha_{2t-n} &  \alpha_{2t-n+1}   &  \cdots      &  \alpha_{2t-1} \\
    \alpha_{2t-n+1}   &  \alpha_{2t-n+2} &  \cdots      &  \alpha_{2t}   \\
    \vdots       &  \vdots       &  \ddots      &  \vdots \\
    \alpha_{2t-1} &  \alpha_{2t} &  \cdots      &  \alpha_{n+2t-2}    
    \end{pmatrix}
    .
\end{align*}
Since the matrices  
\begin{align*}
    \begin{pmatrix}
I_n  &   X        \\
 0    & I_n
    \end{pmatrix}
,\ 
\begin{pmatrix}
I_n  &   0        \\
 X    & I_n
    \end{pmatrix}
\end{align*}
are symplectic matrices for any symmetric matrix $X$,
	and the multiple of two symplectic matrices is a symplectic matrix \cite[Section~8.2.2]{Haya2}, 
% $A$ and $B$ are symmetric matrices,
the matrix 
\begin{align*}
S = \begin{pmatrix}
    I_n + B A^{-1}   &   B        \\
    A^{-1}                &   I_n
    \end{pmatrix}
    = 
    \begin{pmatrix}
    I_n   &   B        \\
     0      &   I_n
    \end{pmatrix}
    \begin{pmatrix}
    I_n   &   0        \\
    A^{-1}                &   I_n
    \end{pmatrix}
\end{align*}
is a symplectic matrix,
where the inverse $A^{-1}$ exists from Proposition \ref{prop:al_ex}.
With the notation $B= (B_1, B_2) \in \mathbb{F}_q^{n\times (2t-n)} \times \mathbb{F}_q^{n\times (2n-2t)}$,
we have 
\begin{align}
% \left[
% \begin{array}{c|c}
% \usebox{0}&\makebox[\wd0]{\large $B$}\\
% \hline
%   \vphantom{\usebox{0}}\makebox[\wd0]{\large $C$}&\makebox[\wd0]{\large $D$}
% \end{array}
% \right]
% \\
S'  \coloneqq (\mathbf{s}_{1}, \ldots,  \mathbf{s}_{2t})= 
% \begin{pmatrix}
%     I_n + B A^{-1}   &   B_1        \\
%    \multirow{2}{*}{A^{-1}}                 &   I     \\
%                           &   O_{2n-2m,2m-n}     \\
%     \end{pmatrix}.
    \left(\begin{array}{c | c}
     \multirow{2}{*}{$I_n + B A^{-1}$}                        &   \multirow{2}{*}{$B_1$}        \\
     \\
     \hline
   \multirow{2}{*}{$A^{-1}$}                 &   I_{2t-n}    \\
                                           &   0     \\
    \end{array}
    \right).
\end{align}

Now, we prove that the row vectors of $S'$ satisfy condition $\mathrm{(a)}$.
Since (i) the right multiplication of invertible matrices 
and (ii) elementary column operations 
do not change the linear independence of the row vectors, 
we manipulate the matrix $S'$ in the following way:
\begin{align*}
& S'=
    \left(\begin{array}{c | c}
     \multirow{2}{*}{$I_n + B A^{-1}$}                        &   \multirow{2}{*}{$B_1$}        \\
     \\
     \hline
   \multirow{2}{*}{$A^{-1}$}                 &   I_{2t-n}    \\
                                           &   0     \\
    \end{array}
    \right)
     \\[1em]
&\xrightarrow{\text{(i)}}
    \left(\begin{array}{c | c}
     \multirow{2}{*}{$I_n + B A^{-1}$}                        &   \multirow{2}{*}{$B_1$}        \\
     \\
     \hline
    \multirow{2}{*}{$A^{-1}$}                 &   I_{2t-n}    \\
                                           &   0     \\
    \end{array}
    \right)
    \begin{pmatrix}
    A    &   0        \\
    0    &   I_{2t-n}
    \end{pmatrix}
    \\
    &= 
    \left(\begin{array}{c | c}
     \multirow{2}{*}{$A + B $}                        &   \multirow{2}{*}{$B_1$}        \\
     \\
     \hline
    \multirow{2}{*}{$I_n$}                 &   I_{2t-n}    \\
                                           &   0     \\
    \end{array}
    \right)
    \\[1em]
&=
    \left(\begin{array}{c c | c}
     \multirow{2}{*}{$A_1+B_1$}   &    \multirow{2}{*}{$A_2 + B_2$}              &   \multirow{2}{*}{$B_1$}        \\
     & & 
     \\
     \hline
     I_{2t-n}  & 0               &   I_{2t-n}    \\
      0                    & I_{2n-2t}       &   0     \\
    \end{array}
    \right)
    \\[1em]
&\xrightarrow{\text{(ii)}}
    \left(\begin{array}{c c | c}
%       \multicolumn{2}{c}{ \multirow{2}{*}{$A + [ 0, B_2] $} }                        &   \multirow{2}{*}{$B_1$}        \\
      \multirow{2}{*}{$A_1$}  & \multirow{2}{*}{$A_2+ B_2$}                         &   \multirow{2}{*}{$B_1$}        \\
     & & \\
     \hline
     0  & 0               &   I_{2t-n}    \\
     0  & I_{2n-2t}       &   0     \\
    \end{array}
    \right)
     \\[1em]
&\xrightarrow{\text{(ii)}}
    \left(\begin{array}{c c c}
      \multirow{2}{*}{$A_1 $}   & \multirow{2}{*}{$B_1 $}                         &   \multirow{2}{*}{$ A_2+ B_2$}        \\
     \\
     \hline
      0  & I_{2t-n}  &   0     \\
      0  & 0         &   I_{2n-2t}     \\
    \end{array}
    \right)
    \eqqcolon S''
    ,
    \label{eq:ch17}
\end{align*}
    where $A =(A_1, A_2) \in \mathbb{F}_q^{n\times (2t-n)} \times \mathbb{F}_q^{n\times (2n-2t)}$.
    By the above transformation,
    the linear independence of the row vectors of $S'$ is equivalent to that of $S''$.
    Let 
    \begin{align}
%     \xrightarrow{\text{($\ast$)}}
     S'''\coloneqq 
    \left(\begin{array}{c c c}
      \multirow{2}{*}{$A_1 $}   & \multirow{2}{*}{$B_1 $}                         &   \multirow{2}{*}{$ A_2+ B_2$}        \\
     \\
     \hline
     I_{2t-n}            & 0         &   0     \\
     0  & I_{2t-n}  &   0     \\
     0  & 0         &   I_{2n-2t}     \\
    \end{array}
    \right)
    \end{align}
    by adding the row vectors $(I_{2t-n}, 0, 0)$ to $S''$. %at ($\ast$).
    If any $2t$ row vectors of $S'''$ are linearly independent, then $S''$ and $S'$ also satisfy the same property.
    Note that we can apply Proposition~\ref{prop:al_ex} to $S'''$ 
    since the matrices $A_1$, $B_1$, $A_2+ B_2$ are written as 
    \begin{align*}
    A_1 &=
    \left(\begin{array}{c  c  c}
    \alpha_0        &   \cdots  &  \alpha_{2t-n-1} \\
    \vdots          &   \ddots  &  \vdots \\
    \alpha_{n-1}    &   \cdots  &  \alpha_{2t-2}    
    \end{array}\right)
    \\
    B_1& = 
    \left(\begin{array}{c  c   c}
    \alpha_{2t-n}   &    \cdots      &  \alpha_{4t-2n - 1} \\
    \vdots          &    \ddots      &  \vdots \\
    \alpha_{2t-1}   &   \cdots      &  \alpha_{4t-n-2}    
    \end{array}\right)
    \\
    A_2+B_2
    &=
    \left(\begin{array}{c  c   c}
    \alpha_{4t-2n}+\alpha_{2t-n}   &  \cdots      &  \alpha_{2t-1} +\alpha_{n-1}\\
    \vdots           &  \ddots      &  \vdots \\
    \alpha_{4t-n-1}+\alpha_{2t-1}  &  \cdots      &  \alpha_{n+2t-2} +\alpha_{2n-2}  
    \end{array}\right)   
    \end{align*}
    and therefore $(A_1 , \   B_1    ,\   A_2+ B_2)$
    satisfies the condition \eqref{def:A}. % and therefore, $S'''$ is in the form of \eqref{def:A234}.
    The application of Proposition~\ref{prop:al_ex} to $S'''$ shows that 
%     by applying Proposition \ref{prop:al_ex} for $S'''$, 
any $2t$ row vectors of $S'''$
are linearly independent.
Thus, the matrix $S'$ also satisfies the same property as $S'''$, which implies condition $\mathrm{(a)}$.
This finishes the proof of the desired statement.

\section{Proof of Lemma~\ref{L1}} \label{append:upper}
Throughout this section, we use the following notation.
For random variables $X$ and $Y$,
	we denote by $p_X$ the probability distribution of $X$,
	by $p_{X|Y}$ the distribution of $X$ conditioned by $Y$,
	by $p_{X|Y=y}$ the distribution of $X$ conditioned by $Y=y$.
We also denote $p_{X=x} = \Pr[X=x]$ and $p_{X=x|Y=y} = \Pr[X=x|Y=y]$ for simplicity.

For the proof of Lemma~\ref{L1},
	we prepare the following lemma.
% the security conditions \eqref{eq:error_probab}, \eqref{eq:serv_sec}, and \eqref{eq:user_sec} give the following bounds.
\begin{lemm} \label{lemm:upper_relate}
The server secrecy $S_{\mathrm{serv}}(\Psi_{\mathrm{QPIR}}^{(\vM)}) \le \beta$ implies 
\begin{align}
&
I(M_k^c;\cA Q|K=k)_{ \rho(M,Q) } \leq \beta .
\label{eq:serv_sec3_remark}
\end{align}
The user $\vT$-secrecy $S_{\mathrm{user}}^{(\vT)}(\Psi_{\mathrm{QPIR}}^{(\vM)}) \le \gamma$ implies 
\begin{align}
    \max_{i\neq k\in\{1,\ldots,\vF\}  , \pi\in\mathsf{S}_{\vN}} d ( p_{Q_{\vT} | K=k}, p_{Q_{\vT} | K=i}) \le \sqrt{2\vF\gamma},
\label{eq:user_sec3_remark}
\end{align}
where $d(\cdot,\cdot)$ is the variational distance $d(p , q) \coloneqq (1/2)\cdot \sum_{j} |p_j-q_j|$ for probability distributions $p,q$.
\end{lemm}

\begin{IEEEproof}
The relation \eqref{eq:serv_sec3_remark} is proved as follows:
\begin{align*}
    &{I(M_k^c;\cA Q|K=k)_{\rho(M,Q)} } \\
        &= I(M_k^c;\cA|Q,K=k)_{\rho(M,Q)} + I(M_k^c;Q|K=k)_{\rho(M,Q)}  \\
        &= I(M_k^c;\cA|Q,K=k)_{\rho(M,Q)}  \label{eq:QmK}  \\
        &= \sum_{q}  p_{Q=q|K=k} \cdot I(M_k^c;\cA|Q=q,K=k)_{\rho(M,Q)} \leq \beta,
\end{align*}
where the equality \eqref{eq:QmK} holds because $Q$ is independent of $M_k^c$.

The relation \eqref{eq:user_sec3_remark} is proved as follows.
For any $\pi\in\mathsf{S}_{\vN}$ and any $k\in\{1,\ldots,\vF\}$,
\begin{align*}
    \gamma & \geq I(K; Q_{\vT} )  
        = D( p_{KQ_{\vT}} \| p_K \times p_{Q_{\vT}} ) \\
    &= \frac{1}{\vF} \sum_{k'} D(p_{Q_{\vT}|K=k'} \| p_{Q_{\vT}} ) 
    \stackrel{\mathclap{(a)}}{\geq} \frac{2}{\vF} \sum_{k'} d^2( p_{Q_{\vT}|K=k'} , p_{Q} )  \\
    &\geq \frac{2}{\vF} d^2( p_{Q_{\vT}|K=k} , p_{Q_{\vT}} ), 
\end{align*}
where the inequality $(a)$ follows from Pinsker inequality.
% Therefore,
% \begin{align}
% \sqrt{\frac{\vF\gamma}{2}} \ge d( p_{Q|K=k''} , p_{Q} ) 
% \end{align}
Thus, for any $i,k\in\{1,\ldots,\vF\}$, we have
\begin{align}
\sqrt{2\vF\gamma} 
    &\geq  d( p_{Q_{\vT}|K=k} , p_{Q_{\vT}} ) + d( p_{Q_{\vT}} , p_{Q_{\vT}|K=i} )  \\
    &\geq d( p_{Q_{\vT}|K=k} , p_{Q_{\vT}|K=i} ) ,
\end{align}
which implies \eqref{eq:user_sec3_remark}.
\end{IEEEproof}

% We will only use the evaluation given in Lemma \ref{lemm:upper_relate} for the proof of Lemma~\ref{L1}
% 	instead of the conditions $S_{\mathrm{serv}}(\Psi_{\mathrm{QPIR}}^{(\vM)}) \le \beta$ and $S_{\mathrm{user}}^{(\vT)}(\Psi_{\mathrm{QPIR}}^{(\vM)}) \le \gamma$.
% We also prepare the following lemma.

% \begin{IEEEproof}

Now, we prove Lemma~\ref{L1}.
Let $k\neq i \in \{1,\ldots, \vF\}$.
Then, we obtain Lemma~\ref{L1} as 
\begin{align}
% &{I(M_k;\cA_{\pi,\vT}|Q,K=k)_{\rho(M,Q_{\vT})} }\\
% &\stackrel{(a)}{=} 
& I(M_k;\cA_{\vT}|Q_{\vT},K=k)_{\rho(M,Q_{\vT})}\\
&\le I(M_k;\cA_{\vT}Q_{\vT}|K=k)_{\rho(M,Q_{\vT})}\\
&\stackrel{\mathclap{(a)}}{\le}  I(M_k;\cA_{\vT}Q_{\vT}|K=i)_{\rho(M,Q_{\vT})} + g(\vM,\gamma) \\
&\le  I(M_i^c;\cA Q|K=i)_{\rho(M,Q)}  + g(\vM,\gamma) \\
&\stackrel{\mathclap{(b)}}{\le}  \beta  + g(\vM,\gamma),
\end{align}
where the inequality $(b)$ follows from \eqref{eq:serv_sec3_remark} of Lemma~\ref{lemm:upper_relate}.

The inequality $(a)$ is derived as follows.
% From \eqref{eq:serv_sec3_remark}, %.Appendix \ref{append:upper},
% the user $\vT$-secrecy ${S}_{\mathrm{user}}^{(\vT)}(\Psi_{\mathrm{QPIR}}^{(\vM)}) \leq \gamma$
% implies 
% \begin{align}
%     \max_{i\neq k\in\{1,\ldots,\vF\}  , \pi\in\mathsf{S}_{\vN}} d ( p_{Q_{\vT} | K=k}, p_{Q_{\vT} | K=i}) \le \sqrt{2\vF\gamma}.
%     \label{eq:user_sec3}
% \end{align}
When we define 
$\tilde{\rho}(M,Q_{\vT}|k) \coloneqq  \sum_{m,q_{\vT}} (1/\vM^{\vF})\cdot 
    p_{Q_{\vT}=q_{\vT} | K=k }
    %\pr [Q_{\vT}=q_{\vT} | K=k ]
    \cdot |m,q_{\vT}\rangle\langle m,q_{\vT}|\otimes \rho(m,q_{\vT})$ for $k\in\{1,\ldots,\vF\}$,
the inequality \eqref{eq:user_sec3_remark} of Lemma~\ref{lemm:upper_relate} implies that
\begin{align}
d(\tilde{\rho}(M,Q_{\vT}|k),\tilde{\rho}(M,Q_{\vT}|i)) \leq \sqrt{2\vF\gamma}
\end{align}
for any $i\neq k \in\{1,\ldots, \vF\}$,
where $d(\cdot,\cdot)$ is the trace distance $d(\rho,\sigma) \coloneqq (1/2)\cdot \Tr |{\rho}-{\sigma}|$ for quantum states $\rho$ and $\sigma$.
Thus, Fannes inequality for mutual information \cite[{Eq.~(5.106)}]{Hay17} implies that
\begin{align*}
&|I(M_k;\cA_{\vT} Q_{\vT}|K=k)_{ \rho(M,Q_{\vT}) } - I(M_k;\cA_{\vT} Q_{\vT}|K=i)_{\rho(M,Q_{\vT})}|    \\
&\le  g(\vM,\gamma),
\end{align*}
which yields the inequality $(a)$.
% \end{IEEEproof}
This finishes the proof of Lemma~\ref{L1}.

\section{Proofs of converse bounds \eqref{ineq:conv1} and \eqref{ineq:conv2}}   \label{append:converse}

In the following proofs, we use the notations given in Step 1 of Section~\ref{lemmProffd}
	and the notation for distributions introduced in the beginning of Appendix~\ref{append:upper}.

\subsection{Proof of \eqref{ineq:conv1}}

Eq.~\eqref{ineq:conv1} is proved as follows.
Fix $K=k$ and $Q=q$.
% Let $Z = (W^c, Q)$.
Applying \cite[(4.66)]{Hay17} with $\rho(q)$ and $Y_{k,q}$,
% to %the choice
% $\rho_{z}=(1/\vM) \sum_{m=0}^{\vM-1} \rho_{m,z}$,
we have
\begin{align}
 &(1-P_{\mathrm{err},k,q}(\Psi_{\mathrm{QPIR}}^{(\vM)}) )^{1+r} \vM^{r} \nonumber \\
 &\le \frac{1}{\vM} \sum_{m_k=1}^{\vM} \Tr \rho(m_k,q)^{1+r} \rho(q)^{-r} \label{eq:AP1}
 \end{align}
for $r \in (0,1)$. 
Then,
% where the inequality $(a)$ is from \cite[(4.66)]{Hay17}.
 \begin{align}
  &  \frac{1}{\vM} \sum_{m_k=1}^{\vM} \Tr \rho(m_k,q)^{1+r} \rho(q)^{-r}
 \le \frac{1}{\vM}  \sum_{m_k=1}^{\vM} \Tr \rho(m_k,q) \rho(q)^{-r}
  \nonumber \\
 & = \Tr \rho(q)^{1-r} \le \max_{\rho}\Tr \rho^{1-r}
 = \max_{p_X} \sum_{x=1}^{\vD^{\vN} } p_{X=x} ^{1-r} \nonumber \\
 &\stackrel{\mathclap{(a)}}{=} 
 \paren*{\prod_{s=1}^{\vN } \dim \cA_s}^r,
 \label{eq:AP2}
\end{align}
where the equation $(a)$ is proved from the fact that 
$\max_{p_X} \sum_{x=1}^{\vD^{\vN}} p_{X=x}^{1-r}$ is achieved by choosing $p_X$ as the uniform distribution since the map $y \mapsto y^{1-r}$ is concave.
Combining \eqref{eq:AP1} and \eqref{eq:AP2}, we have
\begin{align}
(1-{P_{\mathrm{err},k,q}} (\Psi_{\mathrm{QPIR}}^{(\vM)}) )^{1+r} 
	\le \paren*{\frac{\prod_{s=1}^{\vN } \dim \cA_s}{\vM}}^r.
	\label{eq:24}
% 	\vM^{r}
% (1-P)^{1+r}\le ((\Pi dim A)/m)^r.
\end{align}

Let $\{\Psi_{\mathrm{QPIR}}^{(\vM_\ell)}\}_{\ell=1}^{\infty}$ be an arbitrary sequence of QPIR protocols such that
	the QPIR rate of $\Psi_{\mathrm{QPIR}}^{(\vM_\ell)}$ is strictly greater than $1$ for any sufficiently large $\ell$,
	i.e.,
		\begin{align}
	R(\Psi_{\mathrm{QPIR}}^{(\vM_\ell)}) = \frac{\log \vM_\ell }{\log \vD_\ell^\vN} > 1,
	\label{eq:rate_ff}
	\end{align}
where $\vD_\ell$ is the dimension of $\cA_s$ ($\forall s\in\{1,\ldots,\vN\}$) for the protocol $\Psi_{\mathrm{QPIR}}^{(\vM_\ell)}$.
Then, Eq.~\eqref{eq:rate_ff} implies that
$\vD_\ell^{\vN}/\vM_\ell = (\prod_{i=1}^{\vN} \dim \cA_i)/\vM_\ell$ goes to $0$.
Hence, from \eqref{eq:24}, for any $k$ and $q$, $1-P_{\mathrm{err},k,q}(\Psi_{\mathrm{QPIR}}^{(\vM_\ell)})$ approaches zero.
Since 
\begin{align}
	1- P_{\mathrm{err}}(\Psi_{\mathrm{QPIR}}^{(\vM_\ell)})  \leq 1- P_{\mathrm{err},k,q}(\Psi_{\mathrm{QPIR}}^{(\vM_\ell)}),
	\label{EQQ1}
\end{align}	
	we have $1-P_{\mathrm{err}}(\Psi_{\mathrm{QPIR}}^{(\vM_\ell)})\to 0$,
which implies \eqref{ineq:conv1}.

\subsection{Proof of \eqref{ineq:conv2}}

Eq.~\eqref{ineq:conv2} is proved as follows.
Assume that 
$S_{\mathrm{serv}}(\Psi_{\mathrm{QPIR}}^{(\vM)})=0$ and ${S}_{\mathrm{user}}^{(\vT)}(\Psi_{\mathrm{QPIR}}^{(\vM)})=0$.
We consider the case where arbitrary $K=k$, $Q=q$, and $\pi\in\mathsf{S}_{\vN}$ are fixed.
Since Lemma \ref{L1} guarantees that the reduced density $\rho(m_k,q_{\vT})$ on $ \cA_{\vT}$ does not depend on $m_k$, 
	we have $\rho(m_k,q_{\vT}) = \rho(q_{\vT})$.
Applying \cite[(4.66)]{Hay17} with $\rho(q_{\vT}) \otimes (I/\vD^{\vN-\vT})$ and $Y_{k,q}$,
we have
\begin{align}
 &(1-{P_{\mathrm{err},k,q}} (\Psi_{\mathrm{QPIR}}^{(\vM)}) )^{1+r} \vM^{r}\\
 &\le 
  \frac{1}{\vM} \sum_{m_k=1}^{\vM} \Tr \rho(m_k,q)^{1+r} \paren*{\rho(q_{\vT}) \otimes \frac{I}{\vD^{\vN-\vT}}}^{-r}
\label{H4-2}
\end{align}
for any $r \in (0,1)$.
Given $m_k$ and $q$, consider the decomposition 
$\rho(m_k,q)=\sum_x p_{x} |\psi_{m_k,q,x}\rangle \langle\psi_{m_k,q,x}|$.
Let $\rho(q_{\vT},x)$ be the reduced density of $|\psi_{m_k,q,x}\rangle \langle\psi_{m_k,q,x}|$ on $\cA_{\vT}$, i.e., $\rho(q_{\vT}) = \sum_x p_x \rho(q_{\vT},x)$.
Then,
\begin{align}
\lefteqn{ \Tr \rho(m_k,q)^{1+r} \paren*{ \rho(q_{\vT}) \otimes \frac{I}{\vD^{\vN-\vT}}}^{-r}} \nonumber \\
 &\stackrel{\mathclap{(b)}}{\le}  \sum_x p_x 
\Tr (|\psi_{m_k,q,x}\rangle \langle\psi_{m_k,q,x}|)^{1+r} \paren*{ \rho(q_{\vT},x) \otimes \frac{I}{\vD^{\vN-\vT}}}^{-r}  \nonumber \\ 
 &= \sum_x p_x 
\Tr |\psi_{m_k,q,x}\rangle \langle\psi_{m_k,q,x}| \paren*{ \rho(q_{\vT},x) \otimes \frac{I}{\vD^{\vN-\vT}}}^{-r} \nonumber \\
 &= \vD^{r(\vN-\vT)} \sum_x p_x 
\Tr (\rho(q_{\vT},x))^{1-r} 
\stackrel{\mathclap{(c)}}{\le} \vD^{2r(\vN-\vT)},\label{H5-2}
\end{align}
where $(b)$ follows from the application of the inequality
% {\eqref{ineq:rev} follows from the information processing inequality of the quantum relative R\'enyi entropy \cite{} and}
\begin{align*}
\phi(-r|\rho\|\rho) := \log \Tr \rho^{1+r}\rho^{-r} \geq \phi(-r|\kappa(\rho)\|\kappa(\rho))
\end{align*}
for states $\rho,\rho$, TP-CP map $\kappa$, and $r\in(0,1)$ \cite[(5.53)]{Hay17}
to the choice $\rho := \sum_x {p_{x}} |x\rangle\langle x|\otimes |\psi_{m_k,q,x}\rangle\langle \psi_{m_k,q,x}|$,
$\rho := \sum_x {p_{x}} |x\rangle\langle x|\otimes (\rho(q_{\vT},x)\otimes {I}/\vD^{\vN-\vT}  )$ on the composite system $\mathcal{X}\otimes\cA$,
and $\kappa:=\Tr_{\mathcal{X}}$.
The last inequality $(c)$ is shown as follows.
Since $\rho(q_{\vT},x)$ is a reduced state of the pure state $|\psi_{m_k,q,x}\rangle$ in $\cA_{\vT}\otimes \cA_{\vT^c}$,
we have $\rank \rho(q_{\vT},x) \leq \min\{ \dim \cA_{\vT},  \dim \cA_{\vT^c} \} = \vD^{\vN-\vT}$, 
which implies 
$\Tr \rho(q_{\vT},x)^{1-r} \le (\vD^{\vN-\vT})^r$.
Combining \eqref{H4-2} and \eqref{H5-2}, we have
\begin{align}
% (1-P)^{1+r}=(1-P)^{1+r}m^r/m^r \le d^{2r(n-t)}/m^r =(d^{n-t}/m)^r\\
&(1-P_{\mathrm{err},k,q}(\Psi_{\mathrm{QPIR}}^{(\vM)}) )^{1+r} \nonumber \\
&=
(1-P_{\mathrm{err},k,q}(\Psi_{\mathrm{QPIR}}^{(\vM)}) )^{1+r} \vM^{r} / \vM^{r} \nonumber \\
&\le
	\frac{\vD^{2r(\vN-\vT)}}{\vM^{r}}
=\paren*{\frac{\vD^{2(\vN-\vT)}}{\vM} }^{r}
%  \le \vD^{2r(\vN-\vT)}. % = d^{2(\vN-\vT)r}.
 \label{S5}
\end{align}

Let $\{\Psi_{\mathrm{QPIR}}^{(\vM_\ell)}\}_{\ell=1}^{\infty}$ be an arbitrary sequence of QPIR protocols such that 
	the QPIR rate greater than $2(\vN-\vT)/\vN$ for any sufficiently large $\ell$, i.e., 
	\begin{align}
	R(\Psi_{\mathrm{QPIR}}^{(\vM_\ell)}) = \frac{\log \vM_\ell }{\log \vD_\ell^\vN} > \frac{2(\vN-\vT)}{\vN},
	\end{align}
which is equivalent to
	\begin{align}
	\frac{\log \vM_\ell }{\log \vD_\ell^{2(\vN-\vT)}} > 1.
	\label{eq:rate_f}
	\end{align}
Here, $\vD_\ell$ is the dimension of $\cA_s$ ($\forall s\in\{1,\ldots,\vN\}$) for the protocol $\Psi_{\mathrm{QPIR}}^{(\vM_\ell)}$.
From \eqref{eq:rate_f}, $\vD_\ell^{2(\vN-\vT)}/ \vM_\ell $ goes to $0$, and then from \eqref{S5}, the probability $1-P_{\mathrm{err},k,q}(\Psi_{\mathrm{QPIR}}^{(\vM_\ell)})$ approaches $0$.
Since 
	\begin{align}
	1- P_{\mathrm{err}}(\Psi_{\mathrm{QPIR}}^{(\vM_\ell)})  \leq 1- P_{\mathrm{err},k,q}(\Psi_{\mathrm{QPIR}}^{(\vM_\ell)}),
	\label{EQQ2}
	\end{align}
	we have $1-P_{\mathrm{err}}(\Psi_{\mathrm{QPIR}}^{(\vM_\ell)})\to 0$,
which implies \eqref{ineq:conv2}.

\begin{IEEEbiographynophoto}{Seunghoan Song}(GS'20--M'21)
received the B.E. degree from Osaka University, Japan, in 2017 and the M.Math.\ and Ph.D.\ degrees in mathematical science from Nagoya University, Japan, in 2019 and 2020, respectively. 
He is a research fellow of the Japan Society of the Promotion of Science (JSPS) from 2020.
He is currently a JSPS postdoctoral fellow at the Graduate School of Mathematics, Nagoya University.
He awarded the School of Engineering Science Outstanding Student Award at Osaka University in 2017 and 
Graduate School of Mathematics Award for Outstanding Masters Thesis at Nagoya University in 2019.
His research interests include classical and quantum information theory and its applications to secure communication protocols.
\end{IEEEbiographynophoto}

\begin{IEEEbiographynophoto}{Masahito Hayashi}(M'06--SM'13--F'17) was born in Japan in 1971.
He received the B.S.\ degree from the Faculty of Sciences in Kyoto
University, Japan, in 1994 and the M.S.\ and Ph.D.\ degrees in Mathematics from
Kyoto University, Japan, in 1996 and 1999, respectively. He worked in Kyoto University as a Research Fellow of the Japan Society of the
Promotion of Science (JSPS) from 1998 to 2000,
and worked in the Laboratory for Mathematical Neuroscience,
Brain Science Institute, RIKEN from 2000 to 2003,
and worked in ERATO Quantum Computation and Information Project,
Japan Science and Technology Agency (JST) as the Research Head from 2000 to 2006.
He also worked in the Superrobust Computation Project Information Science and Technology Strategic Core (21st Century COE by MEXT) Graduate School of Information Science and Technology, The University of Tokyo as Adjunct Associate Professor from 2004 to 2007.
He worked in the Graduate School of Information Sciences, Tohoku University as Associate Professor from 2007 to 2012.
In 2012, he joined the Graduate School of Mathematics, Nagoya University as Professor.
Also, he was appointed in Centre for Quantum Technologies, National University of Singapore as Visiting Research Associate Professor from 2009 to 2012
and as Visiting Research Professor from 2012 to now.
He worked in Center for Advanced Intelligence Project, RIKEN as
a Visiting Scientist from 2017 to 2020.
He worked in Shenzhen Institute for Quantum Science and Engineering, Southern University of Science and Technology, Shenzhen, China as a Visiting Professor from 2018 to 2020,
and
in Center for Quantum Computing, Peng Cheng Laboratory, Shenzhen, China
as a Visiting Professor from 2019 to 2020.
In 2020, he joined Shenzhen Institute for Quantum Science and Engineering, Southern University of Science and Technology, Shenzhen, China
as Chief Research Scientist. 
In 2011, he received Information Theory Society Paper Award (2011) for ``Information-Spectrum Approach to Second-Order Coding Rate in Channel Coding''.
In 2016, he received the Japan Academy Medal from the Japan Academy
and the JSPS Prize from Japan Society for the Promotion of Science.

In 2006, he published the book ``Quantum Information: An Introduction'' from Springer, whose revised version was published as ``Quantum Information Theory: Mathematical Foundation'' from Graduate Texts in Physics, Springer in 2016.
In 2016, he published other two books ``Group Representation for Quantum Theory'' and ``A Group Theoretic Approach to Quantum Information'' from Springer.
He is on the Editorial Board of {\it International Journal of Quantum Information}
and {\it International Journal On Advances in Security}.
His research interests include classical and quantum information theory and classical and quantum statistical inference.
\end{IEEEbiographynophoto}

\end{document}